\documentclass[journal,comsoc]{IEEEtran}

\usepackage[T1]{fontenc}
\usepackage{graphicx}
\usepackage{cite}
\usepackage{subfig}
\usepackage[dvipsnames,svgnames]{xcolor}
\usepackage{amssymb}
\usepackage{amsmath}
\usepackage{array}
\usepackage{mathtools}
\usepackage{scalerel}
 \usepackage{algorithm}
\usepackage{algorithmic}

\renewcommand{\IEEEQED}{\IEEEQEDopen}
\newtheorem{lemma}{Lemma}

\newtheorem{theorem}{Theorem}

\newtheorem{remark}{Remark}

\begin{document}

\title{Performance of Cell-Free Massive MIMO with Rician Fading and Phase Shifts}

\author{\"{O}zgecan~\"{O}zdogan,~\IEEEmembership{Student Member,~IEEE,}
	Emil Bj\"{o}rnson,~\IEEEmembership{Senior Member,~IEEE,}\\
	and~Jiayi~Zhang,~\IEEEmembership{Member,~IEEE}

\thanks{Manuscript received March 13, 2019; revised June 13, 2019; accepted August 3, 2019. This work was supported in part by ELLIIT, in part by the Swedish Research Council, in part by the National Natural Science Foundation of China (Grant Nos. 61601020 and U1834210), and in part by the Beijing Natural Science Foundation (Grant Nos. 4182049 and L171005). A preliminary version of this manuscript was presented at IEEE SPAWC 2019 \cite{Ozdogan2019}. The associate editor coordinating the review of this article and approving it for publication was R. Brown. (\emph{Corresponding author: \"{O}zgecan \"{O}zdogan.) }}
\thanks{\"{O}. \"{O}zdogan and E. Bj\"{o}rnson are with the Department of Electrical Engineering (ISY), Link\"oping University, Link\"oping SE-581 83, Sweden (e-mail: ozgecan.ozdogan@liu.se; emil.bjornson@liu.se).}
\thanks{J. Zhang is with the School of Electronic and Information Engineering, Beijing Jiaotong University, Beijing 100044, China (e-mail: jiayizhang@bjtu.edu.cn)}
\thanks{Color versions of one or more of the figures in this article are available
	online at http://ieeexplore.ieee.org. }
\thanks{Digital Object Identifier 10.1109/TWC.2019.2935434}}
%

\maketitle

\begin{abstract}
In this paper, we  study the uplink (UL) and downlink (DL) spectral efficiency (SE) of a cell-free massive multiple-input-multiple-output (MIMO) system  over Rician fading channels. The phase of the line-of-sight (LoS) path is modeled as a uniformly distributed random variable to take the phase-shifts due to mobility and phase noise into account. Considering the availability of  prior information at the access points (APs), the phase-aware minimum mean square error (MMSE), non-aware linear MMSE (LMMSE), and least-square (LS) estimators are derived. The MMSE estimator requires perfectly estimated phase knowledge whereas the LMMSE and LS are derived without it. 

In the UL, a two-layer decoding method is investigated in order to mitigate both coherent and non-coherent interference. Closed-form  UL SE expressions with phase-aware MMSE, LMMSE, and LS estimators are derived for maximum-ratio (MR) combining in the first layer and optimal large-scale fading decoding (LSFD) in the second layer. In the DL, two different transmission modes are studied: coherent and non-coherent. Closed-form DL SE expressions for both transmission modes with MR precoding are derived for the three estimators. Numerical results show that the LSFD improves the UL SE performance and coherent transmission mode performs much better than non-coherent transmission in the DL. Besides, the performance loss due to the lack of phase information depends on the pilot length and it is small when the pilot contamination is low.

\end{abstract}

\IEEEpeerreviewmaketitle

\section{Introduction}

Cell-free massive MIMO refers to a distributed MIMO system with a large number of APs that jointly serve a smaller number of user equipments (UEs)  \cite{Truong2013,Ngo2017,Nayebi2017a,Interdonato2018}. The APs cooperate via a fronthaul network \cite{Interdonato2018} to spatially multiplex the UEs on the same time-frequency resource, using network MIMO methods that only require locally obtained channel state information (CSI) \cite{Bjoernson2010}.

 In its canonical form, cell-free massive MIMO uses MR combining because of its low complexity. Due to the fact that MR combining cannot suppress the interference well, some APs receive more interference from other UEs than signal power from the desired UE. The LSFD method is proposed in \cite{Ashikhmin2012} and \cite{Adhikary2017}  to mitigate the interference for co-located massive MIMO systems. This method is generalized for more realistic spatially correlated Rayleigh fading channels with arbitrary first-layer decoders in \cite{vanChien2018,vanChien2019}. The two-layer decoding technique is first adapted to cell-free massive MIMO networks in \cite{Nayebi2016} for a Rayleigh fading scenario. In the concurrent paper \cite{Ngo2018}, the LSFD method is studied in a setup with Rician fading channels where the LoS phase is static.

Joint transmission from multiple APs can be either coherent (same data from all APs) or non-coherent (different data). Only the former has been considered in cell-free massive MIMO, but it requires that the APs are phase-synchronized. A synchronization method is outlined in \cite{Interdonato2018, Rogalin2014a, Perlman2015a} without validation.

In densely deployed systems, like cell-free massive MIMO, the channels typically consist of a combination of a semi-deterministic LoS path and small-scale fading caused by multipath propagation, which can be modeled as Rician fading \cite{Ozdogan2018b,Ngo2018}. A small change in the UE location may result in a significant phase-shift of the LoS component, but no change in amplitude. For instance, if the UE moves half a wavelength away from the AP, the phase of the channel response changes by $\pm\pi $. Similarly, hardware effects such as phase noise may create severe shift in the phase.  These effects are usually neglected in the analysis of Rician fading channels by assuming a LoS path with static phase. Especially in high mobility scenarios, the phase shift in LoS path may have a large impact on system performance. Recently, \cite{Dandrea2019} studied a cell-free network that supports both unmanned aerial vehicles (UAVs) and ground UEs where the channels between AP-UAV pairs have Rician distribution with uniformly distributed phase on the LoS paths.

Each AP needs to learn the channel statistics of each UE that it serves, as well as the statistics of the combined interfering signals, if Bayesian channel estimators are to be used. The large-scale fading coefficients can be estimated with a negligible overhead since the coefficients are deterministic \cite{EmilsBook}; several practical methods to estimate these coefficients using uplink pilots are  presented in \cite[Section IV]{Sanguinetti2019}. However, the phase-shifts are harder to estimate since they change as frequently as the small-scale fading. Depending on the availability of channel statistics, the phase-aware MMSE estimator which requires all prior information, LMMSE estimator with only the large-scale fading parameters, or LS estimator with no prior information can be utilized. In this paper, the specific technical contributions are as follows:
\begin{itemize}
	\item  We consider Rician fading channels between the APs and UEs, where the mean and variance are different for every AP-UE pair. Additionally, the phases of the LoS paths are modeled as independent and identically distributed (i.i.d.) random variables in each coherence block.
	
	\item We derive the phase-aware MMSE, LMMSE, and LS channel estimators and obtain their statistics. In the UL, using the estimates for MR combining in the first-layer, we compute closed-form  UL achievable SE expressions for two-layer decoding scheme. In the DL, we obtain closed-form  DL achievable SE expressions for both coherent  and non-coherent transmission.
\end{itemize}

The conference version of this paper \cite{Ozdogan2019} only considered the DL transmission and used the phase-aware MMSE and LMMSE estimators.

\textit{Reproducible Research}: All the simulation results can be reproduced using the Matlab code and data files available at: https://github.com/emilbjornson/rician-cell-free

\textit{Notation}: Lower and upper case bold letters are used for vectors and matrices.  The transpose and Hermitian transpose of a matrix $\mathbf{A}$ are written as $\mathbf{A}^T$ and $\mathbf{A}^H$. The superscript $(.)^*$ denotes the complex conjugate operation. The $M \times M $-dimensional matrix with the diagonal elements $d_1, d_2, \dots, d_M$ is denoted as $\mathrm{diag}\left(d_1, d_2, \dots, d_M \right) $. The diagonal elements of a matrix $\mathbf{D}$ are extracted  to a $M \times 1$ vector as $\mathrm{diag}(\mathbf{D})=[d_1, d_2, \dots, d_M]^T$. The notation  $\mathbf{X} =[x_{i,j} : i=1,\dots,M, j=1,\dots,N] $ denotes the  $M \times N$ matrix. The expectation of a random variable $X$ is denoted by $\mathbb{E}\left\lbrace X\right\rbrace$.  The expectations are taken with respect to all sources of randomness.

\section{System Model}

We consider a cell-free Massive MIMO system with $M$ APs and $K$ UEs. All APs and UEs are equipped with a single antenna. The multi-antenna AP case can be straightforwardly covered by treating each antenna as a separate AP, if it is assumed that there is no correlation between the small-scale fading coefficients (or phase-shifts) \cite{Mai2018}. However, for a  more realistic analysis, single-antenna results can be generalized to multiple antenna case by taking the spatial correlations between antennas into account. It will result in non-diagonal covariance matrices. The channels are assumed to be constant and frequency-flat in a coherence block of length $\tau_c$  samples (channel uses). The length of each coherence block is determined by the carrier frequency and external factors such as the propagation environment and UE mobility \cite{EmilsBook}. The channel $h_{m,k}$ between UE~$k$ and the AP~$m$ is modeled as
\begin{equation}\label{sec1:1}
h_{m,k} = \bar{h}_{m,k} e^{j\varphi_{m,k}} + g_{m,k},
\end{equation}
where $g_{m,k} \sim	\mathcal{N}_\mathbb{C}\left(0,  \beta_{m,k}\right) $, the mean $\bar{h}_{m,k} \geq 0$ represents the LoS component, and  $\varphi_{m,k}\sim 	\mathcal{U}\left[ -\pi,  \pi\right]$ is the phase-shift. The small-scale fading from non-LoS (NLoS) propagation has a variance $\beta_{m,k}$ that models the large-scale fading, including geometric pathloss and shadowing.  Note that \eqref{sec1:1} is a Rician fading model since $|h_{m,k}|$ is Rice distributed, but $h_{m,k}$ is not Gaussian distributed as in many prior works that neglected the phase shift. We assume that $h_{m,k}$ is an independent random variable for every  $m=1,\dots,M$, $k=1,\dots,K$ and the channel realization $h_{m,k}$ in different coherence blocks are i.i.d.

\begin{figure}[t!]
	\centering
	\includegraphics[scale=0.63]{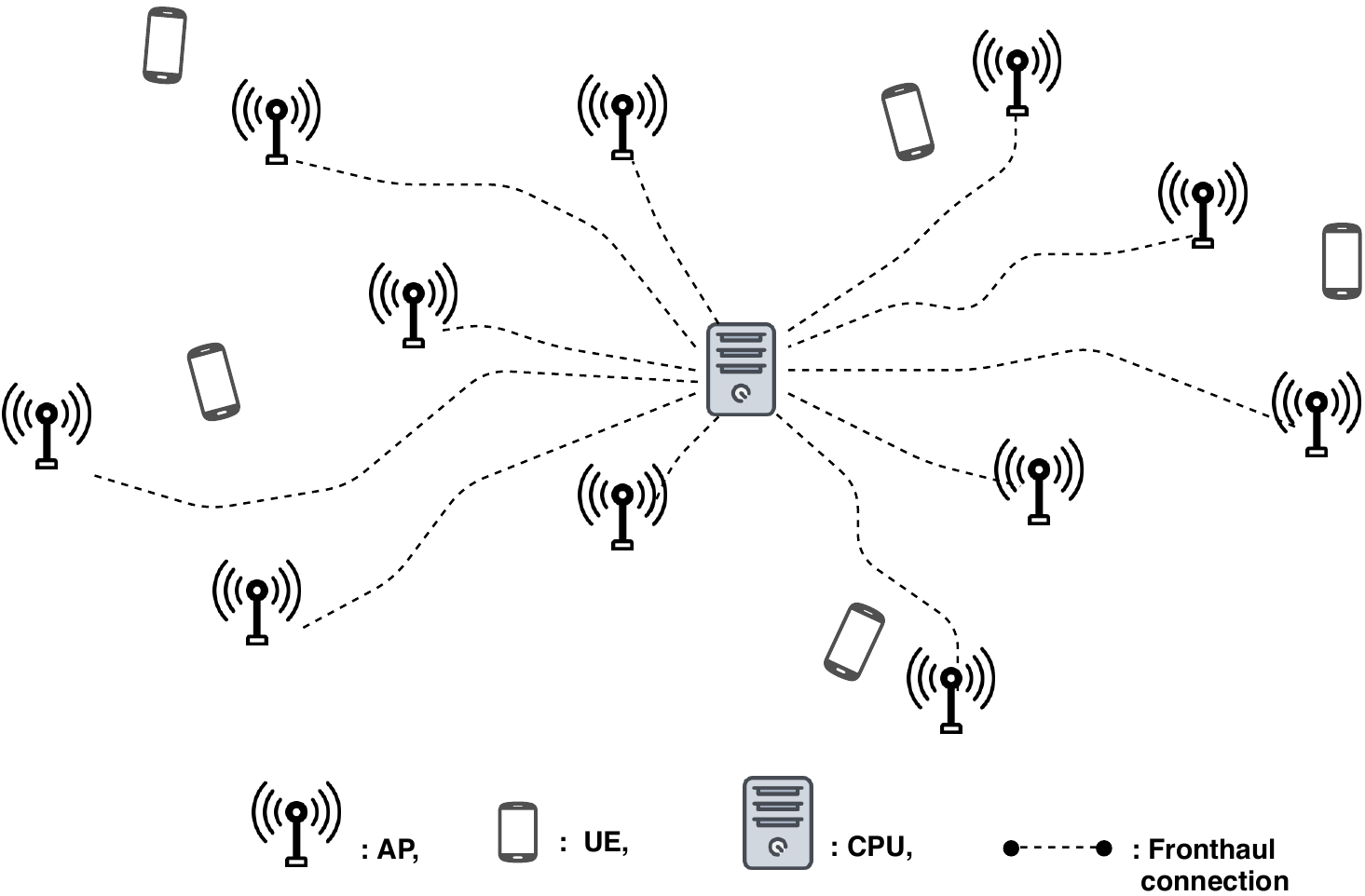}
	\caption{Illustration of a cell-free massive MIMO network.} \vspace{-5mm}
\end{figure}

All APs are connected to a central processing unit (CPU) via a fronthaul network that is error free. The system operates in time division duplex (TDD) mode and the uplink (UL) and DL channels are estimated by exploiting only UL pilot transmission and channel reciprocity.
\section{Uplink Channel Estimation}
\label{Section3}
In each coherence block, $\tau_p$ samples are reserved for UL pilot-based channel estimation, using a set of $\tau_p$ mutually orthogonal pilot sequences. The pilot sequence of UE~$k$ is denoted by $\boldsymbol{\phi}_k \in \mathbb{C}^{\tau_p \times 1}$ and satisfies $\| \boldsymbol{\phi}_k \|^2=\tau_p$.  It is scaled by $\sqrt{\hat{p}_k}$, with $\hat{p}_k$ being the  pilot power, and sent to the APs.  The received  signal $\mathbf{y}^{p, \mathrm{ap}}_m \in \mathbb{C}^{\tau_p \times 1}$  at  AP~$m$ is
\begin{equation}\label{se3:eq1}
\mathbf{y}^{p, \mathrm{ap}}_m = \sum_{k=1}^{K} \sqrt{\hat{p}_k} h_{m,k} \boldsymbol{\phi}_k + \mathbf{n}^p_m,
\end{equation}
where $\mathbf{n}^p_m  \sim \mathcal{N}_\mathbb{C}\left(\boldsymbol{0}_{\tau_p}, \sigma^2_\mathrm{ul} \mathbf{I}_{\tau_p}\right)  $ is additive noise. AP~$m$ computes an inner product between $\mathbf{y}^{p, \mathrm{ap}}_m$ and $\boldsymbol{\phi}_k $ to get sufficient statistics for estimation of $h_{m,k}$. This results in 
\begin{equation}\label{se3:eq2}
y^p_{m,k}=\boldsymbol{\phi}^H_k\mathbf{y}^{p, \mathrm{ap}}_m =  \sum_{l=1}^{K} \sqrt{\hat{p}_{l}} h_{m,l} \boldsymbol{\phi}_k^H\boldsymbol{\phi}_{l}  + \boldsymbol{\phi}_k^H\mathbf{n}^p_m .
\end{equation}

We assume that the number of UEs is large such that  $\tau_p < K $ and we define the set $\mathcal{P}_k$ of UEs that use same pilot sequence as UE~$k$, including itself. Since UEs with different pilots have orthogonal pilots,  $y^p_{m,k}$ in \eqref{se3:eq2} can be rewritten as
\begin{equation}\label{se3:eq3}
y^p_{m,k}= \sqrt{\hat{p}_{k}} \tau_p h_{m,k} + \sum_{l\in \mathcal{P}_k \backslash \{k\}} \sqrt{\hat{p}_{l}}\tau_p h_{m,l} + \boldsymbol{\phi}^H_k\mathbf{n}^p_m,
\end{equation}
where $\boldsymbol{\phi}^H_k\mathbf{n}^p_m  \sim \mathcal{N}_\mathbb{C}(0, \sigma^2_{\mathrm{ul}} \tau_p )$. Based on this signal, we will derive three different channel estimators, and characterize their statistics. In particular, we will explore the impact of the phase-shift in the LoS paths by considering the analytically tractable cases where the phase-shifts are either perfectly known or fully unknown.  In co-located MIMO systems, the phase can be estimated by employing classical algorithms such as estimation of signal parameters via rotational invariance techniques (ESPRIT) \cite{Paulraj1985} and multiple signal classification (MUSIC) \cite{Schmidt1981}. However, these methods either require multiple antennas and/or long sample sequences (snapshots) which are not available in our system setup. In a low mobility scenario in which the APs are equipped with multiple antennas, the phase-shifts can be estimated quite well, but the estimates will anyway be imperfect since the estimation resources are practically limited. Nevertheless, we will consider a phase-aware MMSE estimator that serves as an upper bound on the achievable performance. The practical performance will be somewhere in between the cases with perfectly known and fully unknown phases, which are considered below. We consider two estimators of the latter kind; one Bayesian LMMSE and one classical LS estimator.

\subsection{Phase-aware MMSE Channel Estimator}
If the channel statistics $\bar{h}_{m,k} $, $\beta_{m,k}$ are available and the phase $\varphi_{m,k}$ is somehow perfectly known at AP~$m$, for every $k$, we  can derive the MMSE estimator of $h_{m,k}$ as \cite{KayBookESt}
\begin{equation}\label{se3:eq4}
\hat{h}^{\mathrm{mmse}}_{m,k} = \bar{h}_{m,k} e^{j\varphi_{m,k}}   + \frac{ \sqrt{\hat{p}_{k}} \beta_{m,k}\left( y^p_{m,k} -\bar{y}^p_{m,k}\right) }{\lambda_{m,k}},
\end{equation}
where  $\bar{y}^p_{m,k} = \sum_{l\in \mathcal{P}_k } \sqrt{\hat{p}_{l}} \tau_p  \bar{h}_{m,l} e^{j\varphi_{m,l}}$ and $\lambda_{m,k} =\sum_{l\in \mathcal{P}_k } \hat{p}_{l} \tau_p \beta_{m,{l}} + \sigma^2_{\mathrm{ul}}$. Note that $\hat{h}^{\mathrm{mmse}}_{m,k} $ is a random variable and \eqref{se3:eq4} is a single realization for a specific coherence block. The $y^p_{m,k}$, $\bar{y}^p_{m,k}$ and $\varphi_{m,k}$ change in every coherence block whereas $\bar{h}_{m,k}$ and $\beta_{m,k}$ remain constant for a longer period of time. AP~$m$ estimates channels to all the UEs. The estimation error $\tilde{h}^{\mathrm{mmse}}_{m,k}= {{h}}_{m,k} - \hat{h}^{\mathrm{mmse}}_{m,k}$ has zero mean and the variance
 \begin{equation}\label{se3:eq6}
 c_{m,k}=  \beta_{m,k} - \frac{\hat{p}_{k}\tau_p \beta^2_{m,k}}{\lambda_{m,k}} .
 \end{equation}
 The mean-squared error is $\mathrm{MSE}= \mathbb{E}\lbrace |  {h}_{m,k} - \hat{h}^{\mathrm{mmse}}_{m,k} |^2 \rbrace  =c_{m,k}$. The MMSE estimate  $\hat{h}^{\mathrm{mmse}}_{m,k}$ and  the estimation error $\tilde{h}^{\mathrm{mmse}}_{m,k} $ satisfy
 \begin{align}\label{se3:eq7}
 &\mathbb{E}\left\lbrace \hat{h}^{\mathrm{mmse}}_{m,k} \middle| \varphi_{m,k} \right\rbrace = \bar{h}_{m,k} e^{j\varphi_{m,k}} ,\\
 & \mathrm{Var}\left\lbrace \hat{h}^{\mathrm{mmse}}_{m,k} \middle| \varphi_{m,k} \right\rbrace = \beta_{m,k} -  c_{m,k},\\
 & \mathbb{E}\left\lbrace \tilde{h}^{\mathrm{mmse}}_{m,k}  \right\rbrace = 0 , \ \ \mathrm{Var}\left\lbrace \tilde{h}^{\mathrm{mmse}}_{m,k} \right\rbrace = c_{m,k},
  \end{align}
where $\hat{h}^{\mathrm{mmse}}_{m,k}$  is not Gaussian distributed. Also,  $\hat{h}^{\mathrm{mmse}}_{m,k}$ and $\tilde{h}^{\mathrm{mmse}}_{m,k}$ are uncorrelated but not independent random variables.

The APs do not share these channel estimates with each other, or the CPU, but they are used to perform MR  processing distributively  as in \cite{Ngo2017, Nayebi2015,Nayebi2017a}. Nevertheless, the collection of channel estimates of UE~$k$ from all APs can be written in vector form as
\begin{equation}
\hat{\mathbf{h}}^{\mathrm{mmse}}_k = \boldsymbol{\Phi}_k  \bar{\mathbf{h}}_k + \sqrt{\hat{p}_{k}} \mathbf{R}_k \boldsymbol{\Lambda}_k(\mathbf{y}^p_k - \bar{\mathbf{y}}^p_k),
\end{equation}
where $\bar{\mathbf{h}}_k = [\bar{h}_{1,k}, \dots, \bar{h}_{M,k} ]^T $, $\boldsymbol{\Phi}_k = \mathrm{diag}\left(e^{j\varphi_{1,k}},\dots,e^{j\varphi_{M,k}} \right) $,  $\mathbf{y}^p_k =[y^p_{1,k}, \dots, y^p_{M,k}]^T$, $\bar{\mathbf{y}}^p_k = \left[  \bar{y}^p_{1,k}, \dots, \bar{y}^p_{M,k} \right] ^T  $, $\boldsymbol{\Lambda}_k = \mathrm{diag}\left( \lambda_{1,k},\dots,\lambda_{M,k}\right)^{-1} $, and $\mathbf{R}_k=\mathrm{diag}(\beta_{1,k}, \ldots, \beta_{M,k} )$.

The estimation error $\tilde{\mathbf{h}}^{\mathrm{mmse}}_k= {\mathbf{h}}_k - \hat{\mathbf{h}}^{\mathrm{mmse}}_k$ has zero mean and the covariance matrix 
	\begin{equation}
	\mathbf{C}_{k}=  \mathbf{R}_{k} - \hat{p}_{k} \tau_p \boldsymbol{\Omega}_k,
	\end{equation}
where $ \boldsymbol{\Omega}_k =\mathbf{R}_{k} \boldsymbol{\Lambda}_{k} \mathbf{R}_{k}.$	The mean-squared error is $\mathrm{MSE}= \mathbb{E}\lbrace \|  {\mathbf{h}}_k - \hat{\mathbf{h}}^{\mathrm{mmse}}_k\|^2 \rbrace  =\mathrm{tr}( \mathbf{C}_{k} ) $. The MMSE estimate  $\hat{\mathbf{h}}^{\mathrm{mmse}}_k $ and the estimation error $\tilde{\mathbf{h}}^{\mathrm{mmse}}_k $ satisfy
\begin{align}
	&\mathbb{E}\left\lbrace \hat{\mathbf{h}}^{\mathrm{mmse}}_k \middle| \boldsymbol{\Phi}_k  \bar{\mathbf{h}}_k \right\rbrace =\boldsymbol{\Phi}_k  \bar{\mathbf{h}}_k, \\
	  &	\mathrm{Cov}\left\lbrace \hat{\mathbf{h}}^{\mathrm{mmse}}_k \middle| \boldsymbol{\Phi}_k  \bar{\mathbf{h}}_k \right\rbrace =\mathbf{R}_{k}  - \mathbf{C}_{k},\\
	&\tilde{\mathbf{h}}^{\mathrm{mmse}}_k \sim \mathcal{N}_\mathbb{C}\left( \mathbf{0}_M, \mathbf{C}_{k} \right).
	\end{align}

\subsection{LMMSE Channel Estimator }
\label{Section3B}
If the channel statistics $\bar{h}_{m,k} $, $\beta_{m,k}$  are available  but the phase $\varphi_{m,k}$   is completely unknown at AP~$m$, for every $k$, we  can derive the LMMSE estimator of $h_{m,k}$ as
\begin{equation} \label{lmmse:eq1}
\hat{h}^{\mathrm{lmmse}}_{m,k} =  \frac{\sqrt{\hat{p}_{k}} \beta'_{m,k}  y^p_{m,k} }{\lambda'_{m,k}} ,   
\end{equation}
where $ \beta'_{m,k} = \beta_{m,k}  + \bar{h}^2_{m,k}$ and $\lambda'_{m,k} =\sum_{l\in \mathcal{P}_k } \hat{p}_{l} \tau_p ( \beta_{m,l} + \bar{h}^2_{m,l} ) + \sigma^2_{\mathrm{ul}}$. 
The estimation error $\tilde{h}^{\mathrm{lmmse}}_{m,k}= {{h}}_{m,k} - \hat{h}^{\mathrm{lmmse}}_{m,k}$ has zero mean and the variance
\begin{equation}
c'_{m,k}=  \beta'_{m,k} - \frac{\hat{p}_{k}\tau_p (\beta'_{m,k})^2}{\lambda'_{m,k}}.
\end{equation}
The mean-squared error is $\mathrm{MSE}= \mathbb{E}\lbrace |  {h}_{m,k} - \hat{h}^{\mathrm{lmmse}}_{m,k} |^2 \rbrace  =c'_{m,k}$. The LMMSE estimate  $\hat{h}^{\mathrm{lmmse}}_{m,k}$ and the estimation error $\tilde{h}^{\mathrm{lmmse}}_{m,k} $ are uncorrelated random variables and satisfy
\begin{align}
&\mathbb{E}\left\lbrace \hat{h}^{\mathrm{lmmse}}_{m,k} \right\rbrace =0, \ \ \mathrm{Var}\left\lbrace \hat{h}^{\mathrm{lmmse}}_{m,k} \right\rbrace =\beta'_{m,k}-  c'_{m,k} , \\
&\mathbb{E}\left\lbrace \tilde{h}^{\mathrm{lmmse}}_{m,k} \right\rbrace =0, \ \ \mathrm{Var}\left\lbrace \tilde{h}^{\mathrm{lmmse}}_{m,k} \right\rbrace = c'_{m,k} .
\end{align}

A detailed derivation is given in Appendix \ref{AppendixLMMSE}. We can write \eqref{lmmse:eq1} in vector form as
\begin{equation}
\hat{\mathbf{h}}^{\mathrm{lmmse}}_k =  \sqrt{\hat{p}_{k} } \mathbf{R}'_k \boldsymbol{\Lambda}'_k\mathbf{y}^p_k,
\end{equation}
where  $\mathbf{R}'_k=\mathrm{diag}(\beta'_{1,k} \ldots, \beta'_{M,k} )$,  $\boldsymbol{\Lambda}'_k = \mathrm{diag}( \lambda'_{1,k},  \dots ,\lambda'_{M,k})^{-1}$. The estimation error $\tilde{\mathbf{h}}^{\mathrm{lmmse}}_k={\mathbf{h}}_k - \hat{\mathbf{h}}^{\mathrm{lmmse}}_k$ has zero mean and the covariance matrix 
\begin{equation}
\mathbf{C}'_{k}=  \mathbf{R}'_{k} - \hat{p}_{k} \tau_p \boldsymbol{\Omega}'_k,
\end{equation}
where $\boldsymbol{\Omega}'_k =\mathbf{R}'_{k} \boldsymbol{\Lambda}'_{k} \mathbf{R}'_{k}$.	The mean-squared error is $\mathrm{MSE}= \mathbb{E}\lbrace \|  {\mathbf{h}}_k - \hat{\mathbf{h}}^{\mathrm{lmmse}}_k\|^2 \rbrace  =\mathrm{tr}( \mathbf{C}'_{k} ) $. The MMSE estimate  $\hat{\mathbf{h}}^{\mathrm{lmmse}}_k $ and the estimation error $\tilde{\mathbf{h}}^{\mathrm{lmmse}}_k $ are uncorrelated random variables and satisfy
\begin{align}  
&\mathbb{E}\left\lbrace \hat{\mathbf{h}}^{\mathrm{lmmse}}_k \right\rbrace  = \mathbf{0}_M, \ \ \mathrm{Cov}\left\lbrace \hat{\mathbf{h}}^{\mathrm{lmmse}}_k \right\rbrace = \mathbf{R}'_{k} - \mathbf{C}'_{k} ,\\
&\mathbb{E}\left\lbrace 	\tilde{\mathbf{h}}^{\mathrm{lmmse}}_k\right\rbrace =  \mathbf{0}_M, \ \  \mathrm{Cov}\left\lbrace \tilde{\mathbf{h}}^{\mathrm{lmmse}}_k \right\rbrace= \mathbf{C}'_{k} .
\end{align}

\subsection{LS Channel Estimator with No Prior Information}
If AP $m$ has no prior information regarding the phase or statistics, then we can derive the non-Bayesian LS estimator  that minimizes $| y^p_{m,k} - \sqrt{p_{k}} \tau_p h_{m,k}| ^2 $, which is
\begin{equation}
\hat{h}^{\mathrm{ls}}_{m,k} = \frac{1}{\sqrt{p_{k}} \tau_p} y^p_{m,k}.
\end{equation}

The LS estimator and estimation error satisfy
 \begin{align}
&\mathbb{E}\left\lbrace \hat{h}^{\mathrm{ls}}_{m,k} \right\rbrace =0, \ \ \mathrm{Var}\left\lbrace \hat{h}^{\mathrm{ls}}_{m,k} \right\rbrace = \frac{1}{\hat{p}_k \tau_p}\lambda'_{m,k},\\
&\mathbb{E}\left\lbrace \tilde{h}^{\mathrm{ls}}_{m,k} \right\rbrace =0, \ \ \mathrm{Var}\left\lbrace \tilde{h}^{\mathrm{ls}}_{m,k} \right\rbrace = \frac{1}{\hat{p}_k \tau_p}\lambda'_{m,k} -\beta'_{m,k}.
  \end{align}

In vector form, we have $\hat{\mathbf{h}}^{\mathrm{ls}}_k = \frac{1}{\sqrt{p_{k}} \tau_p} \mathbf{y}^p_k$ for each user $k$. The LS estimator and estimation error satisfy
\begin{align}
&\mathbb{E}\left\lbrace \hat{\mathbf{h}}^{\mathrm{ls}}_k \right\rbrace = \mathbf{0}_M, \ \ 	\mathrm{Cov}\left\lbrace \hat{\mathbf{h}}^{\mathrm{ls}}_k \right\rbrace = \frac{1}{\hat{p}_k \tau_p}(\boldsymbol{\Lambda}'_{k})^{-1}, \\
&\mathbb{E}\left\lbrace \tilde{\mathbf{h}}^{\mathrm{ls}}_k\right\rbrace = \mathbf{0}_M, \ \ 	\mathrm{Cov}\left\lbrace \tilde{\mathbf{h}}^{\mathrm{ls}}_k \right\rbrace =  \frac{1}{\hat{p}_k \tau_p} (\boldsymbol{\Lambda}'_{k})^{-1} \!-\mathbf{R}'_{k} .
\end{align}
A detailed derivation is given in Appendix \ref{AppendixLS}.

\section{Uplink Data Transmission}
Let $\tau_u$ be the number of UL data symbols per coherence block, where $\tau_u \leq \tau_c - \tau_p$. The received signal at  AP~$m$ is 
\begin{equation}
y^\mathrm{ul}_{m} = \sum_{k=1}^{K}  h_{m,k} s_k + n^{\mathrm{ul}}_m,
\end{equation}
where  $ s_k \sim \mathcal{N}_\mathbb{C}\left( 0, p_k \right) $ is the UL signal with power $p_k =\mathbb{E}\left\lbrace|s_k|^2 \right\rbrace $ and $n^{\mathrm{ul}}_m \sim \mathcal{N}_\mathbb{C}(0,\sigma^2_{\mathrm{ul}})$ is additive receiver noise.  AP~$m$ selects the receiver combining scalar as $v_{m,k}= \hat{h}_{m,k} \in \lbrace \hat{h}^{\mathrm{mmse}}_{m,k},\hat{h}^{\mathrm{lmmse}}_{m,k}, \hat{h}^{\mathrm{ls}}_{m,k}\rbrace  $ to locally detect the signal of  UE~$k$. Then, the combined received signal at AP~$m$ is 
\begin{equation}\label{sec4:1}
\tilde{s}_{m,k} = v^*_{m,k} y^\mathrm{ul}_{m} =  {v}^*_{m,k}h_{m,k} s_{k} +  \displaystyle \sum_{\substack{l=1\\{l\neq k}}}^{K}  {v}^*_{m,k}h_{m,l} s_{l} + {v}^*_{m,k} n^{\mathrm{ul}}_m.
\end{equation}

A second layer decoding is performed to mitigate the inter-user interference using  LSFD coefficients.  The single-layer decoded signal in \eqref{sec4:1} is sent to the CPU and it performs the second layer decoding by computing the LSFD weighted signal as
\begin{align}\label{sec4:2}
\hat{s}_k & = \sum_{m=1}^{M} \alpha^*_{m,k} \tilde{s}_{m,k}= \sum_{m=1}^{M} \alpha^*_{m,k} v^*_{m,k} h_{m,k} s_{k} \nonumber \\
&+  \sum_{m=1}^{M} \displaystyle  \sum_{\substack{l=1\\{l\neq k}}}^{K}  \alpha^*_{m,k} {v}^*_{m,k}h_{m,l} s_{l} + \sum_{m=1}^{M} \alpha^*_{m,k} {v}^*_{m,k} n^{\mathrm{ul}}_m. 
\end{align}
where $\alpha_{m,k}$ is the complex LSFD coefficient for AP~$m$ and UE~$k$. The weighting operation reduces the inter-user interference by balancing the received signals from all APs. The CPU computes the LSFD weights based on the large-scale fading coefficients that vary slowly compared to the small-scale coefficients.  Recall that the value of large-scale fading between an AP-UE pair depends mainly on the distance and shadowing. Thus, it gives us a valuable information about the likely quality of the received signal. For instance, if the UE is far away from the AP, potentially the received signal will contain a considerable amount of interference and noise whereas the desired signal is weak. Then, the corresponding LSFD coefficient for this signal will be small to reduce the risk of adding and amplifying more interference into the detection. 
Note that assigning $\alpha_{m,k}=1$ for all APs and UEs corresponds to the single-layer decoding considered in \cite{Ngo2017,Nayebi2017a},\cite{Buzzi2017,Bashar2018}. Similarly, setting $\alpha_{m,k}=0$ means that AP $m$ is not serving UE $k$, which can be utilized to handle practical cases where each UE is served by a subset of all the APs; see \cite{Buzzi2017,Bjornson2019a} for details. The $\alpha_{m,k}$ is deterministic since it only depends on deterministic large-scale fading coefficients.

	\begin{figure}[t!]
	\centering
	\includegraphics[scale=0.58]{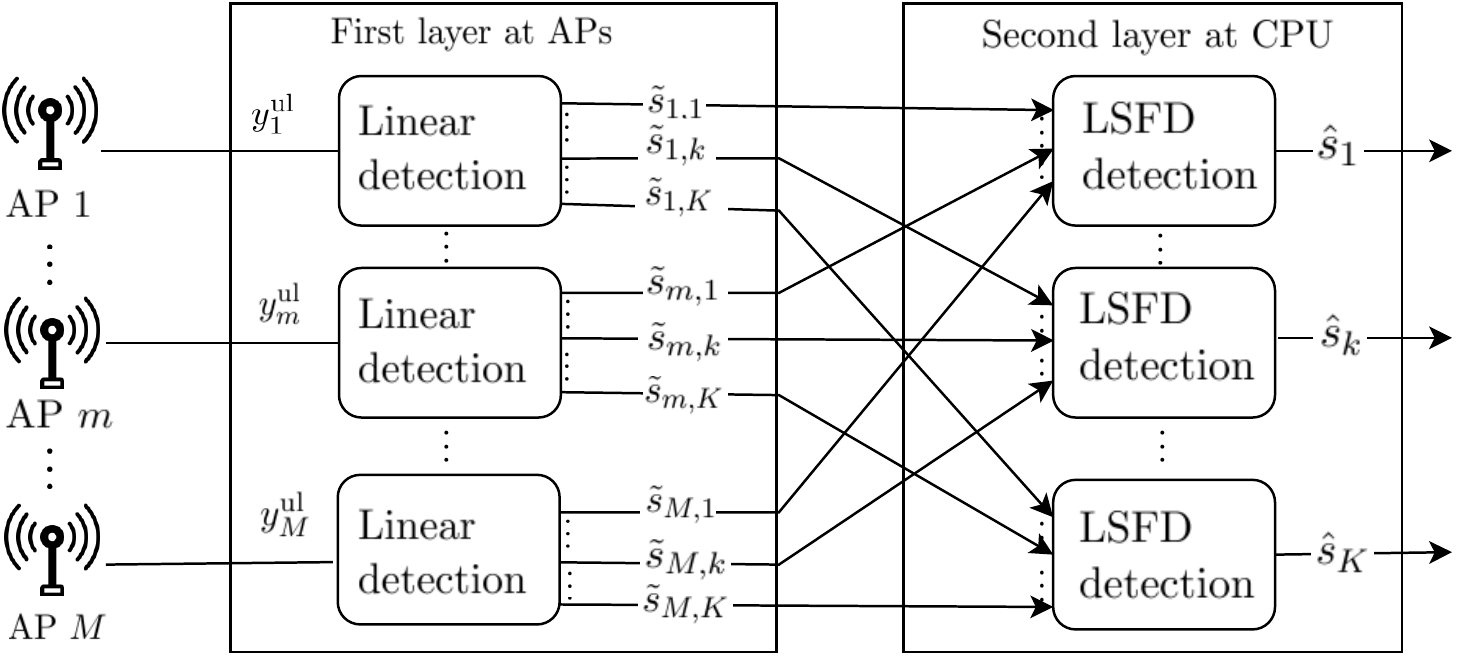}
	\caption{Two-layer decoding technique.} 
\end{figure} 

 Based on \eqref{sec4:2}, the UL ergodic channel capacity of UE~$k$ can be lower bounded using the use-and-then-forget (UatF) bound  \cite[Th.~4.4]{EmilsBook}. The expected values of the desired signal, inter-user interference and noise related terms at all APs  for UE $k$ can be computed respectively as
\begin{align}
&\left|  \sum_{m=1}^{M} \alpha^*_{m,k}\mathbb{E}\left\lbrace {v}^*_{m,k}  {h}_{m,k}\right\rbrace \right| ^2 =	\left| \mathbb{E}\left\lbrace \mathbf{v}^H_k \mathbf{A}^H_k {\mathbf{h}}_k\right\rbrace \right| ^2 \\
& \mathbb{E}\left\lbrace \left|  \sum_{m=1}^{M} \alpha^*_{m,k} {v}^*_{m,k}  {h}_{m,l} \right| ^2 \right\rbrace =	\mathbb{E}\left\lbrace \left| \mathbf{v}^H_k \mathbf{A}^H_k {\mathbf{h}}_{l} \right|^2 \right\rbrace  \\
&\mathbb{E}\left\lbrace  \left| \sum_{m=1}^{M}	 \alpha^*_{m,k} {v}^*_{m,k} \right|^2  \right\rbrace  = \mathbb{E}\left\lbrace\|\mathbf{A}_k\mathbf{v}_k\|^2\right\rbrace  
\end{align}
where $\mathbf{A}_k = \mathrm{diag}\left( \alpha_{1,k}, \dots, \alpha_{M,k} \right) $ and  ${\mathbf{v}}_k =[v_{1,k}, \dots, v_{M,k} ]^T$. Then, a lower bound on the UL ergodic SE with the LSFD receiver is
 \begin{equation}\label{uatf_SE}
 \mathrm{SE}^\mathrm{ul}_k=\frac{\tau_u}{\tau_c}\log_2(1 + \gamma^{\mathrm{ul}}_{k} ),
 \end{equation}
where the effective SINR $\gamma^{\mathrm{ul}}_{k}$ is
\begin{equation}\label{SINR_UL}
\gamma^{\mathrm{ul}}_{k}=\frac{p_k\left| \mathbb{E}\left\lbrace \mathbf{v}^H_k \mathbf{A}^H_k {\mathbf{h}}_k\right\rbrace \right| ^2}{\sum\limits_{l=1}^{K} p_{l} \mathbb{E}\left\lbrace \left| \mathbf{v}^H_k \mathbf{A}^H_k{\mathbf{h}}_{l} \right|^2 \right\rbrace  - p_k\left| \mathbb{E}\left\lbrace \mathbf{v}^H_k \mathbf{A}^H_k {\mathbf{h}}_k\right\rbrace \right| ^2 + \sigma^2_{\mathrm{ul}} \mathbb{E}\left\lbrace\|\mathbf{A}_k \mathbf{v}_k \|^2\right\rbrace  }.
\end{equation}
and the expectations are with respect to all sources of randomness. This bound is referred as UatF bound since the channel estimates are used for combining locally at the APs and then they are unknown at the CPU. The effective SINR of UE $k$ in \eqref{SINR_UL} with the LSFD receiver can be rewritten as
\begin{equation}\label{uatf_SINR}
\gamma^{\mathrm{ul}}_{k}=	\frac{p_k\left| \mathbf{a}^H_{k} \mathbf{b}_{k,k} \right| ^2}{\mathbf{a}^H_{k} \left(  \sum\limits_{l=1}^{K} p_{l} \boldsymbol{\Gamma}^\textsuperscript{(1)}_{k,l} - p_k \mathbf{b}_{k,k} \mathbf{b}^H_{k,k} +  \sigma^2_{\mathrm{ul}} \boldsymbol{\Gamma}^\textsuperscript{(2)}_{k} \right) \mathbf{a}_{k}  },
\end{equation}
where 
\begin{align}
&\mathbf{a}_{k} = \left[ \alpha_{1,k}, \dots,\alpha_{M,k},\right] ^T \in \mathbb{C}^{M \times 1}, \\
&\mathbf{b}_{k,l} = \left[ \mathbb{E}\left\lbrace {v}^*_{1,k}  {h}_{1,l}\right\rbrace, \dots, \mathbb{E}\left\lbrace {v}^*_{M,k}  {h}_{M,l}\right\rbrace\right] ^T \in \mathbb{C}^{M \times 1},\\
&\boldsymbol{\Gamma}^\textsuperscript{(1)}_{k,l} =  \left[ \mathbb{E}\left\lbrace {v}_{m,k}  {h}^*_{m,l} {v}^*_{n,k}  {h}_{n,l}\right\rbrace:\forall m,n\right] \in \mathbb{C}^{M \times M},\\
&\boldsymbol{\Gamma}^\textsuperscript{(2)}_{k} = \mathrm{diag}\left( \mathbb{E}\left\lbrace  |  {v}_{1,k}   | ^2 \right\rbrace , \dots,  \mathbb{E}\left\lbrace  |  {v}_{M,k}  | ^2 \right\rbrace\right) \in  \mathbb{R}^{M \times M}.
\end{align}

In order to maximize the SE of UE $k$, we need to select the LSFD vector that maximizes \eqref{uatf_SINR} as shown in the following Lemma.
\begin{lemma} \label{lemma:LSFD-vector}
For a given set of pilot and data power coefficients, the SE of UE $k$ is	
	\begin{align} \label{LSFD_SE}
		&\mathrm{SE}^\mathrm{ul}_k = \frac{\tau_u}{\tau_c}\\ 
		&\times \log_2\left( 1 + p_k \mathbf{b}^H_{k,k} \left(  \sum\limits_{l=1}^{K} p_{l} \boldsymbol{\Gamma}^\textsuperscript{(1)}_{k,l} - p_k \mathbf{b}_{k,k} \mathbf{b}^H_{k,k} +  \sigma^2_{\mathrm{ul}} \boldsymbol{\Gamma}^\textsuperscript{(2)}_{k} \right)^{-1} \mathbf{b}_{k,k} \right)  \nonumber
	\end{align}
with the LSFD vector
	\begin{equation}\label{LSFD_vec}
	\mathbf{a}_k =\left(  \sum\limits_{l=1}^{K} p_{l} \boldsymbol{\Gamma}^\textsuperscript{(1)}_{k,l} - p_k \mathbf{b}_{k,k} \mathbf{b}^H_{k,k} +  \sigma^2_{\mathrm{ul}} \boldsymbol{\Gamma}^\textsuperscript{(2)}_{k} \right)^{-1} \mathbf{b}_{k,k} 
	\end{equation}

\end{lemma}
\begin{IEEEproof}
	The SINR expression in \eqref{uatf_SINR} is a generalized Rayleigh quotient with respect to $\mathbf{a}_{k}$ and we apply \cite[Lemma B.10]{EmilsBook} to obtain the maximizing $\mathbf{a}_{k}$ in \eqref{LSFD_vec}. Inserting this vector into \eqref{uatf_SE} gives the SE in \eqref{LSFD_SE}.
\end{IEEEproof}

The UL SE of two-layer decoding scheme for Rayleigh fading channels with MMSE estimator when using MR combining is presented in \cite{Nayebi2016}. Lemma~\ref{lemma:LSFD-vector} generalizes this results to cover arbitrary channel distributions and beamforming schemes. Every choice of channel model and beamforming will affect the values of $\mathbf{b}_{k,l}$, $\boldsymbol{\Gamma}^\textsuperscript{(1)}_{k,l}$, and $\boldsymbol{\Gamma}^\textsuperscript{(2)}_{k}$, but the solution structure in Lemma~\ref{lemma:LSFD-vector} remains the same.
In the following subsections,  $\gamma^{\mathrm{ul}}_{k}$ is computed for MR combining when using the phase-aware MMSE, LMMSE, and LS channel estimators derived in Section \ref{Section3}.

\subsection{Uplink Spectral Efficiency with the Phase-aware MMSE Estimator }

\begin{theorem}
		\begin{figure*}[h]
		\normalsize
		
		\begin{equation}\label{wide1}
		\gamma^{\mathrm{ul,mmse}}_k =\frac{p_k |  \mathrm{tr}\left(\mathbf{A}_k \mathbf{Z}_{k}\right)  |^2}{\displaystyle \sum_{l=1}^{K}  p_{l} \mathrm{tr}\left( \mathbf{A}^H_k \mathbf{R}'_{l} \mathbf{Z}_{k} \mathbf{A}_k \right)    + \displaystyle \sum_{l\in \mathcal{P}_k \backslash \{k\}} p_{l} \hat{p}_{k} \hat{p}_{l}  \tau^2_p |\mathrm{tr}\left(   \mathbf{A}_k\mathbf{R}_{l}  \boldsymbol{\Lambda}_{k} \mathbf{R}_{k}\right) |^2 + \mathrm{tr}\left( \mathbf{A}^H_k (\sigma^2_{\mathrm{ul}}\mathbf{Z}_{k} -p_k \mathbf{L}^2_{k})\mathbf{A}_k \right)},
		\end{equation}
		\hrulefill	
	\end{figure*}
	If MR combining with ${v}_{m,k} = \hat{h}^{\mathrm{mmse}}_{m,k} $ is used based on the phase-aware MMSE estimator, then $
	\mathrm{SE}^\mathrm{ul}_k = \frac{\tau_u}{\tau_c}\log_2\left( 1 + \gamma^{\mathrm{ul,mmse}}_k \right)$ with $\gamma^{\mathrm{ul,mmse}}_k $ in \eqref{wide1}, at the top of next page, where	$\mathbf{Z}_{k}  =  \hat{p}_{k} \tau_p \boldsymbol{\Omega}_k +\mathbf{L}_k $ and $\mathbf{L}_k = \mathrm{diag}\left( \bar{h}^2_{1,k}, \dots, \bar{h}^2_{M,k} \right)$. In Rayleigh quotient form, the SINR of UE $k$ is
\begin{equation} \label{gamma_ulmmse}
\gamma^{\mathrm{ul,mmse}}_k =\frac{p_{k}  \mathbf{a}^H_k\mathbf{b}_{k}\mathbf{b}^H_{k}\mathbf{a}_k}{\mathbf{a}^H_k \boldsymbol{\Gamma}_k \mathbf{a}_k\ },
\end{equation}
where	$
	\boldsymbol{\Gamma}_k= \sum\limits_{l=1}^{K} p_{l} \mathbf{R}'_{l} \mathbf{Z}_{k} 
	+ \sum_{l\in \mathcal{P}_k \backslash \{k\}} p_{l} \hat{p}_{k} \hat{p}_{l}  \tau^2_p  \mathbf{z}_{k,l} \mathbf{z}^H_{k,l}  - p_k\mathbf{L}^2_{k} + \sigma^2_{\mathrm{ul}}\mathbf{Z}_{k},$
$\mathbf{b}_{k} =\mathrm{diag}\left( \mathbf{Z}_{k} \right) $, and  $\mathbf{z}_{k,l}  = \mathrm{diag}\left( \mathbf{R}_{l} \boldsymbol{\Lambda}_{k} \mathbf{R}_{k} \right) $. The SE in \eqref{LSFD_SE} with the maximizing LSFD receiver vector $\mathbf{a}_k =\boldsymbol{\Gamma}_k^{-1} \mathbf{b}_{k} $ is
\begin{equation} \label{LSFD_SE_mmse}
\mathrm{SE}^\mathrm{ul}_k=\frac{\tau_u}{\tau_c}\log_2\left( 1 + p_k \mathbf{b}^H_{k} \boldsymbol{\Gamma}_k^{-1}\mathbf{b}_{k} \right).
\end{equation}

\end{theorem}
\begin{IEEEproof}
	The proof is given in Appendix  \ref{AppendixUL_MMSE}.
\end{IEEEproof}

\subsection{Uplink Spectral Efficiency with the LMMSE Estimator }
\begin{theorem}
		\begin{figure*}[h]
		\normalsize
		
		\begin{equation}\label{lmmse:1}
		\gamma^\mathrm{ul,lmmse}_k= \frac{  p_k\hat{p}^2_k \tau^2_p | \mathrm{tr}\left(\mathbf{A}_{k}\boldsymbol{\Omega}'_{k} \right) |^2}{\displaystyle \sum_{l=1}^{K}  p_{l}  \mathbf{T}^\textsuperscript{lmmse,(1)}_{k,l}  + \sum_{l\in \mathcal{P}_k  }  {p}_{l}   \mathbf{T}^\textsuperscript{lmmse,(2)}_{k,l} -  p_k\hat{p}^2_k \tau^2_p | \mathrm{tr}\left(\mathbf{A}_{k}\boldsymbol{\Omega}'_{k} \right) |^2+ \sigma^2_{\mathrm{ul}}  \hat{p}_{k}  \tau_p \mathrm{tr}\left(\mathbf{A}^H_{k}\boldsymbol{\Omega}'_{k}\mathbf{A}_{k} \right)},
		\end{equation}
		\setcounter{equation}{48}
		\begin{equation} \label{gamma_ulmmse_mmsewp}
		\gamma^\mathrm{ul,lmmse}_k = \frac{p_{k}    \mathbf{a}^H_k\mathbf{b}^{\mathrm{lmmse}}_{k} (\mathbf{b}^{\mathrm{lmmse}}_{k} )^H \mathbf{a}_k}{\mathbf{a}^H_k\left( \sum_{l=1}^{K}   p_{l}\boldsymbol{\Gamma}^{\textsuperscript{lmmse,(1)}}_{k} +   \sum_{l\in \mathcal{P}_k  }  p_{l} \boldsymbol{\Gamma}^\textsuperscript{lmmse,(2)}_{k,l} -p_{k} \mathbf{b}^{\mathrm{lmmse}}_{k} (\mathbf{b}^{\mathrm{lmmse}}_{k} )^H + \hat{p}_{k}  \tau_p\boldsymbol{\Omega}'_{k} \sigma^2_{\mathrm{ul}} \right) \mathbf{a}_k\ },
		\end{equation}
		\hrulefill	
	\end{figure*}
	\setcounter{equation}{46}
If MR combining with ${v}_{m,k} = \hat{h}^{\mathrm{lmmse}}_{m,k} $ is used based on the LMMSE estimator, then $\mathrm{SE}^\mathrm{ul}_k = \frac{\tau_u}{\tau_c}\log_2( 1 + \gamma^\mathrm{ul,lmmse}_k )$ with $\gamma^\mathrm{ul,lmmse}_k$ in \eqref{lmmse:1}, at the top of next page, where 
\begin{align}
&\mathbf{T}^\textsuperscript{lmmse,(1)}_{k,l}  = \hat{p}_k \tau_p  \mathrm{tr}\left( \mathbf{A}^H_{k}\mathbf{R}'_{l}\boldsymbol{\Omega}'_{k}\mathbf{A}_{k} \right), \\
&\mathbf{T}^\textsuperscript{lmmse,(2)}_{k,l}  =\hat{p}_k \hat{p}_l \tau^2_p \left[  \mathrm{tr}\left( \mathbf{A}^H_{k}\mathbf{R}^2_{l} \boldsymbol{\Lambda}'_{k} \boldsymbol{\Omega}'_{k} \mathbf{A}_{k} \right) + \left| \mathrm{tr}\left(   \mathbf{A}_k\mathbf{R}'_{l}  \boldsymbol{\Lambda}'_{k} \mathbf{R}'_{k}\right) \right|^2    \right. \nonumber\\
&\left. + 2  \mathrm{tr}\left( \mathbf{A}^H_{k} \boldsymbol{\Lambda}'_{k} \boldsymbol{\Omega}'_{k} \mathbf{L}_{l}\mathbf{R}_{l}\mathbf{A}_{k} \right)  -  \mathrm{tr}\left(   \mathbf{A}^H_k(\mathbf{R}'_{l}  \boldsymbol{\Lambda}'_{k} \mathbf{R}'_{k})^2  \mathbf{A}_k\right) \right] .
\end{align}

The effective SINR in \eqref{lmmse:1} can be reformulated in Rayleigh quotient form as given in \eqref{gamma_ulmmse_mmsewp}, at the top of next page where \setcounter{equation}{49}
\begin{align}\label{lmmse_eq3}
&\boldsymbol{\Gamma}^\textsuperscript{lmmse,(2)}_{k,l}  = \hat{p}_k \hat{p}_l \tau^2_p \\ &\times \left[  \mathbf{R}^2_{l} \boldsymbol{\Lambda}'_{k} \boldsymbol{\Omega}'_{k} + 2   \boldsymbol{\Lambda}'_{k} \boldsymbol{\Omega}'_{k} \mathbf{L}_{l}\mathbf{R}_{l}+
\mathbf{d}_{k,l} \mathbf{d}^H_{k,l} -   ( \mathbf{R}'_{l}  \boldsymbol{\Lambda}'_{k} \mathbf{R}'_{k})^2  \right],\nonumber\\
&\boldsymbol{\Gamma}^\mathrm{lmmse}_k =  \sum_{l=1}^{K}   p_{l}\boldsymbol{\Gamma}^{\textsuperscript{lmmse,(1)}}_{k} +  
\nonumber\\
& \sum_{l\in \mathcal{P}_k  }  p_{l} \boldsymbol{\Gamma}^\textsuperscript{lmmse,(2)}_{k,l}  -p_{k} \mathbf{b}^{\mathrm{lmmse}}_{k,k} (\mathbf{b}^{\mathrm{lmmse}}_{k,k} )^H + \hat{p}_{k}  \tau_p\boldsymbol{\Omega}'_{k} \sigma^2_{\mathrm{ul}} ,
\end{align}
 $\mathbf{b}^{\mathrm{lmmse}}_{k}   = \hat{p}_{k}  \tau_p \mathrm{diag}(  \boldsymbol{\Omega}'_{k} )$, $ \ \boldsymbol{\Gamma}^{\textsuperscript{lmmse,(1)}}_{k,l}=  \hat{p}_{k}  \tau_p  \mathbf{R}'_{l} \boldsymbol{\Omega}'_{k}$, and $\mathbf{d}_{k,l}= \mathrm{diag}\left( \mathbf{R}'_{l} \boldsymbol{\Lambda}'_{k} \mathbf{R}'_{k} \right)= [\frac{\beta'_{1,l} \beta'_{1,k}}{\lambda'_{1,k}},  \dots,  \frac{\beta'_{M,l} \beta'_{M,k}}{\lambda'_{M,k}}  ]^T \in \mathbb{R}^{M \times 1}$.
The SE in \eqref{LSFD_SE} with the maximizing LSFD receiver vector $
\mathbf{a}_k =( \boldsymbol{\Gamma}_k^{\mathrm{lmmse}})^{-1}  \mathbf{b}^{\mathrm{lmmse}}_{k} $ is
\begin{equation}\label{ULSE_LMMSE:LSFD}
\mathrm{SE}^\mathrm{ul}_k=\frac{\tau_u}{\tau_c}\log_2\left( 1 + p_k   (\mathbf{b}^{\mathrm{lmmse}}_{k} )^H \left( \boldsymbol{\Gamma}_k^\mathrm{lmmse}\right)^{-1} \mathbf{b}^{\mathrm{lmmse}}_{k}  \right).
\end{equation}

\end{theorem}

\begin{IEEEproof}
	The proof is given in Appendix  \ref{AppendixUL_LMMSE}.
\end{IEEEproof}

\subsection{Uplink Spectral Efficiency with the LS Estimator }
\label{LS:LSFD}
\begin{theorem}
	\begin{figure*}[h]
		\normalsize
		\setcounter{equation}{52}
		\begin{equation}\label{ls:1}
		\gamma^\mathrm{ul,ls}_k = \frac{  p_k | \mathrm{tr}\left(\mathbf{A}_{k}\mathbf{R}'_{k}  \right) |^2}{\displaystyle \sum_{l=1}^{K} p_{l}\mathbf{T}^\textsuperscript{ls,(1)}_{k,l}  + \sum_{l\in \mathcal{P}_k }  p_{l}  \mathbf{T}^\textsuperscript{ls, (2)}_{k,l} -  p_k | \mathrm{tr}\left(\mathbf{A}_{k}\mathbf{R}'_{k}  \right) |^2+   \frac{\sigma^2_{\mathrm{ul}} }{\hat{p}_k \tau_p} \mathrm{tr}\left( \mathbf{A}^H_k \left( \boldsymbol{\Lambda}'_{k}\right)^{-1}  \mathbf{A}_k \right)},
		\end{equation}
		\setcounter{equation}{55}
			\begin{equation} \label{gamma_ul_ls}
		\gamma^\mathrm{ul,ls}_k=\frac{p_{k}    \mathbf{a}^H_k\mathbf{b}^{\mathrm{ls}}_{k} (\mathbf{b}^{\mathrm{ls}}_{k} )^H \mathbf{a}_k}{\mathbf{a}^H_k\left( \sum_{l=1}^{K}   p_{l} \boldsymbol{\Gamma}^{\textsuperscript{ls,(1)}}_{k,l} +   \sum_{l\in \mathcal{P}_k  } p_{l} \boldsymbol{\Gamma}^{\textsuperscript{ls,(2)}}_{k,l} - p_{k} \mathbf{b}^{\mathrm{ls}}_{k} (\mathbf{b}^{\mathrm{ls}}_{k} )^H +\frac{\sigma^2_{\mathrm{ul}} }{\hat{p}_k \tau_p}   \left( \boldsymbol{\Lambda}'_{k}\right)^{-1}  \right) \mathbf{a}_k\ },
		\end{equation}	
		\hrulefill
	\end{figure*}
	\setcounter{equation}{53}
	If MR combining with ${v}_{m,k} = \hat{h}^{\mathrm{ls}}_{m,k} $ is used based on the LS estimator, then $\mathrm{SE}^\mathrm{ul}_k = \frac{\tau_u}{\tau_c}\log_2\left( 1 + \gamma^\mathrm{ul,ls}_k \right)$ with $\gamma^\mathrm{ul,ls}_k $ in \eqref{ls:1}, at the top of next page, where
	\begin{align}
	&\mathbf{T}^\textsuperscript{ls,(1)}_{k,l}  = \frac{1}{\hat{p}_k \tau_p} \mathrm{tr}\left( \mathbf{A}^H_{k} (\boldsymbol{\Lambda}'_{k})^{-1} \mathbf{R}'_{l}\mathbf{A}_{k}\right) ,\\
	&\mathbf{T}^\textsuperscript{ls, (2)}_{k,l} = \frac{ \hat{p}_{l} }{\hat{p}_k }\\
	&\times \left[ \mathrm{tr}\left( \mathbf{A}^H_{k} (\mathbf{R}^2_{l} + 2\mathbf{L}_{l}\mathbf{R}_{l}) 
	\mathbf{A}_{k}\right) + \left| \mathrm{tr}\left( \mathbf{A}^H_{k} \mathbf{R}'_{l}\right)\right| ^2 - \mathrm{tr}\left( \mathbf{A}^H_{k} (\mathbf{R}'_{l})^2\mathbf{A}_{k} \right)\right]. \nonumber
	\end{align}
	
The effective SINR in \eqref{ls:1} can be reformulated in Rayleigh quotient form as given in
\eqref{gamma_ul_ls} where
\setcounter{equation}{56}
	\begin{align}
	&\boldsymbol{\Gamma}^\textsuperscript{ls,(2)}_{k,l} =\frac{ \hat{p}_{l} }{\hat{p}_k }\left[ \mathbf{R}^2_{l} + 2\mathbf{L}_{l}\mathbf{R}_{l} +\mathbf{b}^{\mathrm{ls}}_{l} (\mathbf{b}^{\mathrm{ls}}_{l})^H -  (\mathbf{R}'_{l})^2   \right], \\
	&\boldsymbol{\Gamma}^\mathrm{ls}_k =   \sum_{l=1}^{K}   p_{l} \boldsymbol{\Gamma}^{\textsuperscript{ls,(1)}}_{k,l} +   \sum_{l\in \mathcal{P}_k  } p_{l} \boldsymbol{\Gamma}^{\textsuperscript{ls,(2)}}_{k,l} - p_{k} \mathbf{b}^{\mathrm{ls}}_{k} (\mathbf{b}^{\mathrm{ls}}_{k} )^H +\frac{\sigma^2_{\mathrm{ul}} \left( \boldsymbol{\Lambda}'_{k}\right)^{-1}}{\hat{p}_k \tau_p}    , 
	\end{align}
$\mathbf{b}^{\mathrm{ls}}_{l}   =  \mathrm{diag}\left( \mathbf{R}'_{l} \right) 
$  and  $\boldsymbol{\Gamma}^{\textsuperscript{ls,(1)}}_{k,l}=  \frac{1}{\hat{p}_k \tau_p}  (\boldsymbol{\Lambda}'_{k})^{-1} \mathbf{R}'_{l}$.   
Note that the SINR expressions with the LS estimator above are proportional to the ones with LMMSE estimator since the LMMSE and LS estimators equal up to a scaling factor that depends on the large-scale fading coefficients. The SE in \eqref{LSFD_SE} with the maximizing LSFD receiver $
	\mathbf{a}_k =\left( \boldsymbol{\Gamma}_k^{\mathrm{ls}}\right)^{-1}  \mathbf{b}^{\mathrm{ls}}_{k} $ is
	\begin{equation}
	\mathrm{SE}^\mathrm{ul}_k=\frac{\tau_u}{\tau_c}\log_2\left( 1 + p_k   (\mathbf{b}^{\mathrm{ls}}_{k} )^H \left( \boldsymbol{\Gamma}_k^\mathrm{ls}\right)^{-1} \mathbf{b}^{\mathrm{ls}}_{k}  \right)
	\end{equation}
and it is equal to the UL SE with LMMSE estimator in \eqref{ULSE_LMMSE:LSFD}. 
\end{theorem}

\begin{IEEEproof}
	The proof is given in Appendix  \ref{AppendixUL_LS}.
\end{IEEEproof}

\begin{remark}
	 Note that the LSFD method only requires the large-scale fading coefficients and not the channel realizations themselves. Hence, it can be implemented along with the LMMSE and MMSE estimators without requiring any additional information. 
	However, the optimal LSFD vector cannot be practically computed when using the LS estimator since the statistics are not known in that case. 
	We have nevertheless included the optimal LSFD vector in that case to demonstrate that it makes the SE expressions with the LMMSE and LS estimators identical, thus LSFD can compensate for the use of a bad channel estimator. We will demonstrate later that the LS estimator (without LSFD) performs poorly in cell-free massive MIMO, thus it is a conservative lower bound on the performance.
\end{remark}

\section{Coherent Downlink Transmission}

Each coherence block contains $\tau_d$ DL data symbols, where $\tau_d=\tau_c-\tau_u-\tau_p$. In this section, we assume that each BS transmits to each UE and sends the same data symbol as the other APs. By setting some transmit powers to zero, the analysis also covers cases where each AP only serves a subset of the UEs. The transmitted signal from AP~$m$ is  
\begin{equation}
{x}_m= \sum_{k=1}^{K} w^\mathrm{coh}_{m,k}  \varsigma_{k},
\end{equation}
where $\varsigma_{k} \sim \mathcal{N}_\mathbb{C}\left( 0, 1\right)$ is the DL data signal to UE~$k$ which is same for all APs. The scalar $w^\mathrm{coh}_{m,k}=\sqrt{\frac{\rho_{m,k}}{\mathbb{E}\left\lbrace |\hat{h}_{m,k}|^2 \right\rbrace}}\hat{h}_{m,k}$ is the coherent beamforming  and $\rho_{m,k} \geq 0 $ is chosen to satisfy the DL power constraint $\mathbb{E}\lbrace |{x}_m|^2 \rbrace \leq \rho^{\mathrm{dl}} $ which implies $	\sum_{l=1}^{K} \rho_{m,l} \leq \rho^{\mathrm{dl}}$. The received signal at the $k$th UE is
\begin{equation}\label{DL:coherent:received}
y^\mathrm{dl}_{k}=\mathbf{h}^H_{k}\mathbf{w}_{k} \varsigma_{k} + \ \mathop{\sum_{l=1 }}^{K}_{l \neq k} \mathbf{h}^H_{k} \mathbf{w}_{l} \varsigma_{l} + {n}^{\mathrm{dl}}_{k},
\end{equation}
where  the receiver noise at UE~$k$ is denoted as $ {n}^{\mathrm{dl}}_{k}   \sim \mathcal{N}_\mathbb{C}\left( 0, \sigma^2_{\mathrm{dl}} \right)$ and  $\mathbf{w}_{k} =\mathbf{D}^{1/2}_k \hat{\mathbf{h}}_{k} $ with $\mathbf{D}_k = \mathrm{diag} \left(  \frac{\rho_{1,k}}{\mathbb{E}\left\lbrace |\hat{h}_{1,k}|^2 \right\rbrace}, \dots, \frac{\rho_{M,k}}{\mathbb{E}\left\lbrace |\hat{h}_{M,k}|^2\right\rbrace}  \right)  $. Note that $\mathbf{D}^{1/2}_k$  is a deterministic matrix whereas $\hat{\mathbf{h}}_{k}$ is a random vector. Based on the signal in \eqref{DL:coherent:received}, the ergodic DL capacity of UE~$k$ is lower bounded using the UatF bound as $\mathrm{SE}^{\mathrm{dl}}_{k}=\frac{\tau_d}{\tau_c}\log_2\left(1 + \gamma^{\mathrm{dl,coh}}_{k} \right) $ $ \mathrm{[bit/ s/ Hz]}$ with 
\begin{equation}\label{coherent:SINR}
\gamma^{\mathrm{dl,coh}}_{k} =\frac{ \left|  \mathbb{E}\left\lbrace \mathbf{w}^H_{k} \mathbf{h}_{k}  \right\rbrace   \right| ^2}{\displaystyle\sum_{l=1}^{K}     \mathbb{E}\left\lbrace \left| \mathbf{w}^H_{l} \mathbf{h}_{k}  \right| ^2 \right\rbrace  -   \left|  \mathbb{E}\left\lbrace \mathbf{w}^H_{k} \mathbf{h}_{k}  \right\rbrace   \right| ^2  +\sigma^2_{\mathrm{dl}} },
\end{equation}
where the expectations are with respect to all sources of randomness \cite[Th.~4.6]{EmilsBook}. Next, the effective SINR $\gamma^{\mathrm{dl,coherent}}_{k}$ is computed when using the phase-aware MMSE, LMMSE, and LS estimators. 
\subsection{Coherent Downlink Spectral Efficiency with the Phase-aware MMSE Estimator}

\begin{theorem}
    If MR precoding with $\mathbf{w}_{k} = \mathbf{D}^{1/2}_k \hat{\mathbf{h}}^\mathrm{mmse}_{k}$ is used based on the phase-aware MMSE estimator, then the expectations in \eqref{coherent:SINR} are computed as 
\begin{align}\label{coherent:mmse:1}
&\mathbb{E}\left\lbrace \mathbf{w}^H_{k} \mathbf{h}_{k}  \right\rbrace =   \mathrm{tr}\left(  \mathbf{D}^{1/2}_k\mathbf{Z}_{k} \right),\\
&\mathbb{E}\left\lbrace \left| \mathbf{w}^H_l  {\mathbf{h}}_{k} \right|^2 \right\rbrace = \hat{p}_{l} \tau_p \mathrm{tr}\left( \mathbf{D}_l\mathbf{R}_{k}  \boldsymbol{\Omega}_l\right) 
+ \hat{p}_{l} \tau_p \mathrm{tr}\left( \mathbf{D}_l\boldsymbol{\Omega}_l \mathbf{L}_{k}\right)   + \mathrm{tr}\left( \mathbf{D}_l \mathbf{S}_{l,k} \right)  \nonumber \\
& +   \mathrm{tr}\left( \mathbf{D}_l\mathbf{R}_{k} \mathbf{L}_{l} \right) +\begin{cases}
\hat{p}_{k} \hat{p}_{l}  \tau^2_p \left| \mathrm{tr}\left(   \mathbf{D}^{1/2}_l \mathbf{R}_{k}  \boldsymbol{\Lambda}_{l} \mathbf{R}_{l}\right) \right| ^2 , & \!\!\!\!\! l \in \mathcal{P}_{k} \backslash \{k\} \\
\hat{p}^2_{l} \tau^2_p \left| \mathrm{tr}\left(   \mathbf{D}^{1/2}_l \boldsymbol{\Omega}_l\right) \right| ^2 + \mathrm{tr}\left( \mathbf{D}^{1/2}_l \mathbf{L}_{l}\right) ^2 \\
+ 2\hat{p}_l \tau_p \mathrm{tr}\left( \mathbf{D}^{1/2}_l \boldsymbol{\Omega}_l\right) \mathrm{tr}\left( \mathbf{D}^{1/2}_l \mathbf{L}_{l}\right),  & l =k  \\
0 & l \notin \mathcal{P}_{k}.
\end{cases}
\end{align}

Inserting these expressions into \eqref{coherent:SINR} gives the DL SINR at UE~$k$ in \eqref{wide2}, at the top of next page.

	\begin{figure*}[h]
	\normalsize
	\begin{equation}\label{wide2}
	\gamma^{\mathrm{dl,mmse}}_k =\frac{ \left|  \mathrm{tr}\left(\mathbf{D}^{1/2}_k \mathbf{Z}_{k}\right)  \right| ^2}{\displaystyle \sum_{l=1}^{K}   \mathrm{tr}\left( \mathbf{D}_l \mathbf{R}'_{k} \mathbf{Z}_{l} \right)    + \displaystyle \sum_{l\in \mathcal{P}_k \backslash \{k\}} \hat{p}_{k} \hat{p}_{l}  \tau^2_p \left| \mathrm{tr}\left(   \mathbf{D}^{1/2}_l \mathbf{R}_{k}  \boldsymbol{\Lambda}_{l} \mathbf{R}_{l}\right) \right| ^2  - \mathrm{tr}\left(\mathbf{D}_k \mathbf{L}^2_{k}\right)  + \sigma^2_{\mathrm{dl} }}.
	\end{equation}
	\setcounter{equation}{77}
	\begin{equation}\label{wide3}
	\gamma^\mathrm{dl,mmse}_k	\frac{\mathrm{tr }\left(\mathbf{D}_k \mathbf{Z}_k \right)  }{ \displaystyle \sum_{l =1}^{K} \mathrm{tr}\left( \mathbf{D}_l \mathbf{R}'_{k} \right)  + \sum_{l\in \mathcal{P}_k \backslash \{k\}}   \hat{p}_k \hat{p}_{l}\tau^2_p  \mathrm{tr }\left( \mathbf{D}_{l} ( \mathbf{R}_{k} \boldsymbol{\Lambda}_{l} \mathbf{R}_{l})^2 \mathbf{Z}_l^{-1}\right) - \mathrm{tr}\left( \mathbf{D}_k \mathbf{L}^2_k \mathbf{Z}_k^{-1}\right)  + \sigma^2_\mathrm{dl} },
	\end{equation}
	\setcounter{equation}{65}
	\hrulefill
\end{figure*}

\end{theorem}
\begin{IEEEproof}
	The proof is given in Appendix  \ref{AppendixUL_MMSE}.
\end{IEEEproof}

\subsection{Coherent Downlink Spectral Efficiency with the LMMSE Estimator}

\begin{theorem}
	If MR precoding with $\mathbf{w}_{k} = \mathbf{D}^{1/2}_k \hat{\mathbf{h}}^\mathrm{lmmse}_{k}$ is used based on the LMMSE estimator, then the expectations in \eqref{coherent:SINR} are computed as 
	\begin{align}
	&\mathbb{E}\left\lbrace \mathbf{w}^H_{k} \mathbf{h}_{k}  \right\rbrace =   \hat{p}_{k} \tau_p\mathrm{tr}\left(  \mathbf{D}^{1/2}_k\boldsymbol{\Omega}'_{k} \right),\\
	&\mathbb{E}\left\lbrace \left| \mathbf{w}^H_l  {\mathbf{h}}_{k} \right|^2 \right\rbrace = \hat{p}_{l}  \tau_p \mathrm{tr}\left( \mathbf{D}_{l} \mathbf{R}'_{k}  \boldsymbol{\Omega}'_{l} \right)\\
	 & +\hat{p}_{l} \hat{p}_{k}  \tau^2_p \begin{cases} 
	\mathrm{tr}\left( \mathbf{D}_{l}\boldsymbol{\Omega}'_{l} \boldsymbol{\Lambda}'_{l} \left( \mathbf{R}^2_{k} + 2    \mathbf{L}_{k}\mathbf{R}_{k} \right)\right) \nonumber \\+ \left| \mathrm{tr}\left(   \mathbf{D}^{1/2}_l\mathbf{R}'_{k}  \boldsymbol{\Lambda}'_{l} \mathbf{R}'_{l}\right) \right|^2 
	-  \mathrm{tr}\left(   \mathbf{D}_l(\mathbf{R}'_{k}  \boldsymbol{\Lambda}'_{l} \mathbf{R}'_{l})^2 \right), & l \in \mathcal{P}_{k} \\
	0 & l \notin \mathcal{P}_{k}.
	\end{cases}
	\end{align}
	
	Inserting these expressions into \eqref{coherent:SINR} gives the DL SINR at UE~$k$ as 
	\begin{align}
	&\gamma^\mathrm{dl,lmmse}_k =\\
	& \frac{  \hat{p}^2_k \tau^2_p | \mathrm{tr}(\mathbf{D}^{1/2}_{k}\boldsymbol{\Omega}'_{k} ) |^2}{\displaystyle \sum_{l=1}^{K}    \hat{p}_{l}  \tau_p \mathrm{tr}\left( \mathbf{D}_{l} \mathbf{R}'_{k}  \boldsymbol{\Omega}'_{l} \right)  + \sum_{l\in \mathcal{P}_k  }     \mathbf{T}^\textsuperscript{dl,lmmse,(2)}_{k,l}\!\!\!\!\!\! -  \hat{p}^2_k \tau^2_p | \mathrm{tr}\left(\mathbf{D}_{k}\boldsymbol{\Omega}'_{k} \right) |^2+ \sigma^2_{\mathrm{dl}}   },\nonumber
	\end{align}
	where $\mathbf{T}^\textsuperscript{dl,lmmse,(2)}_{k,l}  = \hat{p}_k \hat{p}_l \tau^2_p \left[ \mathrm{tr}\left( \mathbf{D}_{l}\boldsymbol{\Omega}'_{l} \boldsymbol{\Lambda}'_{l} \left( \mathbf{R}^2_{k} + 2    \mathbf{L}_{k}\mathbf{R}_{k} \right)\right) \right. 
	+ \left.\left| \mathrm{tr}\left(   \mathbf{D}^{1/2}_l\mathbf{R}'_{k}  \boldsymbol{\Lambda}'_{l} \mathbf{R}'_{l}\right) \right|^2 
	-  \mathrm{tr}\left(   \mathbf{D}_l(\mathbf{R}'_{k}  \boldsymbol{\Lambda}'_{l} \mathbf{R}'_{l})^2 \right) \right] .$
\end{theorem}
\begin{IEEEproof}
	The proof is given in Appendix  \ref{AppendixUL_LMMSE}.
\end{IEEEproof}
\subsection{Coherent Downlink Spectral Efficiency with the LS Estimator}

\begin{theorem}
	If MR precoding with $\mathbf{w}_{k} = \mathbf{D}^{1/2}_k \hat{\mathbf{h}}^\mathrm{ls}_{k}$ is used based on the LS estimator, then the expectations in \eqref{coherent:SINR} are computed as 
	\begin{align}\label{coherent:ls:1}
	&\mathbb{E}\left\lbrace \mathbf{w}^H_{k} \mathbf{h}_{k}  \right\rbrace =   \mathrm{tr}\left(  \mathbf{D}^{1/2}_k\mathbf{R}'_{k} \right),\\
	&\mathbb{E}\left\lbrace \left| \mathbf{w}^H_l  {\mathbf{h}}_{k} \right|^2 \right\rbrace = \frac{1}{\hat{p}_{l}  \tau_p }\mathrm{tr}\left( \mathbf{D}_{l} (\boldsymbol{\Lambda}'_{l})^{-1}\mathbf{R}'_{k}   \right) \\
	&+\frac{\hat{p}_{k}}{\hat{p}_{l}} \begin{cases} 
	\mathrm{tr}\left( \mathbf{D}_{l} \left( \mathbf{R}^2_{k} + 2    \mathbf{L}_{k}\mathbf{R}_{k} \right)\right) + \left| \mathrm{tr}\left(   \mathbf{D}^{1/2}_l\mathbf{R}'_{k}  \right) \right|^2 \nonumber \\
	-  \mathrm{tr}\left(   \mathbf{D}_l (\mathbf{R}'_{k})^2 \right), & l \in \mathcal{P}_{k} \\
	0 & l \notin \mathcal{P}_{k}.
	\end{cases}
	\end{align}
	
	Inserting these expressions into \eqref{coherent:SINR} gives the DL SINR at UE~$k$ as
	\begin{align}\label{coherent:ls:3}
	&\gamma^\mathrm{dl,ls}_k \nonumber\\
	&= \frac{ \left|  \mathrm{tr}\left(\mathbf{D}^{1/2}_k\mathbf{R}'_{k}  \right) \right| ^2}{\displaystyle \sum_{l=1}^{K}  \frac{1}{\hat{p}_{l}  \tau_p }\mathrm{tr}\left( \mathbf{D}_{l} (\boldsymbol{\Lambda}'_{l})^{-1}\mathbf{R}'_{k}   \right)  + \sum_{l\in \mathcal{P}_k }    \mathbf{T}^\textsuperscript{dl,ls, (2)}_{k,l} -   \left|  \mathrm{tr}\left( \mathbf{D}^{1/2}_k\mathbf{R}'_{k} \right) \right| ^2+   \sigma^2_{\mathrm{dl}} }, \nonumber
	\end{align}
	where
	\begin{align}
	&\mathbf{T}^\textsuperscript{dl,ls, (2)}_{k,l}  = \frac{ \hat{p}_{k} }{\hat{p}_l }\\
	&\times \left[ \mathrm{tr}\left( \mathbf{D}_{l} \left( \mathbf{R}^2_{k} + 2    \mathbf{L}_{k}\mathbf{R}_{k} \right)\right) + \left| \mathrm{tr}\left(   \mathbf{D}^{1/2}_l\mathbf{R}'_{k}  \right) \right|^2 -  \mathrm{tr}\left(   \mathbf{D}_l (\mathbf{R}'_{k})^2 \right) \right]. \nonumber
	\end{align}

\end{theorem}

\begin{IEEEproof}
	The proof is given in Appendix  \ref{AppendixUL_LS}.
\end{IEEEproof}

\section{Non-coherent Downlink Transmission}

\label{noncoherent:section}
In this section, we consider the alternative case in which each AP is allowed to transmit to each UE but sends a different data symbol than the other APs, to alleviate need for phase-synchronizing the APs. The transmitted signal from AP~$m$ is  
\begin{equation}
{x}_m= \sum_{k=1}^{K} w_{m,k}  \varsigma_{m,k},
\end{equation}
where $\varsigma_{m,k} \sim \mathcal{N}_\mathbb{C}\left( 0, \rho_{m,k}\right)$ is the DL data signal to UE~$k$. The beamforming scalar $w_{m,k}=\frac{\hat{h}_{m,k} }{\sqrt{\mathbb{E}\left\lbrace |\hat{h}_{m,k}|^2 \right\rbrace} } $ is a scaled version of the channel estimate and $\rho_{m,k}$ is chosen to satisfy the DL power constraint $\mathbb{E}\lbrace |{x}_m|^2 \rbrace \leq \rho^{\mathrm{dl}} $ which implies $	\sum_{l=1}^{K} \rho_{m,l}  \leq \rho^{\mathrm{dl}}$.  The received signal at the $k$th UE is
\begin{align}\label{sec1eq2}
&y^\mathrm{dl}_{k}=\sum_{n=1}^{M}h^*_{n,k} w_{n,k} \varsigma_{n,k} + \ \mathop{\sum_{l=1 }}^{K}_{l \neq k} \sum_{n=1}^{M}h^*_{n,k} w_{n,l} \varsigma_{n,l} + {n}^{\mathrm{dl}}_{k},
\end{align}
where  the receiver noise at UE~$k$ is denoted  $ {n}^{\mathrm{dl}}_{k}   \sim \mathcal{N}_\mathbb{C}\left( 0, \sigma^2_{\mathrm{dl}} \right)$. 
\begin{lemma}
	Based on the signal in \eqref{sec1eq2}, if the UE~$k$ detects the $M$ signals using successive interference cancellation (with arbitrary decoding order)  a lower bound on the DL sum SE of UE~$k$ is $
	\mathrm{SE}^\mathrm{dl,nc}_{k}= \frac{\tau_d}{\tau_c} \log_2 (1 + \gamma^\mathrm{dl,nc}_{k})$ 	where the effective SINR is
	\begin{align}\label{DLSINR_k}
	&\gamma^\mathrm{dl,nc}_{k} = \nonumber\\
	&\frac{\sum_{n=1}^{M} \rho_{n,k} \left| \mathbb{E}\left\lbrace h^*_{n,k} w_{n,k} \right\rbrace \right| ^2}{ \displaystyle \sum_{n=1}^{M}  \mathop{\sum_{l=1 }}^{K}\rho_{n,l}  \mathbb{E}\left\lbrace | h^*_{n,k} w_{n,l}|^2 \right\rbrace  - \sum_{n=1}^{M} \rho_{n,k}  \left| \mathbb{E}\left\lbrace h^*_{n,k} w_{n,k} \right\rbrace \right| ^2  + \sigma^2_{\mathrm{dl}}}.
	\end{align}
\end{lemma}
\begin{IEEEproof}
	The proof is given in Appendix \ref{AppendixDL}.
\end{IEEEproof}
From the effective SINR above, we notice that the numerator contains the sum of the squared contributions from different APs, which is different from the coherent case where the summation is inside the square. Hence, the coherent case gives a larger signal term, but it requires that all APs are synchronized and co-operate to achieve a coherent beamforming gain that is analogous to co-located massive MIMO. 
The numerical comparisons of these DL transmission modes are presented in  Section~\ref{numerical:results}. In the following subsections, the effective  $\gamma^{\mathrm{dl,nc}}_{k}$ is calculated for MR precoding when using the phase-aware MMSE, LMMSE, and LS  estimators. 

\subsection{Non-coherent Downlink Spectral Efficiency with the Phase-aware MMSE Estimator}

	If MR precoding with ${w}_{m,k} = \frac{\hat{h}^{\mathrm{mmse}}_{m,k} }{\sqrt{\mathbb{E}\left\lbrace |\hat{h}^{\mathrm{mmse}}_{m,k}|^2\right\rbrace  }}$ is used based on the MMSE estimator, then the expectations in \eqref{DLSINR_k} are computed as
\begin{align}
&\mathbb{E}\left\lbrace {h}^*_{n,k} {w}_{n,k}  \right\rbrace = \sqrt{\hat{p}_k \tau_p \beta^2_{n,k} \lambda^{-1}_{n,k} + \bar{h}^2_{n,k}},\\
&\mathbb{E} \left\lbrace  |  \hat{h}^\mathrm{mmse}_{n,l}  |^2  \right\rbrace = \hat{p}_{l} \tau_p \beta^2_{n,l} \lambda^{-1}_{n,l} + \bar{h}^2_{n,l},\\
& \mathbb{E}\left\lbrace  |  {h}^*_{n,k} {w}_{n,l}   | ^2  \right\rbrace =  \beta'_{n,k} \nonumber \\
&+\begin{cases}
\frac{\hat{p}_k \hat{p}_{l}\tau^2_p \beta^2_{n,k} \beta^2_{n,l}  \lambda^{-2}_{n,l}}{\hat{p}_{l} \tau_p \beta^2_{n,l} \lambda^{-1}_{n,l} + \bar{h}^2_{n,l}}, &  l \in \mathcal{P}_{k} \backslash \{k\} \\
\frac{\hat{p}^2_k \tau^2_p \beta^4_{n,k}  \lambda^{-2}_{n,k} + 2 \hat{p}_k \tau_p \beta^2_{n,k} \lambda^{-1}_{n,k} \bar{h}^2_{n,k}}{\hat{p}_{k} \tau_p \beta^2_{n,k} \lambda^{-1}_{n,k} + \bar{h}^2_{n,k}}, & l = k \\
0, & l \notin \mathcal{P}_{k} .
\end{cases}
\end{align}
 Substituting these expressions into \eqref{DLSINR_k} gives the DL SINR at UE $k$ as given in \eqref{wide3}, at the top of this page where $\mathbf{D}_k= \mathrm{diag}\left( \rho_{1,k}, \dots, \rho_{M,k}\right) $ is the downlink power matrix.

\begin{IEEEproof}
	The proof is also given in Appendix  \ref{AppendixUL_MMSE}.
\end{IEEEproof}

\subsection{Non-Coherent Downlink Spectral Efficiency with the LMMSE and LS Estimator}
	If MR precoding with ${w}_{m,k} = \frac{\hat{h}^{\mathrm{lmmse}}_{m,k} }{\sqrt{\mathbb{E}\left\lbrace |\hat{h}^{\mathrm{lmmse}}_{m,k}|^2\right\rbrace  }}$ is used based on the LMMSE estimator without phase information, then the expectations in \eqref{DLSINR_k} are computed as
\setcounter{equation}{78}
\begin{align}\label{nc_eq1}
&\mathbb{E}\left\lbrace  {h}^*_{n,k} {w}_{n,k}\right\rbrace = \mathbb{E}\left\lbrace  |  {w}_{n,k}   | ^2 \right\rbrace =\sqrt{\hat{p}_k \tau_p  (\beta'_{n,k})^2 (\lambda'_{n,k})^{-1}},\\
&\mathbb{E}\left\lbrace  |   {h}^*_{n,k} {w}_{n,l}  | ^2  \right\rbrace =  \beta'_{n,k} \nonumber \label{nc_eq2}\\
&+\begin{cases}
\hat{p}_{k}  \tau_p   (\lambda'_{n,l})^{-1} \left( \beta^2_{n,k} +2 \bar{h}^2_{n,k} \beta_{n,k}  \right)  , &  l \in \mathcal{P}_{k} \\
0 & l \notin \mathcal{P}_{k} ,
\end{cases}
\end{align}
where $\mathbb{E} \left\lbrace  |  \hat{h}^\mathrm{lmmse}_{m,k}  |^2  \right\rbrace = \hat{p}_{k} \tau_p  (\beta'_{m,k})^2  (\lambda'_{m,k})^{-1}$. If MR precoding with ${w}_{m,k} = \frac{\hat{h}^{\mathrm{ls}}_{m,k} }{\sqrt{\mathbb{E}\left\lbrace |\hat{h}^{\mathrm{ls}}_{m,k}|^2\right\rbrace  }}$ is used based on the LS estimator without any prior information regarding the phase or statistics, computing expectations in \eqref{DLSINR_k}  gives the exactly same results as \eqref{nc_eq1} and \eqref{nc_eq2} where  $\mathbb{E}\left\lbrace |\hat{h}^{\mathrm{ls}}_{m,k}|^2\right\rbrace  = \frac{1}{\hat{p}_k\tau_p} \lambda'_{m,k} $. Note that this is due to the precoding normalization at each AP:
	\begin{align}
	{w}_{m,k} &= \frac{\hat{h}^{\mathrm{lmmse}}_{m,k} }{\sqrt{\mathbb{E}\left\lbrace |\hat{h}^{\mathrm{lmmse}}_{m,k}|^2\right\rbrace  }} = \frac{\sqrt{\hat{p}_{k}} \beta'_{m,k} (\lambda'_{m,k})^{-1} y^p_{m,k} }{\sqrt{\hat{p}_{k} \tau_p  (\beta'_{m,k})^2  (\lambda'_{m,k})^{-1}}} \nonumber \\
	&=\frac{\frac{1}{\sqrt{p_{k}} \tau_p} y^p_{m,k}}{\sqrt{\frac{1}{\hat{p}_k\tau_p} \lambda'_{m,k}}} =  \frac{\hat{h}^{\mathrm{ls}}_{m,k} }{\sqrt{\mathbb{E}\left\lbrace |\hat{h}^{\mathrm{ls}}_{m,k}|^2\right\rbrace  }}.
	\end{align}

	Substituting \eqref{nc_eq1} and \eqref{nc_eq2} into \eqref{DLSINR_k} gives the same DL SINR at UE $k$ with the LMMSE and LS estimators
	\begin{align}
	&\gamma^\mathrm{dl,lmmse}_k=\gamma^\mathrm{dl,ls}_k=\\
	&\frac{  \hat{p}_k \tau_p  \mathrm{tr}(\mathbf{D}_{k}\boldsymbol{\Omega}'_{k} ) }{\displaystyle \sum_{l=1}^{K}   \mathrm{tr}( \mathbf{D}_{l}\mathbf{R}'_{k} )  + \sum_{l\in \mathcal{P}_k }     \hat{p}_{k}\tau_p   \mathrm{tr}(\mathbf{D}_{l} \boldsymbol{\Lambda}'_{l}  (\mathbf{R}^2_{k} + 2 \mathbf{R}_{k}\mathbf{L}_{k}) ) - \hat{p}_{k}\tau_p ( \mathbf{D}_{k} \boldsymbol{\Omega}'_{k}) + \sigma^2_{\mathrm{dl}}   }.
	\end{align}

\begin{IEEEproof}
	The proof is also given in Appendix  \ref{AppendixUL_LMMSE} and Appendix   \ref{AppendixUL_LS}.
\end{IEEEproof}

	In contrast to the coherent DL transmission case, the SE is the same for both LMMSE and LS estimator because there is no cooperation between APs. Thus, the expectations are calculated for each antenna individually.

\section{Numerical Results}
\label{numerical:results}
\begin{figure*}[h!]
	\begin{minipage}{.46\textwidth}
		\includegraphics[scale=0.45]{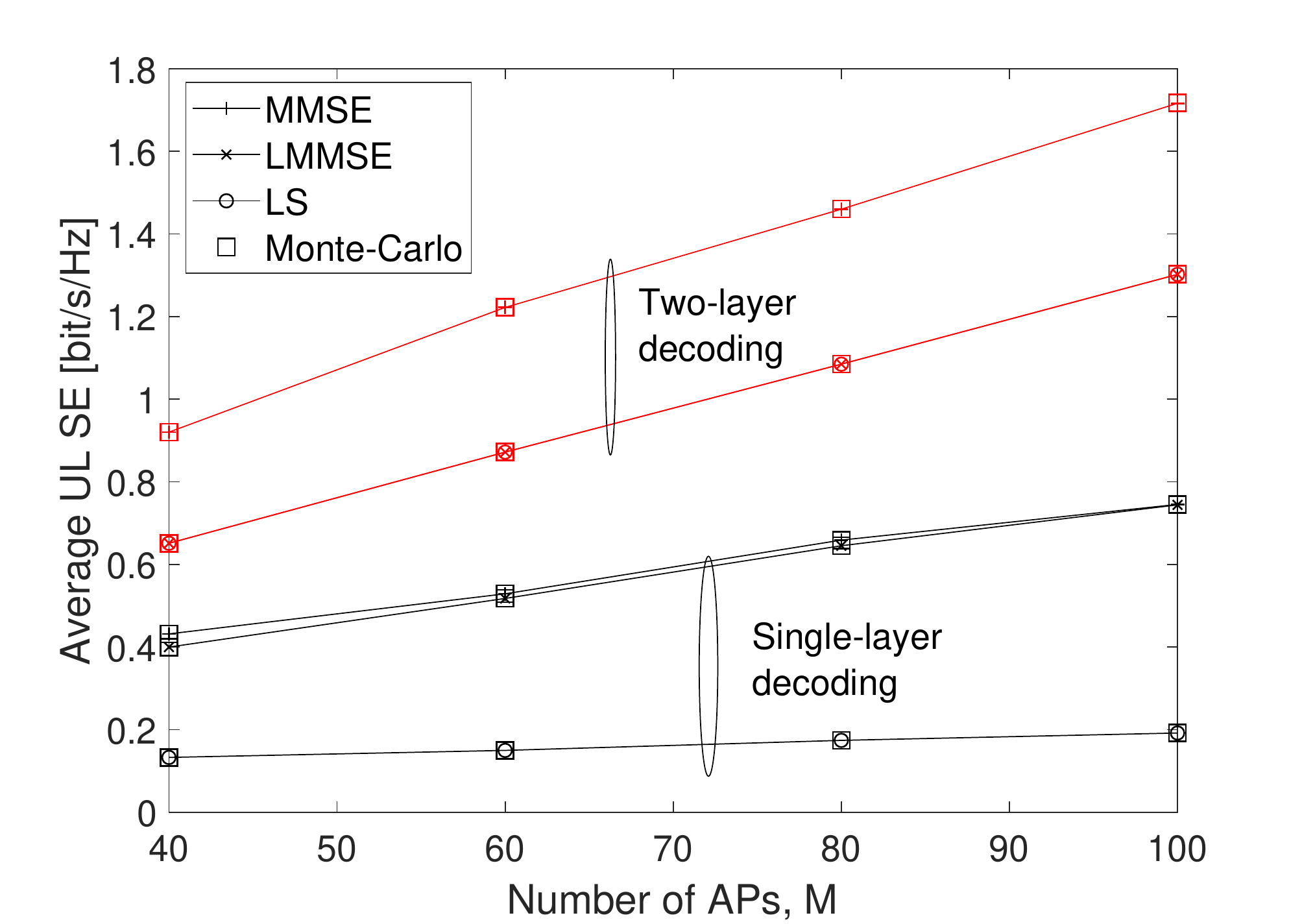}
		\caption{Average UL SE for different number of APs.  $K = 40$, $\tau_p = 5$. } \label{fig1}
	\end{minipage}\qquad
	\begin{minipage}{.46\textwidth}
		\includegraphics[scale=0.45]{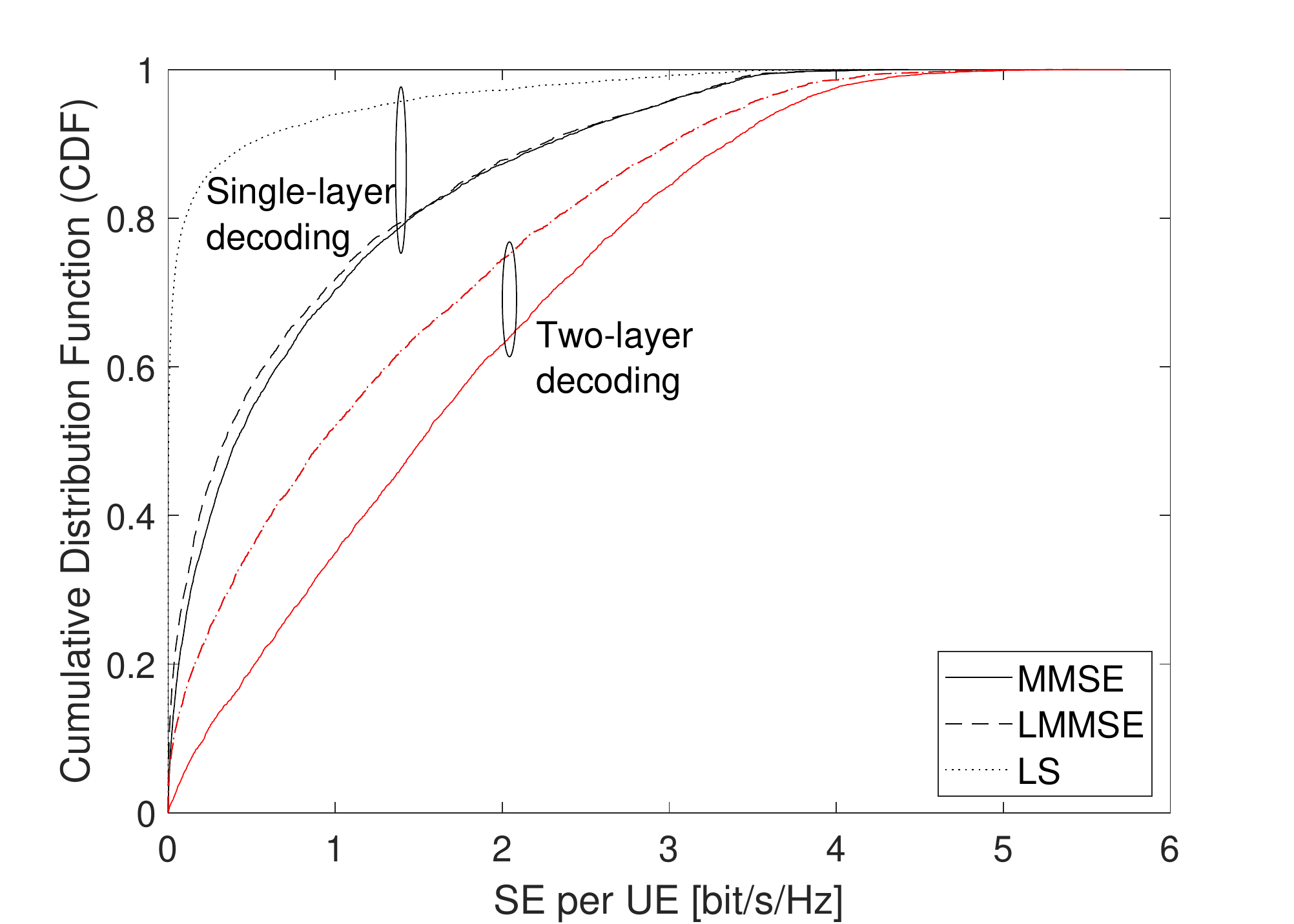}
		\caption{CDF of UL SE for $M=100$, $K=40$ and $\tau_p = 5$.  } \label{fig2}
	\end{minipage}
\end{figure*}

\begin{figure*}[t!]
	\begin{minipage}{.46\textwidth}
		\includegraphics[scale= 0.45]{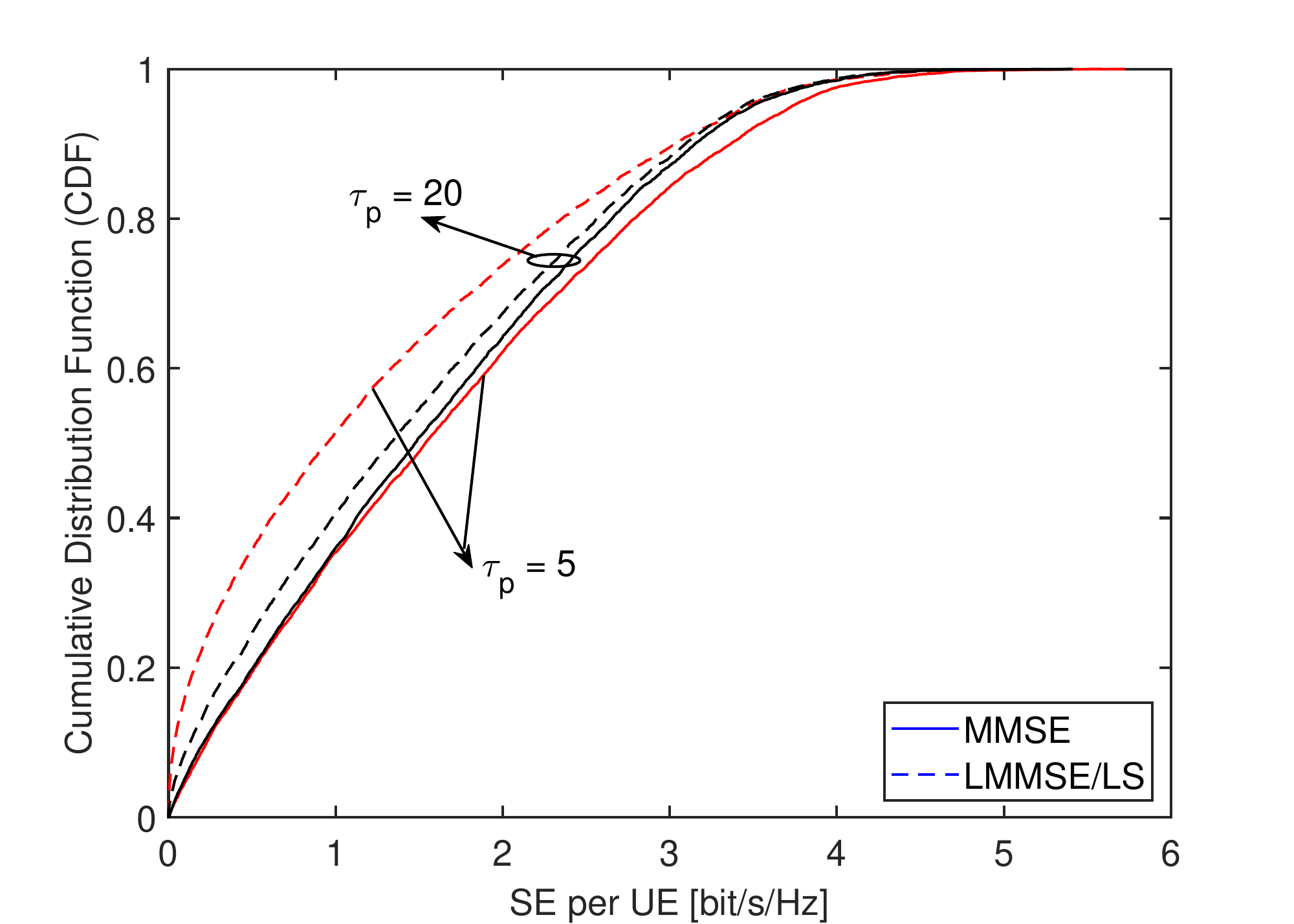}
		\caption{CDF of UL SE for $M=100$, $K=40$ and $\tau_p = [5, 20]$. Two-layer decoding.  } \label{fig3}
	\end{minipage}\qquad
	\begin{minipage}{.46\textwidth}
		\includegraphics[scale=0.45]{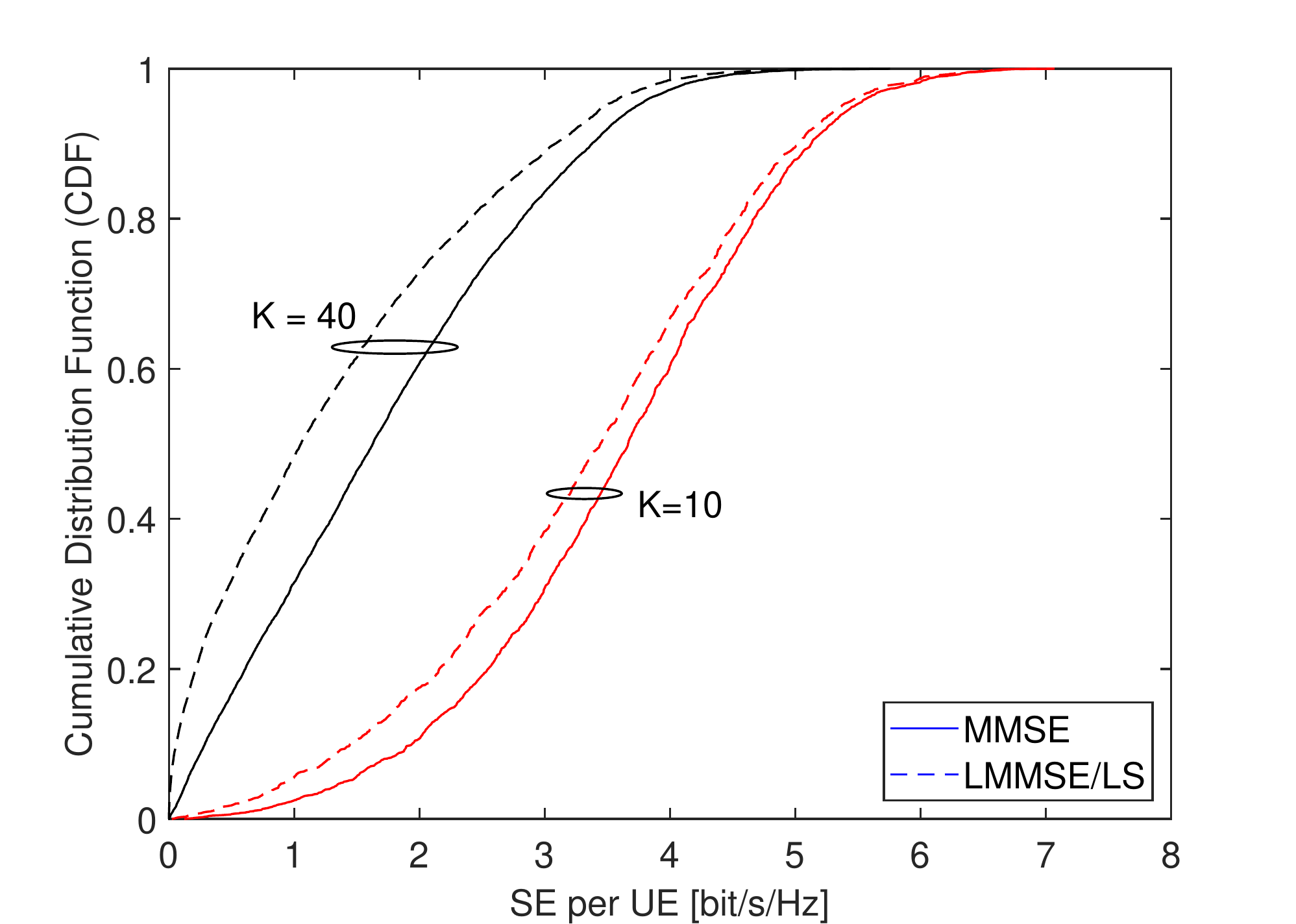}
		\caption{CDF of UL SE for $M=100$, $\tau_p = 5$ and $K =[10, 40]$. Two-layer decoding. } \label{fig4}
	\end{minipage}
\end{figure*}

In this section, the closed-form SE expressions are validated
and evaluated by simulating a cell-free massive MIMO network. We have $M$ APs and $K=40$ UEs that are independently and uniformly distributed within a square of size $1 \times 1$ $\textrm{km}^2$ with a wrap-around setup. Maximum ratio precoding/combining is used at UL/DL in all simulations. The pathloss is computed based on the COST 321 Walfish-Ikegami model for micro-cells in \cite{3gpp} with AP height $12.5$ m and UE height $1.5$ m.  All AP-UE pairs have a LoS path and the  path-loss (PL)  is modeled (in dB) as
\begin{equation}
\mathrm{PL}_{m,k} =-30.18 -26  \log_{10} \! {\left(\frac{d_{m,k}}{1\,\textrm{m}}\right)}  + F_{m,k},  
\end{equation}
where $d_{m,n}$ is the distance between AP~$m$ and UE~$k$ and  $F_{m,k}$ is the shadow fading coefficient. The Rician $\kappa$-factor is calculated as $\kappa_{m,k}=10^{1.3 - 0.003d_{m,k}}$ \cite{3gpp} .

 We assume correlated shadow fading as in \cite{Ngo2017} with $F_{m,k}= \sqrt{\delta} a_m + \sqrt{1-\delta} b_k$ where $a_m \sim \mathcal{N}(0,\sigma^2_\mathrm{sf})$ and $b_k \sim \mathcal{N}(0,\sigma^2_\mathrm{sf})$ are independent and $\delta$ is the shadow fading parameter, $0\leq \delta \leq 1$. The random variables $a_m $ and $b_k$ model the shadow fading effect from blocking objects in the vicinity of the AP~$m$ and UE~$k$, respectively. The covariance functions for arbitrary AP and UE pairs are given as
	$\mathbb{E}\left\lbrace a_m a_n\right\rbrace  = 2^\frac{-d_{m,n}}{d_{dc}}$, 
	and $\mathbb{E}\left\lbrace b_k b_l\right\rbrace  = 2^\frac{-d_{k,l}}{d_{dc}}$,
where $d_{m,n}$ denote the distance between AP~$m$ and AP~$n$ and $d_{k,l}$  is the distance between UE~$k$ and UE~$l$. The $d_{dc}$ is decorrelation distance which depends on the environment. We set the parameters $d_{dc} =100$ m, $\sigma_\mathrm{sf}=8$ and $\delta=0.5$ in the simulation. With this model, the large scale coefficient of $h_{m,k}$ are
\begin{equation}\label{betalos}
\bar{h}_{m,k} =\sqrt{\frac{\kappa_{m,k}}{\kappa_{m,k} +1}} \sqrt{\mathrm{PL}_{m,k}}, \ \beta_{m,k}= {\frac{1}{\kappa_{m,k} +1}} \mathrm{PL}_{m,k}.
\end{equation}

\begin{figure*}
	\begin{minipage}[b]{.46\textwidth}
		\includegraphics[scale =0.48]{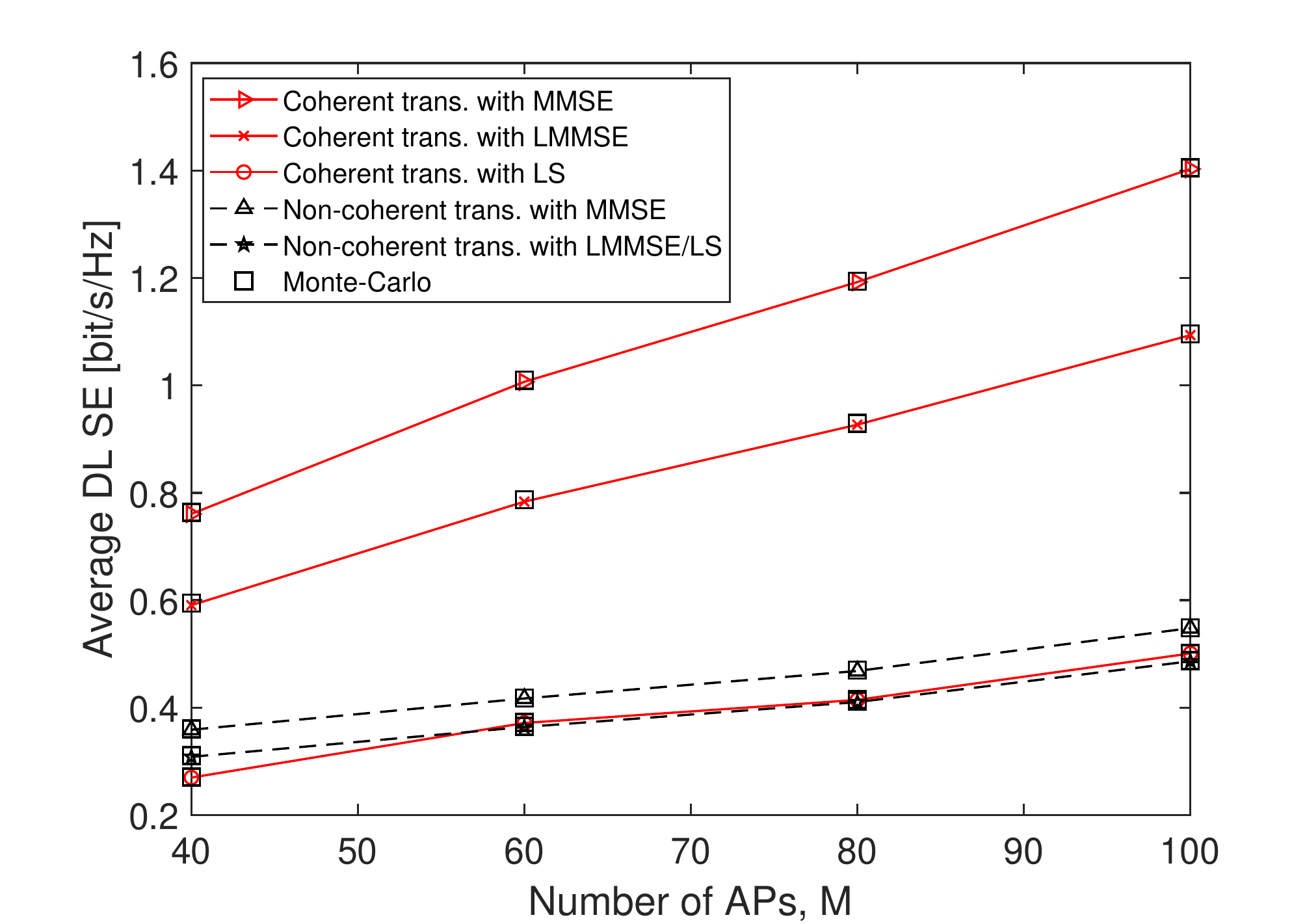}
		\caption{Average DL SE versus different number of APs for coherent and non-coherent transmission. $K = 40$, $\tau_p = 5$. }\label{fig5}
	\end{minipage}\qquad
	\begin{minipage}[b]{.46\textwidth}
		\includegraphics[scale=0.48]{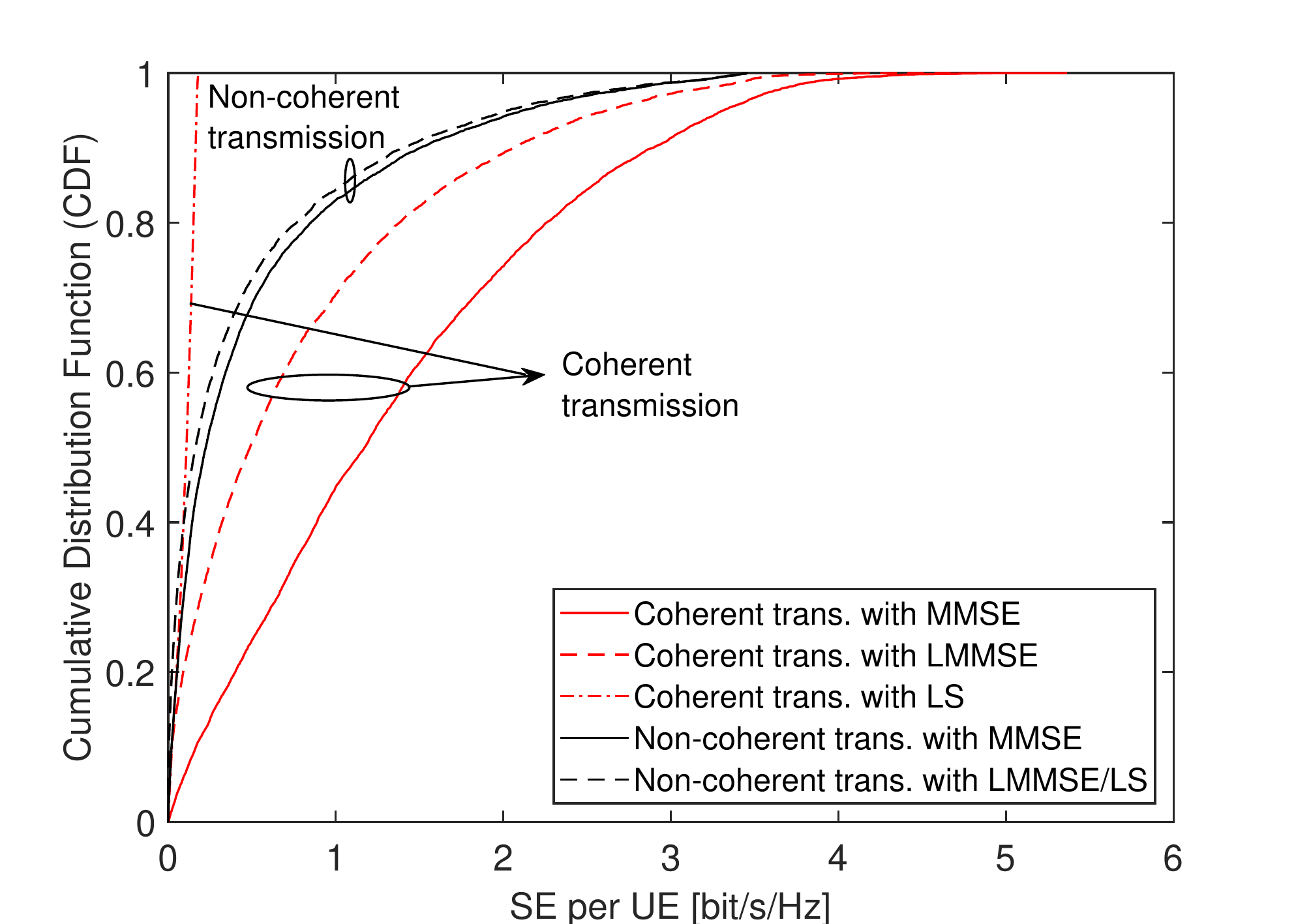}
		\caption{CDF of DL SE for coherent and non-coherent transmission with $M = 100$, $K = 40$ and $\tau_p = 5$. } \label{fig6}
	\end{minipage}
\end{figure*}

\begin{figure}[t!]
	\includegraphics[scale=0.45]{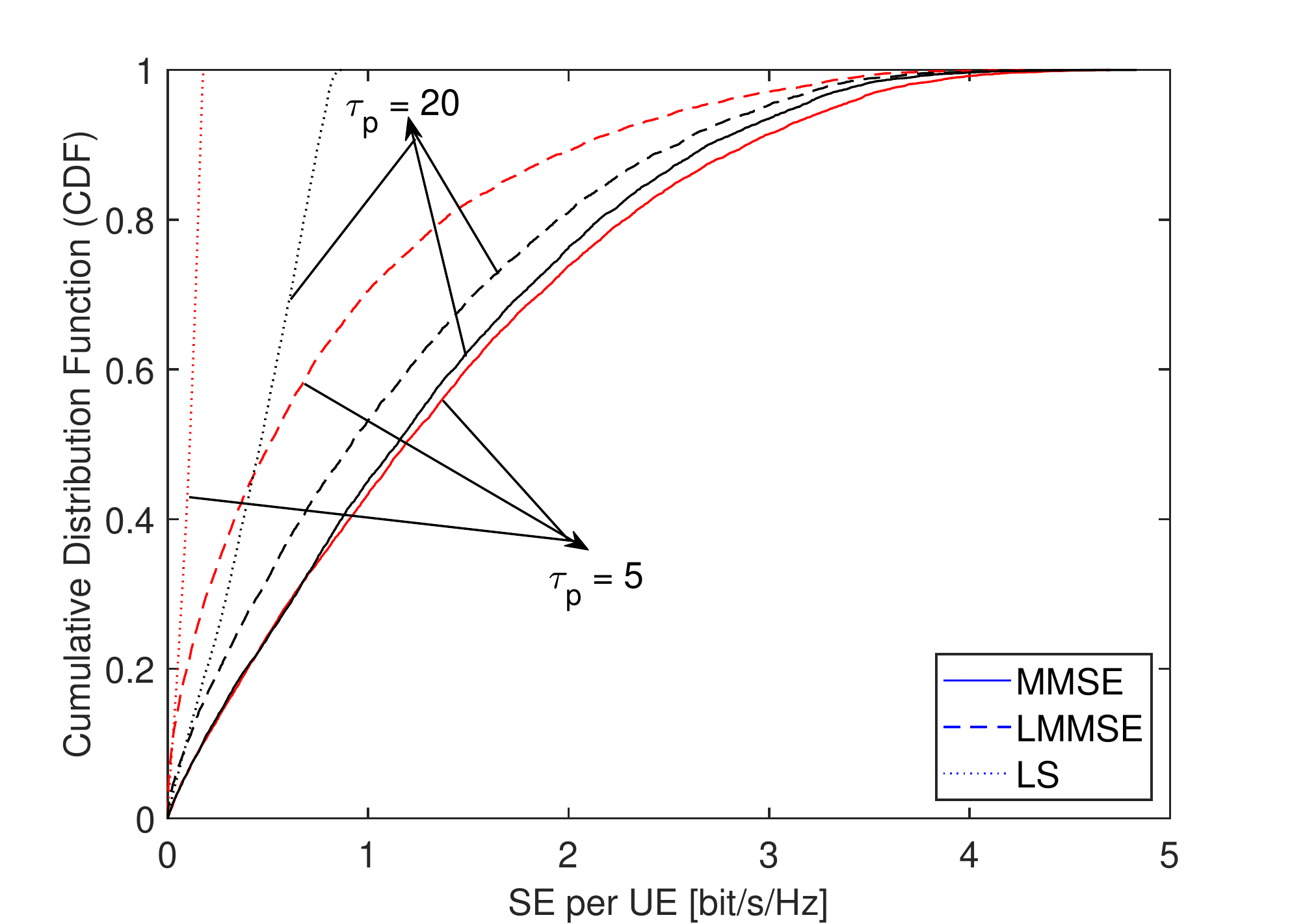}
	\caption{CDF of DL SE with coherent transmission for pilot lengths where $M = 100$, $K = 40$, $\tau_p = [5, 20]$. }\label{fig7}
\end{figure}

\begin{figure}[t!]
	\includegraphics[scale=0.45]{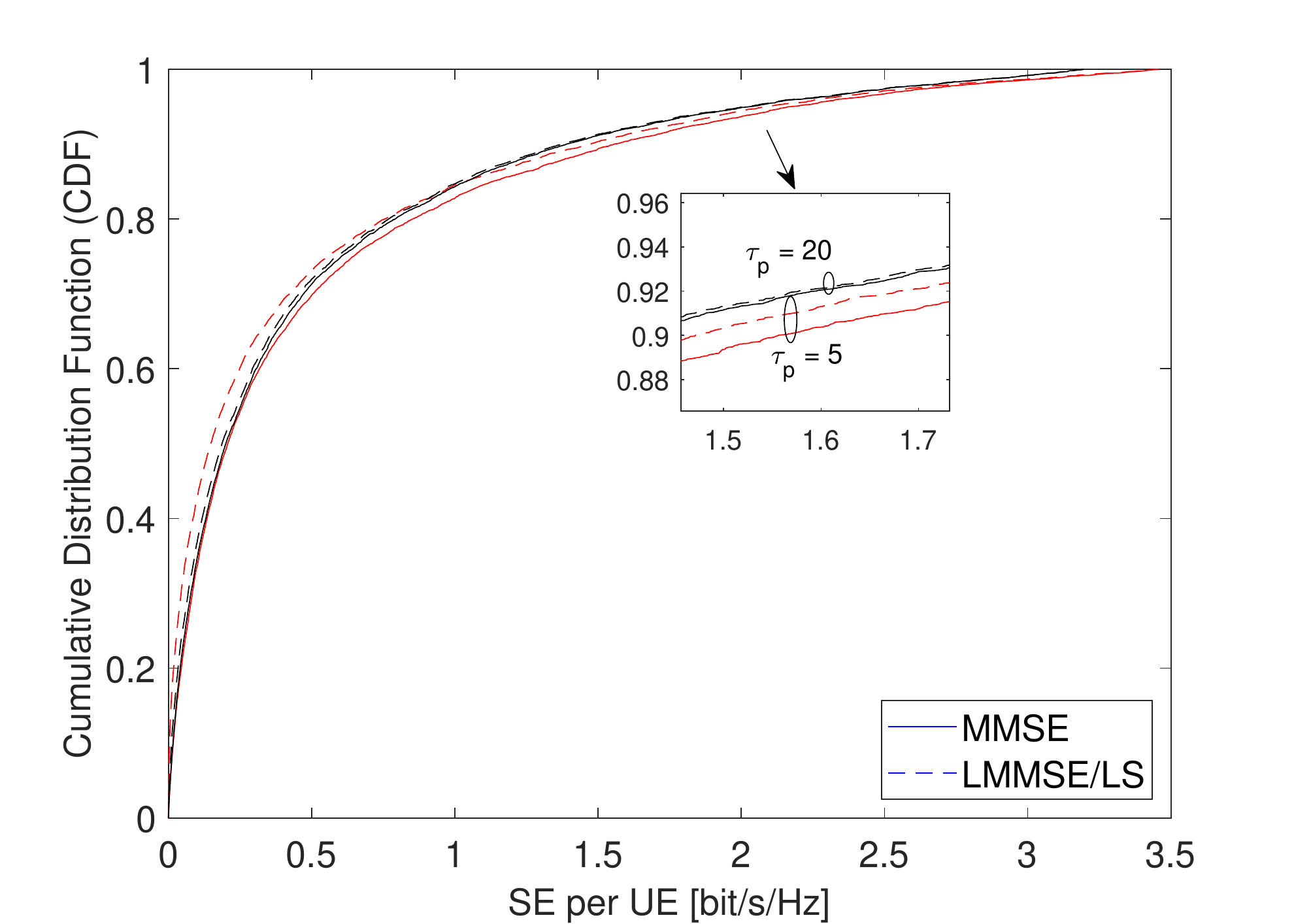}
	\caption{CDF of DL SE with non-coherent transmission for pilot lengths where $M = 100$, $K = 40$, $\tau_p = [5, 20]$. } \label{fig8}
\end{figure}

We consider communication over a 20\,MHz channel and the total receiver noise power is $-94$ dBm. Each coherence block consists of $\tau_c$ = 200 samples and  $\tau_p$ pilots. The pilots of first $\tau_p$ UEs are allocated randomly. The rest of UEs sequentially pick their pilots that give least interference to UEs in the current pilot set. In UL transmission, we set $\tau_u= \tau_c -\tau_p$ and  $\tau_d= \tau_c -\tau_p$ in DL transmission, which means that each coherence block is either used for only UL data or only DL data.

In the UL, all UEs transmit with the same power $200$ mW.  In the two-layer decoding scheme, we utilized the maximizing LSFD vectors for each estimator that is computed by only using large-scale fading coefficients. In single-layer decoding, the LSFD vectors are $\mathbf{a}_k =[1,\dots, 1], \forall k $ which simply correspond to conventional MR combining. The same DL power is assigned to UE~$k$ for both coherent and non-coherent transmission.  The power is allocated proportional to the channel quality of UE by using the matrices for the coherent and non-coherent case respectively as $
\mathbf{D}_k =\frac{\rho^\mathrm{dl}}{K} \mathrm{diag}\left\lbrace 	\frac{ \eta_{1,k}}{ \mathbb{E}\left\lbrace |\hat{h}_{1,k}|^2 \right\rbrace}, \dots, \frac{ \eta_{M,k}}{\mathbb{E}\left\lbrace |\hat{h}_{M,k}|^2 \right\rbrace} \right\rbrace $
and $
\mathbf{D}_k =\frac{\rho^\mathrm{dl}}{K} \mathrm{diag}\left\lbrace \eta_{1,k} , \dots , \eta_{M,k}\right\rbrace$
where $\eta_{m,k} =\frac{\beta_{m,k} + \bar{h}^2_{m,k}}{\sum_{l=1}^{K} \beta_{m,l} + \bar{h}^2_{m,l}}$ is the power fraction parameter ($0 \leq \eta_{m,k} \leq 1$) and $\rho^\mathrm{dl} = K \times 200$ mW.

\begin{figure}[h]
	\includegraphics[scale=0.45]{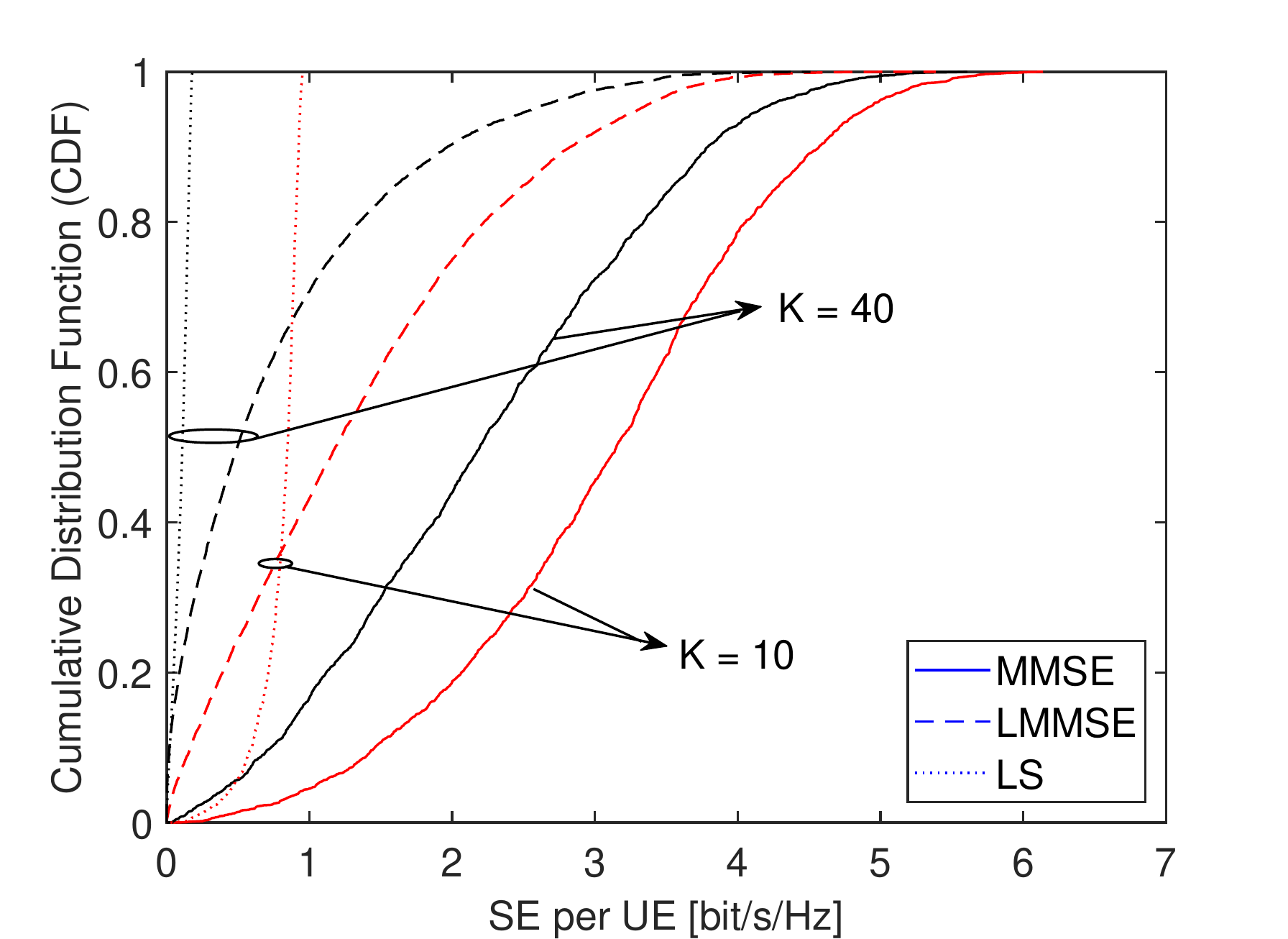}
\caption{CDF of DL SE with coherent transmission for different number of UEs where $M = 100$, $K = [10, 40]$, $\tau_p = 5$. }\label{fig9}
\end{figure}

\begin{figure}[h]
		\includegraphics[scale=0.45]{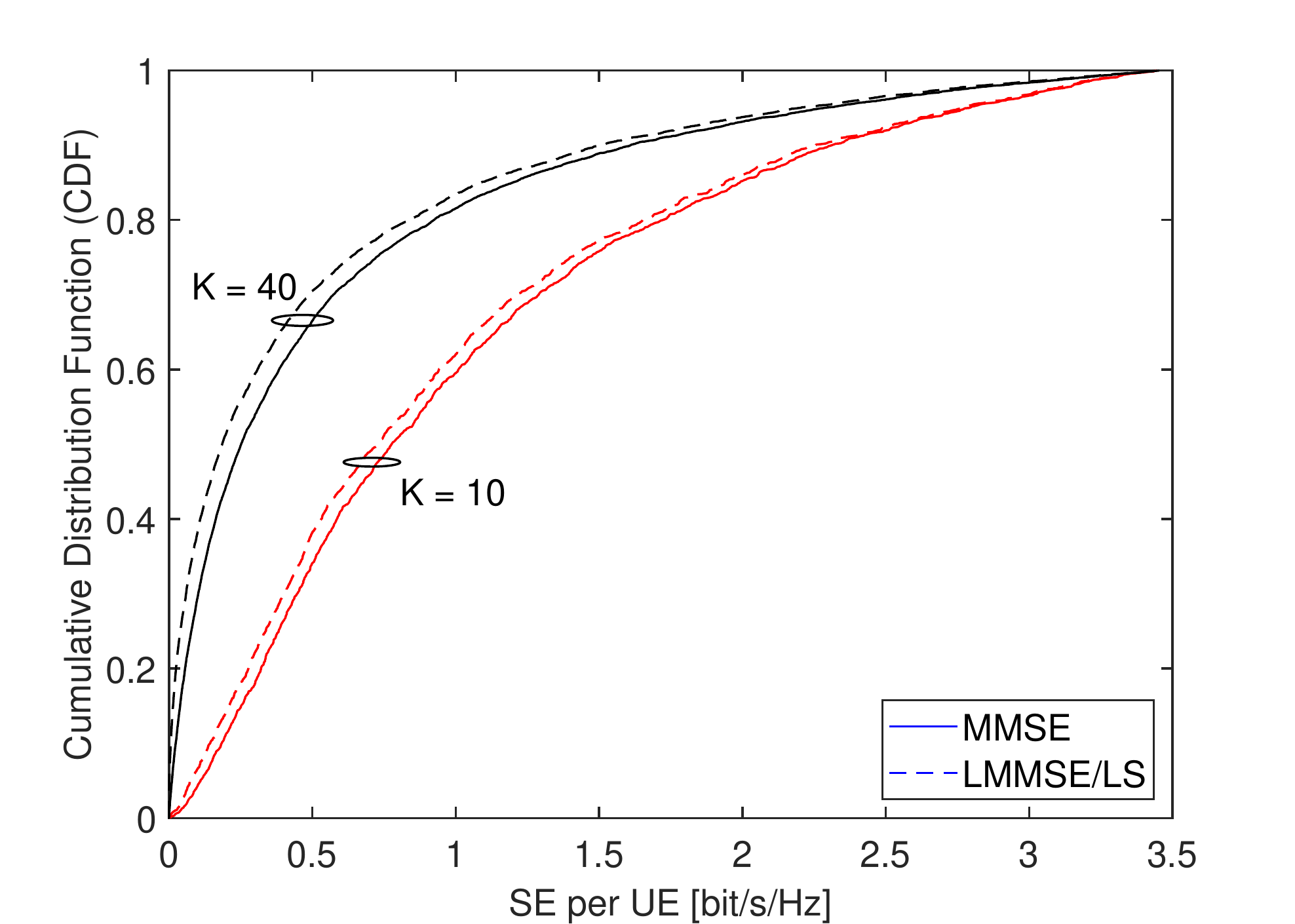}
	\caption{CDF of DL SE with non-coherent transmission for different number of UEs where $M = 100$, $K = [10, 40]$, $\tau_p = 5$. } \label{fig10}
\end{figure}

Fig.~\ref{fig1} shows the average UL SE as function of the number of APs with the phase-aware MMSE, LMMSE, and LS estimators. The average is taken over different UE locations and shadow fading realizations. The ``$\square$'' markers are generated by Monte Carlo simulations. The fact that the markers overlap with the curves validates of our analytical results.  Fig.~\ref{fig2} compares the cumulative distribution functions (CDFs) of the UL SE of estimators with two-layer and single-layer decoding. The randomness is due to random UE locations and shadow fading realizations. From Fig.~\ref{fig1}  and Fig.~\ref{fig2}, we observe that two-layer decoding gives improvements in all the cases. Especially, the LS estimator benefits the most from using LSFD since it is more vulnerable to interference. Also, the LS and LMMSE estimators coincide if the maximizing LSFD vector is used as discussed in Section \ref{LS:LSFD}.

Fig.~\ref{fig3} shows the CDFs of the SEs with different estimators with two-layer decoding for different pilot lengths. In case of LMMSE and LS estimation, increasing the pilot length reduces the pilot contamination and improves the estimation quality. Thus, the CDF curve of LMMSE/LS estimators gets closer to the phase-aware MMSE estimator with $\tau_p =20$. However, the MMSE estimator can already suppress the interference and therefore increasing the pilot length reduces the pre-log factor in the SE expression more than the SINR is improved, thereby reducing the performance slightly. Moreover, the gaps between phase-aware MMSE and LMMSE estimators for different pilot lengths indicate that the amount of performance loss due to lack of phase knowledge depends on the degree of pilot contamination. The gap becomes larger when the pilot length decrease. The loss is $6.9\%$ when $\tau_p =20$ is used and $24.8 \%   $ when $\tau_p = 5$.

 Fig.~\ref{fig4} compares the CDF of the SEs with different estimators with two-layer decoding for different number of UEs.  As the number of UEs decrease while the pilot length is kept the same, the interference and pilot contamination decrease and each UE gets higher SEs.

Fig.~\ref{fig5} shows the average DL SE as a function of the number of APs with coherent and non-coherent transmission for the phase-aware MMSE, LMMSE, and LS estimators. The average is taken over different UE locations and shadow fading realizations. Fig.~\ref{fig6} shows the CDF of DL SE with phase-aware MMSE, LMMSE, and LS estimators for coherent and non-coherent transmission.  These figures demonstrate that the coherent transmission provides higher SE in almost  all cases except when using  the LS estimator. Also, the performance gap between the phase-aware MMSE and LMMSE/LS estimator is higher in the coherent case since it is sensitive to phase errors.  

Fig.~\ref{fig7} shows the CDF of DL SE with phase-aware MMSE, LMMSE and LS estimators for different pilot lengths with coherent transmission. Similar to Fig.~\ref{fig3}, the LMMSE and LS estimators benefit from increased pilot length where the MMSE estimator does not. Also, we observe that the performance gap due to the lack of phase knowledge between phase-aware MMSE and LMMSE estimators are $42.6 \%$ and $13.4 \%$ for $\tau_p = 5$ and $\tau_p =20$ respectively. In Fig.~\ref{fig8}, we consider the same setting as in Fig.~\ref{fig7}  but with non-coherent transmission. The effect of phase knowledge is small in the non-coherent transmission as expected. Quantitatively, it is $10.9 \%$ for $\tau_p = 5$ and $2.4 \%$ for $\tau_p = 20$.

Fig.~\ref{fig9} and  Fig.~\ref{fig10} show  the CDF of DL SE with phase-aware MMSE, LMMSE and LS estimators for different number of UEs with coherent transmission and non-coherent transmission respectively. In both figures, the performance is better when the number of UEs in the network is lower in all cases.

\section{Conclusion}
This paper studied the SE of a cell-free massive MIMO system  over Rician fading channels. The phase of the LoS path is modeled as a uniformly distributed random variable  to take phase-shifts due to mobility  and phase noise into account. To determine the importance of knowing the phase, the phase-aware MMSE, non-aware LMMSE, and LS estimators were derived. In the UL, a two-layer decoding method was studied in order to mitigate the both coherent and non-coherent interference. We observed that the LSFD method provides a substantial gain in UL SE for all estimators, and should therefore always be used in cell-free massive MIMO. Furthermore, the performance losses as a result of unavailable phase knowledge depends on the pilot length. If there is no strongly interfering user, the LoS and NLoS paths  can be jointly estimated without having to know the phase. In the tested scenarios, we observed  $6.9 \%$ performance loss for low pilot contamination and $24.8 \%$ for high pilot contamination.

In the DL part, coherent and non-coherent transmission were studied.
We noticed that the losses from the lack of phase knowledge are  $13.4 \%$ and $2.4 \%$ for coherent and non-coherent transmission respectively in the tested scenarios with low pilot contamination. When we reduced the pilot length, it loss increased up to $42.6 \%$.
Therefore, the pilot length should be adjusted by taking the phase shifts into account in high mobility or low-quality hardware scenarios. In order to deal with the cases where there is not enough pilots, methods to explicitly estimate the phases could be considered in future work. Besides, the coherent transmission performs much better than the non-coherent one, which is the reason of its wide use in the cell-free literature.

\appendices
\section{Derivation of the LMMSE Estimator }\label{AppendixLMMSE}
The estimation is performed based on the received pilot signal in \eqref{se3:eq2} where $h_{m,k}$ is the desired part. The LMMSE estimator and the mean-square error are respectively given in \cite[Chapter 12]{KayBookESt} as
\begin{align}\label{lmmse_eq1}
&\hat{h}^\mathrm{lmmse}_{m,k} = \frac{\mathbb{E}\left\lbrace h_{m,k} (y^p_{m,k})^* \right\rbrace }{\mathbb{E}\left\lbrace |y^p_{m,k}|^2 \right\rbrace}y^p_{m,k}, \\
&c'_{m,k}= \mathbb{E}\left\lbrace | h_{m,k}|^2  \right\rbrace - \frac{\left| \mathbb{E}\left\lbrace h_{m,k} (y^p_{m,k})^* \right\rbrace\right| ^2 }{\mathbb{E}\left\lbrace |y^p_{m,k}|^2 \right\rbrace}.
\end{align}
Computing the following expectations and substituting them into \eqref{lmmse_eq1} gives the LMMSE estimator and the corresponding MSE as given in Section \ref{Section3B}:
\begin{align}
&\mathbb{E}\left\lbrace h_{m,k} (y^p_{m,k})^* \right\rbrace = \sqrt{\hat{p}_{k}} \tau_p\mathbb{E}\left\lbrace |h_{m,k}|^2  \right\rbrace  +  \mathbb{E}\left\lbrace h_{m,k}(\boldsymbol{\phi}^H_k\mathbf{n}^p_m)^* \right\rbrace \nonumber\\ 
&+  \sum_{l\in \mathcal{P}_k \backslash \{k\}} \sqrt{\hat{p}_{l}}\tau_p \mathbb{E}\left\lbrace h_{m,k} h^*_{m,l}\right\rbrace  
= \sqrt{\hat{p}_{k}} \tau_p (\beta_{m,k} + \bar{h}^2_{m,k}), \\
&\mathbb{E}\left\lbrace |y^p_{m,k}|^2 \right\rbrace =  \sum_{l\in \mathcal{P}_k } \hat{p}_{l}\tau^2_p\mathbb{E}\left\lbrace |h_{m,l}|^2 \right\rbrace  + \mathbb{E}\left\lbrace |\boldsymbol{\phi}^H_k\mathbf{n}^p_m |^2 \right\rbrace \nonumber\\
& = \sum_{l \in \mathcal{P}_k } \hat{p}_{l}\tau^2_p (\beta_{m,l} + \bar{h}^2_{m,k}) \ +  \ \sigma^2_\mathrm{ul} \tau_p= \tau_p \lambda'_{m,k}
\end{align}
The mean and  variance of the estimator are $
	\mathbb{E} \left\lbrace \hat{h}^\mathrm{lmmse}_{m,k} \right\rbrace = \sqrt{\hat{p}_{k}}  \frac{\beta'_{m,k}}{ \lambda'_{m,k} } 	\mathbb{E} \left\lbrace y^p_{m,k} \right\rbrace = 0$ and $ \mathrm{Var} \left\lbrace \hat{h}^\mathrm{lmmse}_{m,k} \right\rbrace = \mathbb{E}\left\lbrace \|\hat{h}^\mathrm{lmmse}_{m,k} \|^2 \right\rbrace 
= \hat{p}_{k} \frac{(\beta'_{m,k})^2}{ (\lambda'_{m,k})^{2 } }
  	\mathbb{E} \left\lbrace |y^p_{m,k}|^2 \right\rbrace = \hat{p}_{k} \tau_p (\beta'_{m,k})^2 (\lambda'_{m,k})^{-1 }. $

\section{Derivation of the LS Estimator }\label{AppendixLS}
 The mean and variance of the LS estimator can be computed as
 \begin{align}
 	&\mathbb{E}\left\lbrace \hat{h}^\mathrm{ls}_{m,k} \right\rbrace = \frac{1}{\sqrt{\hat{p}_k} \tau_p} \mathbb{E}\left\lbrace y^p_{m,k}\right\rbrace =0, \\
 	& \mathrm{Var}\left\lbrace \hat{h}^\mathrm{ls}_{m,k} \right\rbrace = \frac{1}{\hat{p}_k \tau^2_p} \mathbb{E}\left\lbrace |y^p_{m,k}|^2\right\rbrace =\frac{1}{\hat{p}_k \tau_p}\lambda'_{m,k}.
 \end{align}
  The estimation error mean and variance are 
  \begin{align}
  &\mathbb{E}\left\lbrace \tilde{h}^\mathrm{ls}_{m,k} \right\rbrace =  \mathbb{E}\left\lbrace h_{m,k} -\frac{1}{\sqrt{\hat{p}_k} \tau_p}y^p_{m,k}\right\rbrace =0,\\
  &\mathrm{Var}\left\lbrace \tilde{h}^\mathrm{ls}_{m,k} \right\rbrace = \mathbb{E}\left\lbrace\left|  h_{m,k}\right|^2\right\rbrace  + \frac{1}{\hat{p}_k \tau^2_p} \mathbb{E}\left\lbrace |y^p_{m,k}|^2\right\rbrace   \\
  &  - \frac{1}{\sqrt{\hat{p}_k} \tau_p}\mathbb{E}\left\lbrace  h_{m,k} (y^p_{m,k})^*\right\rbrace- \frac{1}{\sqrt{\hat{p}_k} \tau_p}\mathbb{E}\left\lbrace  (h_{m,k})^* y^p_{m,k}\right\rbrace \nonumber\\
  &= \beta_{m,k} + \bar{h}^2_{m,k} + \frac{\lambda'_{m,k}}{\hat{p}_k \tau_p} -2(\beta_{m,k} + \bar{h}^2_{m,k} ) = \frac{\lambda'_{m,k}}{\hat{p}_k \tau_p} -\beta'_{m,k} .\nonumber
  \end{align}

\section{Proof of UL and DL SE with MMSE estimator }\label{AppendixUL_MMSE}
The expectations in \eqref{SINR_UL} and \eqref{DLSINR_k} are calculated here. We begin with $\mathbb{E}\left\lbrace {v}^*_{m,k}  h_{m,k}\right\rbrace = \mathbb{E}\left\lbrace (\hat{h}^\mathrm{mmse}_{m,k})^* \hat{h}^\mathrm{mmse}_{m,k}\right\rbrace = \hat{p}_k \tau_p \beta^2_{m,k} \lambda^{-1}_{m,k} + \bar{h}^2_{m,k}$ which gives
\begin{align}
& \mathbb{E}\left\lbrace \mathbf{v}^H_k \mathbf{A}^H_k {\mathbf{h}}_k\right\rbrace   =\sum_{m=1}^{M} \alpha^*_{m,k}\mathbb{E}\left\lbrace {v}^*_{m,k}  {h}_{m,k}\right\rbrace  \\
&=  \sum_{m=1}^{M} \alpha^*_{m,k} \left( \hat{p}_k \tau_p \beta^2_{m,k} \lambda^{-1}_{m,k} + \bar{h}^2_{m,k}\right) =  \mathrm{tr}\left(\hat{p}_{k} \tau_p \mathbf{A}_k\boldsymbol{\Omega}_k+ \mathbf{A}_k\mathbf{L}_{k}\right).\nonumber
\end{align}
Similarly, we compute the second term as
\begin{align}
& \mathbb{E}\left\lbrace\|\mathbf{A}_k\mathbf{v}_k\|^2\right\rbrace =   \\
& \sum_{m=1}^{M}	|  \alpha^*_{m,k}| ^2 \mathbb{E}\left\lbrace |  {v}^*_{m,k} |^2 \right\rbrace  =  \mathrm{tr}\left(\hat{p}_{k} \tau_p \mathbf{A}_k\boldsymbol{\Omega}_k \mathbf{A}^H_k + \mathbf{A}_k\mathbf{L}_{k}\mathbf{A}^H_k\right). \nonumber
\end{align}

The last expectation in the denominator of \eqref{SINR_UL} is written as 
\begin{align}
&\mathbb{E}\left\lbrace \left| \mathbf{v}^H_k \mathbf{A}^H_k {\mathbf{h}}_{l} \right|^2 \right\rbrace = \mathbb{E}\left\lbrace \left|  \sum_{m=1}^{M} \alpha^*_{m,k} {v}^*_{m,k}  {h}_{m,l} \right| ^2 \right\rbrace \\
&=\sum_{m=1}^{M} \sum_{n=1}^{M} \alpha_{m,k}\alpha^*_{n,k}  \mathbb{E}\left\lbrace \left(  {v}^*_{m,k}  {h}_{m,l} \right)^* \left(  {v}^*_{n,k}  {h}_{n,l} \right)\right\rbrace . \nonumber 
\end{align}

The $\mathbb{E}\left\lbrace \left(   v^*_{m,k}  {h}_{m,l} \right)^* \left(   v^*_{n,k}  {h}_{n,l} \right)\right\rbrace$ is computed for all possible AP and UE combinations. We utilize the independence of channel estimates at different APs. The first case is  $m\neq n$ and $l \notin \mathcal{P}_k$ and $\mathbb{E}\left\lbrace \left(  {v}^*_{m,k}  {h}_{m,l} \right)^* \left(  {v}^*_{n,k}  {h}_{n,l} \right)\right\rbrace  = 0$ since $ {v}^*_{n,k}$ and  $ {h}_{n,l}$ are independent and both have zero mean. For  $m\neq n$ and $l \in \mathcal{P}_k \backslash \{k\}$, we obtain 
\begin{align}\label{proof:mmse:1}
\mathbb{E}\left\lbrace (  {v}^*_{m,k}  {h}_{m,l} )^* (  {v}^*_{n,k}  {h}_{n,l} )\right\rbrace  = \mathbb{E}\left\lbrace (\hat{h}^\mathrm{mmse}_{m,k})^* \hat{h}^\mathrm{mmse}_{m,l}\right\rbrace \mathbb{E}\left\lbrace (\hat{h}^\mathrm{mmse}_{n,k})^* \hat{h}^\mathrm{mmse}_{n,l}\right\rbrace,
\end{align}
where 
\begin{align}
&\mathbb{E}\left\lbrace (\hat{h}^\mathrm{mmse}_{m,k})^* \hat{h}^\mathrm{mmse}_{m,l}\right\rbrace =\mathbb{E}\left\lbrace \left( \sqrt{\hat{p}_k} \beta_{m,k} \lambda^{-1}_{m,k}(y^p_{m,k} - \bar{y}^p_{m,k}) + \bar{h}_{m,k}e^{j \varphi_{m,k}}\right)^* \right. \nonumber \\
&\times \left. \left( \sqrt{\hat{p}_{l}} \beta_{m,l} \lambda^{-1}_{m,k}(y^p_{m,k} - \bar{y}^p_{m,k}) + \bar{h}_{m,l} e^{j \varphi_{m,l}}\right)\right\rbrace \nonumber\\
&= \sqrt{\hat{p}_k \hat{p}_{l}} \tau_p\beta_{m,l} \beta_{m,k} \lambda^{-1}_{m,k} ,
\end{align}
since $\mathbb{E}\left\lbrace \bar{h}_{m,k} \bar{h}_{m,l}e^{-j \varphi_{m,k}} e^{j \varphi_{m,l}}\right\rbrace = 0$ and $\mathbb{E}\left\lbrace ( \sqrt{\hat{p}_k} \beta_{m,k} \lambda^{-1}_{m,k}\right.\\ \left.(y^p_{m,k} - \bar{y}^p_{m,k}))^*  \bar{h}_{m,l} e^{j\varphi_{m,l}}\right\rbrace = 0 $. Repeating the same calculation for AP $n$ and putting it into \eqref{proof:mmse:1} gives
\begin{align}
\mathbb{E}\left\lbrace (  {v}^*_{m,k}  {h}_{m,l} )^* (  {v}^*_{n,k}  {h}_{n,l} )\right\rbrace  = \hat{p}_k \hat{p}_{l} \tau^2_p\beta_{m,l} \beta_{m,k} \lambda^{-1}_{m,k} \beta_{n,l} \beta_{n,k} \lambda^{-1}_{n,k}.
\end{align}
Another case is  $m\neq n$ and $l = k$ and we obtain
\begin{align}
&\mathbb{E}\left\lbrace \left(  {v}^*_{m,k}  {h}_{m,k} \right)^* \left(  {v}^*_{n,k}  {h}_{n,k} \right)\right\rbrace = \hat{p}^2_k  \tau^2_p\beta^2_{m,k}  \lambda^{-1}_{m,k} \beta^2_{n,k} \lambda^{-1}_{n,k}  \nonumber \\
&+  \bar{h}^2_{m,k} \bar{h}^2_{n,k} + \hat{p}_k  \tau_p\beta^2_{m,k}  \lambda^{-1}_{m,k}  \bar{h}^2_{n,k} + \hat{p}_k  \tau_p\beta^2_{n,k}  \lambda^{-1}_{n,k}  \bar{h}^2_{m,k}.
\end{align}
Similarly for $m =n$ and  $l = k$, we calculate the following equations
\begin{align}
&\mathbb{E}\left\lbrace  |  \hat{h}^*_{m,k}  \tilde{h}_{m,k} | ^2  \right\rbrace = \left( \hat{p}_k \tau_p \beta^2_{m,k} \lambda^{-1}_{m,k} + \bar{h}^2_{m,k}\right) c_{m,k} \label{mmse:proof:1},\\
&\mathbb{E}\left\lbrace  |  \hat{h}_{m,k}   | ^4 \right\rbrace = \mathbb{E}\left\lbrace  \left|  \left( \sqrt{\hat{p}_k \tau_p} \beta_{m,k} \lambda^{-1/2}_{m,k}w^* + \bar{h}_{m,k}e^{-j \varphi_{m,k}}\right) \right. \right. \nonumber\\
& \times \left. \left. \left( \sqrt{\hat{p}_k \tau_p} \beta_{m,k} \lambda^{-1/2}_{m,k}w + \bar{h}_{m,k}e^{j \varphi_{m,k}}\right) \right|  ^2 \right\rbrace = \hat{p}^2_{k}  \tau^2_p \beta^4_{m,k} \lambda^{-2}_{m,k} \mathbb{E}\left\lbrace  \left| w   \right|^4\right\rbrace \nonumber \\
& + 4\hat{p}_{k} \tau_p\beta^2_{m,k} \lambda^{-1}_{m,k} \bar{h}^2_{m,k} \mathbb{E}\left\lbrace  \left| w   \right|^2\right\rbrace   + \bar{h}^4_{m,k} =2\hat{p}^2_{k}  \tau^2_p \beta^4_{m,k} \lambda^{-2}_{m,k} \nonumber \\
& + 3\hat{p}_{k} \tau_p\beta^2_{m,k} \lambda^{-1}_{m,k} \bar{h}^2_{m,k}  + (\beta_{m,k} - c_{m,k}) \bar{h}^2_{m,k}  + \bar{h}^4_{m,k} . \label{mmse:proof:2}
\end{align}
where $w \sim \mathcal{N}_\mathbb{C} (0, 1)$.  Combining \eqref{mmse:proof:1} and \eqref{mmse:proof:2} gives the result for $m =n$ and  $l = k$ as
\begin{align}
&\mathbb{E}\left\lbrace  |  \hat{h}^*_{m,k}  {h}_{m,k} | ^2  \right\rbrace = \mathbb{E}\left\lbrace  |  \hat{h}_{m,k}   | ^4 \right\rbrace + \mathbb{E}\left\lbrace  |  \hat{h}^*_{m,k}  \tilde{h}_{m,k} | ^2  \right\rbrace  =\hat{p}_k \tau_p \beta^3_{m,k} \lambda^{-1}_{m,k} \nonumber \\
&+  \beta_{m,k} \bar{h}^2_{m,k} + 3\hat{p}_k \tau_p \beta^2_{m,k}\lambda^{-1}_{m,k}\bar{h}^2_{m,k} + \bar{h}^4_{m,k}  +\hat{p}^2_k \tau^2_p \beta^4_{m,k}  \lambda^{-2}_{m,k}.
\end{align}

Putting all the equations together for $l = k$, we obtain
\begin{align}
&\mathbb{E}\left\lbrace \left| \mathbf{v}^H_k \mathbf{A}^H_k {\mathbf{h}}_{k} \right|^2 \right\rbrace =\sum_{m=1}^{M} \sum_{n=1}^{M} \alpha_{m,k}\alpha^*_{n,k}  \mathbb{E}\left\lbrace \left(  {v}^*_{m,k}  {h}_{m,k} \right)^* \left(  {v}^*_{n,k}  {h}_{n,k} \right)\right\rbrace \nonumber \\
& =\sum_{m=1}^{M} \sum_{n=1}^{M} \alpha_{m,k}\alpha^*_{n,k}\left( \hat{p}^2_k  \tau^2_p\beta^2_{m,k}  \lambda^{-1}_{m,k} \beta^2_{n,k} \lambda^{-1}_{n,k}  +  \bar{h}^2_{m,k} \bar{h}^2_{n,k}  \right. \nonumber \\
&+ \left. \hat{p}_k  \tau_p\beta^2_{m,k}  \lambda^{-1}_{m,k}  \bar{h}^2_{n,k} + \hat{p}_k  \tau_p\beta^2_{n,k}  \lambda^{-1}_{n,k}  \bar{h}^2_{m,k}\right)  \nonumber \\
&+ \sum_{m=1}^{M} |\alpha_{m,k}|^2\left( \hat{p}_k \tau_p \beta^3_{m,k} \lambda^{-1}_{m,k} +  \beta_{m,k} \bar{h}^2_{m,k} + \hat{p}_k \tau_p \beta^2_{m,k}\lambda^{-1}_{m,k}\bar{h}^2_{m,k}  \right. \nonumber \\ 
& \left. +\hat{p}^2_k \tau^2_p \beta^4_{m,k}  \lambda^{-2}_{m,k}  \right) = \hat{p}_{k} \tau_p \mathrm{tr}\left( \mathbf{A}^H_k\mathbf{R}_{k}  \mathbf{A}_k \boldsymbol{\Omega}_k \right) + \mathrm{tr}\left( \mathbf{A}_k \mathbf{L}_{k}\right) ^2   \nonumber \\
&+   \mathrm{tr}\left( \mathbf{A}^H_k\mathbf{R}_{k} \mathbf{A}_k\mathbf{L}_{k} \right) +\hat{p}^2_{k} \tau^2_p |\mathrm{tr}\left(   \mathbf{A}_k\boldsymbol{\Omega}_k\right) |^2 + \hat{p}_{k} \tau_p \mathrm{tr}\left( \mathbf{A}^H_k\boldsymbol{\Omega}_k \mathbf{A}_k\mathbf{L}_{k}\right) \nonumber \\
&+ 2\hat{p}_k \tau_p \mathrm{tr}\left( \mathbf{A}_k \boldsymbol{\Omega}_k\right) \mathrm{tr}\left( \mathbf{A}_k \mathbf{L}_{k}\right).
\end{align}

The remaining cases are computed as follows. For $ l \notin \mathcal{P}_{k} $ and $m =n$, $
\mathbb{E}\left\lbrace  | {v}^*_{m,k}  {h}_{m,l} |^2 \right\rbrace  = ( \hat{p}_k \tau_p \beta^2_{m,k} \lambda^{-1}_{m,k} + \bar{h}^2_{m,k}) ( \beta_{m,l} + \bar{h}^2_{m,l}).$ For $ l \in \mathcal{P}_{k} \backslash \{k\}$ and $m =n$, we obtain 
\begin{align}
&\mathbb{E}\left\lbrace   \left| {v}^*_{m,k}  {h}_{m,l}  \right|^2 \right\rbrace  = \mathbb{E}\left\lbrace  \left| \hat{h}^*_{m,k}  \hat{h}_{m,l}   \right|^2\right\rbrace + \mathbb{E}\left\lbrace   \left|\hat{h}^*_{m,k}  \tilde{h}_{m,l} \right|^2 \right\rbrace ,\\
&\mathbb{E}\left\lbrace   \left|\hat{h}^*_{m,k}  \tilde{h}_{m,l} \right|^2 \right\rbrace =\left( \hat{p}_k \tau_p \beta^2_{m,k} \lambda^{-1}_{m,k} + \bar{h}^2_{m,k}\right) c_{m,l}.
\end{align}
To calculate $\mathbb{E}\left\lbrace  \left| \hat{h}^*_{m,k}  \hat{h}_{m,l}   \right|^2\right\rbrace $, we rewrite the MMSE estimator in \eqref{se3:eq4} for UEs $l \in \mathcal{P}_k  \backslash \{k\}$ as 
\begin{align}
&	\hat{h}_{m,k} =  \sqrt{\hat{p}_k} \beta_{m,k} \lambda^{-1}_{m,k}(y^p_{m,k} - \bar{y}^p_{m,k}) + \bar{h}_{m,k}e^{j \varphi_{m,k}} \nonumber \\
&= \sqrt{\hat{p}_k} \beta_{m,k} \lambda^{-1}_{m,k}(\sqrt{\tau_p}\frac{1}{\sqrt{\tau_p}} \lambda^{1/2}_{m,k}\lambda^{-1/2}_{m,k}  )(y^p_{m,k} - \bar{y}^p_{m,k}) + \bar{h}_{m,k}e^{j \varphi_{m,k}} \nonumber \\
&= \sqrt{\hat{p}_k \tau_p} \beta_{m,k} \lambda^{-1/2}_{m,k}w + \bar{h}_{m,k}e^{j \varphi_{m,k}} 
\end{align}
and similarly $\hat{h}_{m,l}  = \sqrt{\hat{p}_{l} \tau_p} \beta_{m,l} \lambda^{-1/2}_{m,k}w + \bar{h}_{m,l} e^{j \varphi_{m,l}}$ where $(y^p_{m,k} - \bar{y}^p_{m,k}) \sim \mathcal{N}_\mathbb{C}\left( 0, \tau_p \lambda_{m,k} \right)  $ and $w \sim \mathcal{N}_\mathbb{C}\left( 0, 1\right) $. After this computation, we directly compute
\begin{align}
&\mathbb{E}\left\lbrace  \left| \hat{h}^*_{m,k}  \hat{h}_{m,l}   \right|^2\right\rbrace = \mathbb{E}\left\lbrace  \left|  \left( \sqrt{\hat{p}_k \tau_p} \beta_{m,k} \lambda^{-1/2}_{m,k}w^* + \bar{h}_{m,k}e^{-j \varphi_{m,k}}\right) \right. \right. \nonumber \\
& \times \left. \left. \left( \sqrt{\hat{p}_{l} \tau_p} \beta_{m,l} \lambda^{-1/2}_{m,k}w + \bar{h}_{m,l}e^{j \varphi_{m,l}}\right) \right|^2\right\rbrace \nonumber \\
	&= \hat{p}_{k} \hat{p}_{l} \tau^2_p \left(  \beta_{m,l} \beta_{m,k} \lambda^{-1}_{m,k}\right) ^2 \mathbb{E}\left\lbrace  \left| w   \right|^4\right\rbrace   \nonumber \\
	&+ \hat{p}_{k} \tau_p\beta^2_{m,k} \lambda^{-1}_{m,k} \bar{h}^2_{m,l} \mathbb{E}\left\lbrace  | w   |^2\right\rbrace + \hat{p}_{l} \tau_p\beta^2_{m,l} \lambda^{-1}_{m,k} \bar{h}^2_{m,k} \mathbb{E}\left\lbrace  \left| w   \right|^2\right\rbrace  \nonumber \\
	&+ \bar{h}^2_{m,l}\bar{h}^2_{m,k} = 2\hat{p}_{k} \hat{p}_{l} \tau^2_p  \beta^2_{m,l} \beta^2_{m,k} \lambda^{-2}_{m,k} \nonumber \\
	&+ \hat{p}_{k} \tau_p\beta^2_{m,k} \lambda^{-1}_{m,k} \bar{h}^2_{m,l} + \hat{p}_{l} \tau_p\beta^2_{m,l} \lambda^{-1}_{m,k} \bar{h}^2_{m,k} + \bar{h}^2_{m,l}\bar{h}^2_{m,k} \nonumber \\
	&=2\hat{p}_{k} \tau_p  \beta^2_{m,k} \lambda^{-1}_{m,k} (\beta_{m,l} - c_{m,l})+ \hat{p}_{k} \tau_p\beta^2_{m,k} \lambda^{-1}_{m,k} \bar{h}^2_{m,l} \nonumber \\
	&+ (\beta_{m,l} - c_{m,l})\bar{h}^2_{m,k} + \bar{h}^2_{m,l}\bar{h}^2_{m,k}.
\end{align}

Then, it yields
\begin{align}
&\mathbb{E}\left\lbrace  |  {v}^*_{m,k}  {h}_{m,l} | ^2  \right\rbrace = \hat{p}_{k} \tau_p \beta_{m,l} \beta^2_{m,k} \lambda^{-1}_{m,k} +  \beta_{m,l} \bar{h}^2_{m,k} + \bar{h}^2_{m,k} \bar{h}^2_{m,l}  \nonumber  \\ 
& +\hat{p}_{k}\tau_p \beta^2_{m,k}\lambda^{-1}_{m,k}\bar{h}^2_{m,l} +\begin{cases}
\hat{p}_{k} \hat{p}_{l} \tau^2_p \beta^2_{m,l} \beta^2_{m,k} \lambda^{-2}_{m,k}, &  l \in \mathcal{P}_{k} \backslash \{k\} \\
0, &  l \notin \mathcal{P}_{k}.
\end{cases}
\end{align}

Finally, arranging all these equations gives
\begin{align}
\mathbb{E}\left\lbrace \left| \mathbf{v}^H_k \mathbf{A}^H_k {\mathbf{h}}_{l} \right|^2 \right\rbrace & = \hat{p}_{k} \tau_p \mathrm{tr}\left( \mathbf{A}^H_k\mathbf{R}_{l}  \mathbf{A}_k \boldsymbol{\Omega}_k\right) 
+ \hat{p}_{k} \tau_p \mathrm{tr}\left( \mathbf{A}^H_k\boldsymbol{\Omega}_k \mathbf{A}_k\mathbf{L}_{l}\right) \nonumber \\
&+   \mathrm{tr}\left( \mathbf{A}^H_k\mathbf{R}_{l} \mathbf{A}_k\mathbf{L}_{k} \right)  + \mathrm{tr}\left( \mathbf{A}^H_k\mathbf{S}_{k,l} \mathbf{A}_k\right)  \\
&+\begin{cases}
	\hat{p}_{k} \hat{p}_{l}  \tau^2_p |\mathrm{tr}\left(   \mathbf{A}_k\mathbf{R}_{l}  \boldsymbol{\Lambda}_{k} \mathbf{R}_{k}\right) |^2 , &  l \in \mathcal{P}_{k} \backslash \{k\} \\
	\hat{p}^2_{k} \tau^2_p |\mathrm{tr}\left(   \mathbf{A}_k\boldsymbol{\Omega}_k\right) |^2 + \mathrm{tr}\left( \mathbf{A}_k \mathbf{L}_{k}\right) ^2 \\
	+ 2\hat{p}_k \tau_p \mathrm{tr}\left( \mathbf{A}_k \boldsymbol{\Omega}_k\right) \mathrm{tr}\left( \mathbf{A}_k \mathbf{L}_{k}\right),  & l =k \\
	0, & l \notin \mathcal{P}_{k}.
	\end{cases} \nonumber 
\end{align}
where $\mathbf{S}_{k,l}= \mathbf{L}_k \mathbf{L}_l$ if $l \neq k$ and zero matrix otherwise. 
Inserting the required results into  \eqref{SINR_UL} and \eqref{DLSINR_k} gives the results for UL and DL SEs respectively.

\section{Proof of UL and DL SE with LMMSE Estimator }\label{AppendixUL_LMMSE} 

In this part, we note that the LMMSE estimate is not Gaussian distributed. We begin with computing the first term 
\begin{align}\label{lmmse:proof:1}
&\mathbb{E}\left\lbrace {v}^*_{m,k}  h_{m,k}\right\rbrace = \mathbb{E}\left\lbrace (\hat{h}^\mathrm{lmmse}_{m,k})^* (\hat{h}^\mathrm{lmmse}_{m,k} + \tilde{h}^\mathrm{lmmse}_{m,k})\right\rbrace \nonumber \\
&=  \mathbb{E}\left\lbrace |\hat{h}^\mathrm{lmmse}_{m,k} |^2 \right\rbrace + \mathbb{E}\left\lbrace (\hat{h}^\mathrm{lmmse}_{m,k})^* \tilde{h}^\mathrm{lmmse}_{m,k}\right\rbrace \nonumber \\
&\overset{(a)}{=} \hat{p}_k \tau_p(\beta'_{m,k})^2 (\lambda'_{m,k} )^{-1} + \mathbb{E}\left\lbrace (\hat{h}^\mathrm{lmmse}_{m,k})^* \right\rbrace \mathbb{E}\left\lbrace\tilde{h}^\mathrm{lmmse}_{m,k}\right\rbrace \nonumber\\
&\overset{(b)}{=}\hat{p}_k \tau_p(\beta'_{m,k})^2 (\lambda'_{m,k} )^{-1}
\end{align}
where (a) uses the fact that the estimate and estimation error are uncorrelated and (b) uses that both estimates have zero mean. Calculating \eqref{lmmse:proof:1} for each AP $m$ gives
\begin{align}\label{lmmse:proof:2}
&\mathbb{E}\left\lbrace \mathbf{v}^H_k \mathbf{A}^H_k {\mathbf{h}}_k\right\rbrace     =\sum_{m=1}^{M} \alpha^*_{m,k}\mathbb{E}\left\lbrace {v}^*_{m,k}  {h}_{m,k}\right\rbrace \nonumber \\
&=  \sum_{m=1}^{M} \alpha^*_{m,k}  \hat{p}_k \tau_p(\beta'_{m,k})^2 (\lambda'_{m,k} )^{-1} = \hat{p}_{k} \tau_p  \mathrm{tr}\left(\mathbf{A}_k \boldsymbol{\Omega}'_k  \right).
\end{align}
Similarly, we obtain
\begin{align} \label{lmmse:proof:3}
&\mathbb{E}\left\lbrace\|\mathbf{A}_k\mathbf{v}_k\|^2\right\rbrace = \mathbb{E}\left\lbrace \left|  \sum_{m=1}^{M} \alpha_{m,k} {v}_{m,k}  \right| ^2 \right\rbrace = \mathbb{E}\left\lbrace \sum_{m=1}^{M} \alpha^*_{m,k} {v}^*_{m,k}    \sum_{n=1}^{M} \alpha_{n,k} {v}_{n,k}   \right\rbrace \nonumber \\
&\overset{(a)}{=} \sum_{m=1}^{M} |\alpha_{m,k}|^2 \mathbb{E}\left\lbrace  |\hat{h}^\mathrm{lmmse}_{m,k} |^2     \right\rbrace= \hat{p}_{k} \tau_p \mathrm{tr}\left( \mathbf{A}_k \boldsymbol{\Omega}'_k \mathbf{A}^H_k \right),
\end{align}
where (a) utilizes the independence of the zero mean channel estimates at different APs.

The last expectation in the denominator of \eqref{SINR_UL} is written as
\begin{align}\label{lmmse:proof:4}
&	\mathbb{E}\left\lbrace \left| \mathbf{v}^H_k \mathbf{A}^H_k {\mathbf{h}}_{l} \right|^2 \right\rbrace= \sum_{m=1}^{M} \sum_{n=1}^{M} \alpha_{m,k}\alpha^*_{n,k}  \mathbb{E}\left\lbrace \left(   v^*_{m,k}  {h}_{m,l} \right)^* \left(   v^*_{n,k}  {h}_{n,l} \right)\right\rbrace .
\end{align}
Then, we need to compute $\mathbb{E}\left\lbrace \left(   v^*_{m,k}  {h}_{m,l} \right)^* \left(   v^*_{n,k}  {h}_{n,l} \right)\right\rbrace$ for all possible AP and UE combinations. If $ m \neq n $ and $ l \notin \mathcal{P}_k$ then all the terms in the expectation are uncorrelated or independent and have zero mean which gives  $\mathbb{E}\left\lbrace \left(   v^*_{m,k}  {h}_{m,l} \right)^* \left(   v^*_{n,k}  {h}_{n,l} \right)\right\rbrace = 0$. However, for $m \neq n$ and $ l \in  \mathcal{P}_k$, we have $
	\mathbb{E}\left\lbrace \left(   v^*_{m,k}  {h}_{m,l} \right)^* \left(   v^*_{n,k}  {h}_{n,l} \right)\right\rbrace = \mathbb{E}\left\lbrace   v_{m,k}  {h}^*_{m,l}  \right\rbrace \mathbb{E}\left\lbrace   v^*_{n,k}  {h}_{n,l}  \right\rbrace,$
since the pilot contaminating estimators at the same AP are not independent. We calculate
\begin{align}\label{lmmse:proof:6}
&\mathbb{E}\left\lbrace   \hat{h}^\mathrm{lmmse}_{m,k}  {h}^*_{m,l}  \right\rbrace = \mathbb{E}\left\lbrace   \hat{h}^\mathrm{lmmse}_{m,k}  (\hat{h}^\mathrm{lmmse}_{m,l} + \tilde{h}^\mathrm{lmmse}_{m,l} )^* \right\rbrace \\
&= \mathbb{E}\left\lbrace   \hat{h}^\mathrm{lmmse}_{m,k}  (\hat{h}^\mathrm{lmmse}_{m,l} )^* \right\rbrace \nonumber\\
&=\mathbb{E}\left\lbrace \left(  \sqrt{\hat{p}_{k}} \beta'_{m,k} (\lambda'_{m,k})^{-1 }   	 y^p_{m,k}\right)  \left(  \sqrt{\hat{p}_{l}} \beta'_{m,l} (\lambda'_{m,k})^{-1 }   	 y^p_{m,k}\right)^* \right\rbrace \nonumber  \\
&= \sqrt{\hat{p}_{k}\hat{p}_{l}} \beta'_{m,k} \beta'_{m,l} (\lambda'_{m,k})^{-2 }  \mathbb{E}\left\lbrace |y^p_{m,k}|^2\right\rbrace =\sqrt{\hat{p}_{k}\hat{p}_{l}} \tau_p \frac{\beta'_{m,k} \beta'_{m,l} }{\lambda'_{m,k}} . \nonumber 
\end{align}

Repeating the same process for  $\mathbb{E}\left\lbrace   v^*_{n,k}  {h}_{n,l}  \right\rbrace$ gives the final result for  $m \neq n$ and $ l \in  \mathcal{P}_k$  as $
\mathbb{E}\left\lbrace \left(   v^*_{m,k}  {h}_{m,l} \right)^* \left(   v^*_{n,k}  {h}_{n,l} \right)\right\rbrace = \hat{p}_{k}\hat{p}_{l} \tau^2_p \beta'_{m,k} \beta'_{m,l} (\lambda'_{m,k})^{-1 }\beta'_{n,k} \beta'_{n,l} (\lambda'_{n,k})^{-1 }. $ Another case is $ m = n$ and $ l \notin \mathcal{P}_k$ in which the channel estimators $\hat{h}^\mathrm{lmmse}_{m,k} $ and $\hat{h}^\mathrm{lmmse}_{m,l} $ are independent. We directly compute $\mathbb{E}\left\lbrace  | v^*_{m,k}  {h}_{m,l} |^2\right\rbrace = \mathbb{E}\left\lbrace  | (\hat{h}^\mathrm{lmmse}_{m,k})^* \hat{h}^\mathrm{lmmse}_{m,l} |^2\right\rbrace + \mathbb{E}\left\lbrace  | (\hat{h}^\mathrm{lmmse}_{m,k})^* \tilde{h}^\mathrm{lmmse}_{m,l} |^2\right\rbrace $ using uncorrelated zero mean estimators and estimation errors. The first and second terms are derived respectively as
\begin{align}\label{lmmse:proof:8}
&\mathbb{E}\left\lbrace  | (\hat{h}^\mathrm{lmmse}_{m,k})^* \hat{h}^\mathrm{lmmse}_{m,l} |^2\right\rbrace \nonumber \\
&= \mathbb{E}\left\lbrace  \left|  \left(  \sqrt{\hat{p}_{k}} \beta'_{m,k} (\lambda'_{m,k})^{-1 }   	 y^p_{m,k}\right)^* \left(  \sqrt{\hat{p}_{l}} \beta'_{m,l} (\lambda'_{m,l})^{-1 }   	 y^p_{m,l}\right)  \right| ^2\right\rbrace \nonumber \\
& = \hat{p}_{k} \hat{p}_{l} (\beta'_{m,k})^2  (\beta'_{m,l})^2 (\lambda'_{m,k})^{-2 } (\lambda'_{m,l})^{-2 }  \mathbb{E}\left\lbrace  |   y^p_{m,k} | ^2 	  \right\rbrace \mathbb{E}\left\lbrace  |   y^p_{m,l} | ^2 	  \right\rbrace \nonumber \\
& = \hat{p}_{k} \hat{p}_{l} \tau^2_p (\beta'_{m,k})^2  (\beta'_{m,l})^2 (\lambda'_{m,k})^{-1} (\lambda'_{m,l})^{-1 } ,\\
&\mathbb{E}\left\lbrace  | (\hat{h}^\mathrm{lmmse}_{m,k})^* \tilde{h}^\mathrm{lmmse}_{m,l} |^2\right\rbrace = \mathbb{E}\left\lbrace  | \hat{h}^\mathrm{lmmse}_{m,k}  |^2\right\rbrace \mathbb{E}\left\lbrace  |  \tilde{h}^\mathrm{lmmse}_{m,l} |^2\right\rbrace \nonumber \\
&= \hat{p}_k \tau_p(\beta'_{m,k})^2 (\lambda'_{m,k} )^{-1} c'_{m,l}.\label{lmmse:proof:9}
\end{align}
Rewriting the first term and combining \eqref{lmmse:proof:8} and \eqref{lmmse:proof:9} leads to $\mathbb{E}\left\lbrace  | v^*_{m,k}  {h}_{m,l} |^2\right\rbrace = \hat{p}_{k}  \tau_p (\beta'_{m,k})^2   (\lambda'_{m,k})^{-1} (\beta'_{m,l}- c'_{m,l}) + \hat{p}_k \tau_p(\beta'_{m,k})^2 (\lambda'_{m,k} )^{-1} c'_{m,l} =\hat{p}_{k}  \tau_p \beta'_{m,l} (\beta'_{m,k})^2   (\lambda'_{m,k})^{-1} $. The last case is  $ m = n$ and $ l \in \mathcal{P}_k$. In this case the channel estimators $\hat{h}^\mathrm{lmmse}_{m,k} $ and $\hat{h}^\mathrm{lmmse}_{m,l} $ are no longer independent. We obtain
\begin{align} \label{lmmse:proof:11}
&\mathbb{E}\left\lbrace  | (\hat{h}^\mathrm{lmmse}_{m,k})^* {h}^\mathrm{lmmse}_{m,l} |^2\right\rbrace  = \mathbb{E}\left\lbrace  \left|  \left(  \sqrt{\hat{p}_{k}} \beta'_{m,k} (\lambda'_{m,k})^{-1 }   	 y^p_{m,k}\right)^* h_{m,l}  \right| ^2\right\rbrace \nonumber \\
& = \hat{p}_{k} (\beta'_{m,k})^2 (\lambda'_{m,k})^{-2 }   \left( \hat{p}_{l} \tau^2_p \mathbb{E}\left\lbrace  \left|    h_{m,l}  \right| ^4\right\rbrace  + \mathbb{E}\left\lbrace \left|  (\boldsymbol{\phi}^H_k\mathbf{n}^p_m)^* h_{m,l}  \right| ^2 \right\rbrace \nonumber \right. \\
&\left.+   \mathbb{E}\left\lbrace \left|  \sum_{z \in \mathcal{P}_k \backslash \{l\}} \sqrt{\hat{p}_{z}} \tau_p h^*_{m,z} h_{m,l}  \right| ^2 \right\rbrace  \right) =\hat{p}_{k} (\beta'_{m,k})^2 (\lambda'_{m,k})^{-2 } \nonumber \\
&\times \left[ \hat{p}_{l} \tau^2_p \left( \beta^2_{m,l} + 2\bar{h}^2_{m,l} \beta_{m,l} \right) + \tau_p \lambda'_{m,k} \beta'_{m,l}  \right] \nonumber \\
& = \hat{p}_{k} \tau_p \beta'_{m,l} (\beta'_{m,k})^2 (\lambda'_{m,k})^{-1 } + \hat{p}_{k}  \hat{p}_{l} \tau^2_p (\beta'_{m,k})^2  (\lambda'_{m,k})^{-2} \beta^2_{m,l} \nonumber \\
&+ 2 \hat{p}_{k}  \hat{p}_{l} \tau^2_p (\beta'_{m,k})^2  (\lambda'_{m,k})^{-2} \bar{h}^2_{m,l} \beta_{m,l}
\end{align}
where $\mathbb{E}\left\lbrace  \left|    h_{m,l}  \right| ^4\right\rbrace= \mathbb{E}\left\lbrace  \left|   g_{m,l} + \bar{h}_{m,l} e^{j \varphi_{m,l}} \right| ^4\right\rbrace=\left( 2\beta^2_{m,l} + 4\bar{h}^2_{m,l} \beta_{m,l} +\bar{h}^4_{m,l} \right)$. Combining all the cases gives \eqref{lmmse:proof:13}, at the top of next page. Using matrix notation, we can reformulate the equation above as

	\begin{figure*}[h]
	\normalsize
	\begin{align} \label{lmmse:proof:13}
	&	\mathbb{E}\left\lbrace \left| \mathbf{v}^H_k \mathbf{A}^H_k {\mathbf{h}}_{l} \right|^2 \right\rbrace= \sum_{m=1}^{M}  |\alpha_{m,k}|^2 \mathbb{E}\left\lbrace \left|    v^*_{m,k}   {h}_{m,l} \right|^2 \right\rbrace  + \mathop{ \sum_{m=1}^{M} \sum_{n=1}^{M}}_{m \neq n} \alpha_{m,k}\alpha^*_{n,k}  \mathbb{E}\left\lbrace \left(   v^*_{m,k}  {h}_{m,l} \right)^* \left(   v^*_{n,k}  {h}_{n,l} \right)\right\rbrace  \\
	&=  \sum_{m=1}^{M}  \hat{p}_{k}  \tau_p |\alpha_{m,k}|^2  \frac{\beta'_{m,l} (\beta'_{m,k})^2}{\lambda'_{m,k}}  + \begin{cases}
	\displaystyle \sum_{m=1}^{M}  |\alpha_{m,k}|^2 2 \hat{p}_{k}  \hat{p}_{l} \tau^2_p (\beta'_{m,k})^2  (\lambda'_{m,k})^{-2} \bar{h}^2_{m,l} \beta_{m,l} + |\alpha_{m,k}|^2 \hat{p}_{k}  \hat{p}_{l} \tau^2_p (\beta'_{m,k})^2  (\lambda'_{m,k})^{-2} \beta^2_{m,l} \\
	+ \displaystyle\mathop{\sum_{m=1}^{M} \sum_{n=1}^{M}}_{m \neq n}\alpha_{m,k}\alpha^*_{n,k} 
	\hat{p}_{k}\hat{p}_{l} \tau^2_p \beta'_{m,k} \beta'_{m,l} (\lambda'_{m,k})^{-1 }\beta'_{n,k} \beta'_{n,l} (\lambda'_{n,k})^{-1 }, &  l \in \mathcal{P}_{k} \\
	0, & l \notin \mathcal{P}_{k} .
	\end{cases}\nonumber
	\end{align}	
	\hrulefill
	\setcounter{equation}{131}
	\begin{align}\label{wide4}
	\sum_{n=1}^{M} \mathrm{SE}^\mathrm{dl}_{n,k} = \frac{\tau_d}{\tau_c}\log_2 \left( \prod_{n=1}^{M} \left( 1 + \gamma^\mathrm{dl}_{n,k}\right)  \right)  =\frac{\tau_d}{\tau_c}\log_2 \left( \prod_{n=1}^{M}  \frac{\displaystyle\sum_{n=1}^{M} \sum_{l=1 }^{K}\rho_{n,l}  \mathbb{E}\left\lbrace\left|h^*_{n,k} w_{n,l} \right|^2 \right\rbrace  - \sum_{n=1}^{m-1} \rho_{n,k} \left| \mathbb{E}\left\lbrace h^*_{n,k} w_{n,k} \right\rbrace \right| ^2 + \sigma^2_{\mathrm{dl}} }{ \displaystyle \sum_{n=1}^{M} \sum_{l=1 }^{K}\rho_{n,l} \mathbb{E}\left\lbrace | h^*_{n,k} w_{n,l}|^2 \right\rbrace  - \sum_{n=1}^{m} \rho_{n,k} \left| \mathbb{E}\left\lbrace h^*_{n,k} w_{n,k} \right\rbrace \right| ^2 + \sigma^2_{\mathrm{dl}}  }\right),
	\end{align}
	\begin{equation}\label{wide5}
	\mathrm{SE}^\mathrm{dl}_k =\frac{\tau_d}{\tau_c}\log_2 \left(  \frac{\displaystyle\sum_{n=1}^{M} \sum_{l=1 }^{K}\rho_{n,l}  \mathbb{E}\left\lbrace\left|h^*_{n,k} w_{n,l} \right|^2 \right\rbrace  + \sigma^2_{\mathrm{dl}} }{ \displaystyle \sum_{n=1}^{M} \sum_{l=1 }^{K}\rho_{n,l} \mathbb{E}\left\lbrace | h^*_{n,k} w_{n,l}|^2 \right\rbrace  - \sum_{n=1}^{M} \rho_{n,k} \left| \mathbb{E}\left\lbrace h^*_{n,k} w_{n,k} \right\rbrace \right| ^2 + \sigma^2_{\mathrm{dl}}  }\right).
	\end{equation}
	\setcounter{equation}{119}
	\hrulefill	
\end{figure*}

\begin{align}\label{lmmse:proof:14}
&	\mathbb{E}\left\lbrace \left| \mathbf{v}^H_k \mathbf{A}^H_k {\mathbf{h}}_{l} \right|^2 \right\rbrace = \hat{p}_{k}  \tau_p \mathrm{tr}\left( \mathbf{A}^H_{k} \mathbf{R}'_{l} \mathbf{A}_{k} \boldsymbol{\Omega}'_{k} \right)  \\
	& +\hat{p}_{k} \hat{p}_{l}  \tau^2_p \begin{cases} 
\mathrm{tr}\left( \mathbf{A}^H_{k}\mathbf{R}^2_{l} \boldsymbol{\Lambda}'_{k} \boldsymbol{\Omega}'_{k} \mathbf{A}_{k} \right) + 2  \mathrm{tr}\left( \mathbf{A}^H_{k} \boldsymbol{\Lambda}'_{k} \boldsymbol{\Omega}'_{k} \mathbf{L}_{l}\mathbf{R}_{l}\mathbf{A}_{k} \right)\\ 
 \left| \mathrm{tr}\left(   \mathbf{A}_k\mathbf{R}'_{l}  \boldsymbol{\Lambda}'_{k} \mathbf{R}'_{k}\right) \right|^2 -  \mathrm{tr}\left(   \mathbf{A}^H_k(\mathbf{R}'_{l}  \boldsymbol{\Lambda}'_{k} \mathbf{R}'_{k})^2  \mathbf{A}_k\right)&  l \in \mathcal{P}_{k} \\
0 & l \notin \mathcal{P}_{k} 
\end{cases}\nonumber
\end{align}
and this finishes the proof. 

\section{Proof of UL and DL SE with the LS Estimator }\label{AppendixUL_LS}
The first two terms to calculate are $
\mathbb{E}\left\lbrace {v}^*_{m,k}  h_{m,k}\right\rbrace =  \frac{\mathbb{E}\left\lbrace (y^p_{m,k})^* h_{m,k} \right\rbrace}{\sqrt{\hat{p}_k} \tau_p} =  \beta'_{m,k}$ and $\mathbb{E}\left\lbrace |{v}_{m,k}|^2  \right\rbrace =\frac{\mathbb{E}\left\lbrace |y^p_{m,k}|^2\right\rbrace}{\hat{p}_k \tau^2_p}  =\frac{\lambda'_{m,k}}{\hat{p}_k \tau_p}$  as given in Appendix \ref{AppendixLS}. Moreover, we calculate
\begin{align}
&\mathbb{E}\left\lbrace \mathbf{v}^H_k \mathbf{A}^H_k {\mathbf{h}}_k\right\rbrace \nonumber  \\
& =\sum_{m=1}^{M} \alpha^*_{m,k}\mathbb{E}\left\lbrace {v}^*_{m,k}  {h}_{m,k}\right\rbrace =  \sum_{m=1}^{M} \alpha^*_{m,k} \beta'_{m,k}  =  \mathrm{tr}\left(\mathbf{A}_k\mathbf{R}'_{k}  \right). \\
&\mathbb{E}\left\lbrace\|\mathbf{A}_k\mathbf{v}_k\|^2\right\rbrace = \sum_{m=1}^{M}	|  \alpha^*_{m,k}| ^2 \mathbb{E}\left\lbrace |  {v}^*_{m,k} |^2 \right\rbrace  = \frac{\mathrm{tr}\left( \mathbf{A}^H_k \left( \boldsymbol{\Lambda}'_{k}\right)^{-1}  \mathbf{A}_k \right)}{\hat{p}_k \tau_p} .
\end{align}

The third term is $	\mathbb{E}\left\lbrace \left| \mathbf{v}^H_k \mathbf{A}^H_k {\mathbf{h}}_{l} \right|^2 \right\rbrace$ and can be expanded as
\begin{align}
&\mathbb{E}\left\lbrace \left|  \sum_{m=1}^{M} \alpha^*_{m,k} {v}^*_{m,k}  {h}_{m,l} \right| ^2 \right\rbrace \nonumber\\
&=\sum_{m=1}^{M} \sum_{n=1}^{M} \alpha_{m,k}\alpha^*_{n,k}  \mathbb{E}\left\lbrace \left(  {v}^*_{m,k}  {h}_{m,l} \right)^* \left(  {v}^*_{n,k}  {h}_{n,l} \right)\right\rbrace. 
\end{align}
Begin with $m=n$ and $l \in \mathcal{P}_k $ and similar to \eqref{lmmse:proof:11}, we obtain
\begin{align}
&\mathbb{E}\left\lbrace  \left| {v}^*_{m,k}  {h}_{m,l} \right|^2 \right\rbrace =  \frac{1}{\hat{p}_k \tau^2_p}\mathbb{E}\left\lbrace \left| y^p_{m,k} {h}_{m,l}  \right|^2 \right\rbrace \nonumber \\
&=  \frac{1}{\hat{p}_k \tau^2_p} \left[ \hat{p}_{l} \tau^2_p \left( \beta^2_{m,l} + 2\bar{h}^2_{m,l} \beta_{m,l} \right) + \tau_p \lambda'_{m,k} \beta'_{m,l}  \right].
\end{align}

For  $m =n$ and $l \notin \mathcal{P}_k$, the $y^p_{m,k}$ and ${h}_{m,l} $ are independent random variables. We can compute directly
\begin{align}
&\mathbb{E}\left\lbrace  \left| {v}^*_{m,k}  {h}_{m,l} \right|^2 \right\rbrace =  \frac{1}{\hat{p}_k \tau^2_p}\mathbb{E}\left\lbrace \left| (y^p_{m,k})^* {h}_{m,l}  \right|^2 \right\rbrace  \nonumber \\
&=  \frac{1}{\hat{p}_k \tau^2_p}\mathbb{E}\left\lbrace \left| y^p_{m,k} \right|^2 \right\rbrace  \mathbb{E}\left\lbrace \left| {h}_{m,l}  \right|^2 \right\rbrace =\frac{1}{\hat{p}_k \tau_p}  \lambda'_{m,k} \beta'_{m,l}.
\end{align}
For  $m \neq n$ and $l \in \mathcal{P}_k$, we have $\mathbb{E}\left\lbrace \left(  {v}^*_{m,k}  {h}_{m,l} \right)^* \left(  {v}^*_{n,k}  {h}_{n,l} \right)\right\rbrace  = \frac{1}{\hat{p}_k \tau^2_p} \mathbb{E}\left\lbrace  (y^p_{m,k})^* {h}_{m,l}   \right\rbrace \mathbb{E}\left\lbrace  (y^p_{n,k})^* {h}_{n,l}   \right\rbrace =  \frac{\hat{p}_{l}}{\hat{p}_k } \beta'_{m,l} \beta'_{n,l}$
and finally for  $m \neq n$ and $l \notin \mathcal{P}_k$ $\mathbb{E}\left\lbrace \left(  {v}^*_{m,k}  {h}_{m,l} \right)^* \left(  {v}^*_{n,k}  {h}_{n,l} \right)\right\rbrace= 0 $ since all the terms are mutually independent and have zero mean. The final form is 
\begin{align}
&\mathbb{E}\left\lbrace \left|  \sum_{m=1}^{M} \alpha^*_{m,k} {v}^*_{m,k}  {h}_{m,l} \right| ^2 \right\rbrace = \sum_{m=1}^{M} |\alpha_{m,k}|^2 \frac{\lambda'_{m,k}\beta'_{m,l}}{\hat{p}_k \tau_p} \nonumber \\
&+ \frac{\hat{p}_{l}}{\hat{p}_k }\begin{cases} \sum_{m=1}^{M} |\alpha_{m,k}|^2  \left( \beta^2_{m,l} + 2\bar{h}^2_{m,l} \beta_{m,l} \right) \\ + \displaystyle
\mathop{\sum_{m=1}^{M} \sum_{n=1}^{M}}_{ m \neq n} \alpha_{m,k}\alpha^*_{n,k}  \beta'_{m,l} \beta'_{n,l}, & l \in \mathcal{P}_k  \\
0, &  l \notin \mathcal{P}_k. 
\end{cases}
\end{align}
In matrix form, we obtain
\begin{align}
&\mathbb{E}\left\lbrace \left| \mathbf{v}^H_k \mathbf{A}^H_k {\mathbf{h}}_{l} \right|^2 \right\rbrace = \frac{1}{\hat{p}_k \tau_p} \mathrm{tr}\left( \mathbf{A}^H_{k} (\boldsymbol{\Lambda}'_{k})^{-1} \mathbf{R}'_{l}\mathbf{A}_{k}\right) + \nonumber \\ &\frac{\hat{p}_{l}}{\hat{p}_k } \begin{cases}
\mathrm{tr}\left( \mathbf{A}^H_{k} (\mathbf{R}^2_{l} + 2\mathbf{L}_{l}\mathbf{R}_{l}) 
 \mathbf{A}_{k}\right) \\+\mathrm{tr}\left( \mathbf{A}^H_{k} \mathbf{R}'_{l}\right)^2 - \mathrm{tr}\left( \mathbf{A}^H_{k} (\mathbf{R}'_{l})^2\mathbf{A}_{k} \right), &  l \in \mathcal{P}_k \\
0, &  l \notin \mathcal{P}_k.
\end{cases}
\end{align}

\section{Proof of DL SE with Non-coherent Transmission}\label{AppendixDL}

 At the beginning of the detection process, UE $k$  does not know any of the transmitted signals. It first detects the signal from AP~$1$ by using the average channel $\mathbb{E}\left\lbrace h^*_{1,k} w_{1,k}\right\rbrace $ only. The received signal can be written as 
\begin{align} 
&y^\mathrm{dl}_{1,k} =y^\mathrm{dl}_{k}= \mathbb{E}\left\lbrace h^*_{1,k} w_{1,k} \right\rbrace \varsigma_{1,k}  + \left( h^*_{1,k} w_{1,k} - \mathbb{E}\left\lbrace h^*_{1,k} w_{1,k} \right\rbrace\right)  \varsigma_{1,k} \nonumber \\
&+   \sum_{n=2}^{M}h^*_{n,k} w_{n,k} \varsigma_{n,k} + \ \mathop{\sum_{l=1 }}^{K}_{l \neq k} \sum_{n=1}^{M}h^*_{n,k} w_{n,l} \varsigma_{n,l} + {n}^{\mathrm{dl}}_{k}.
\end{align}
	where the first term is the desired signal over known deterministic channel while other terms are treated as uncorrelated noise. Sequentially, UE $k$ detects signal from AP $m$ by subtracting the first $m-1$ signals:
\begin{align}\label{y_mk}
&y^\mathrm{dl}_{m,k} =y^\mathrm{dl}_{k} - \sum_{n=1}^{m-1}\mathbb{E}\left\lbrace h^*_{n,k} w_{n,k} \right\rbrace \varsigma_{n,k}= \mathbb{E}\left\lbrace h^*_{m,k} w_{m,k} \right\rbrace \varsigma_{m,k}  \nonumber \\
&+\left( h^*_{m,k} w_{m,k} - \mathbb{E}\left\lbrace h^*_{m,k} w_{m,k} \right\rbrace\right)  \varsigma_{m,k} + \mathop{\sum_{l=1 }}^{K}_{l \neq k} \sum_{n=1}^{M}h^*_{n,k} w_{n,l} \varsigma_{n,l} + {n}^{\mathrm{dl}}_{k}\nonumber \\
&+ \sum_{n=1}^{m-1}\left( h^*_{n,k} w_{n,k} - \mathbb{E}\left\lbrace h^*_{n,k} w_{n,k} \right\rbrace\right)  \varsigma_{n,k}+  \sum_{n=m+1}^{M}h^*_{n,k} w_{n,k} \varsigma_{n,k}.  
\end{align}
The first term in \eqref{y_mk} is equivalent to having a deterministic channel $h=\mathbb{E}\left\lbrace h^*_{m,k} w_{m,k} \right\rbrace$   and $\varsigma_{m,k}$ is the desired signal. The other terms are uncorrelated noise $\upsilon_{m,k}$ with power
\begin{align}
&\mathbb{E}\left\lbrace |\upsilon_{m,k}|^2\right\rbrace = \rho_{m,k} \mathbb{E}\left\lbrace |   h^*_{m,k} w_{m,k} - \mathbb{E}\left\lbrace h^*_{m,k} w_{m,k} \right\rbrace  |^2 \right\rbrace   \\
&+ \sum_{n=1}^{m-1} \rho_{n,k}\mathbb{E}\left\lbrace| h^*_{n,k} w_{n,k} - \mathbb{E}\left\lbrace  h^*_{n,k} w_{n,k} \right\rbrace |^2  \right\rbrace + \sigma^2_\mathrm{dl} \nonumber \\
&  +  \sum_{n=m+1}^{M}   \rho_{n,k} \mathbb{E}\left\lbrace |h^*_{n,k} w_{n,k} |^2 \right\rbrace +  \mathop{\sum_{l=1 }}^{K}_{l \neq k} \sum_{n=1}^{M}  \rho_{n,l} \mathbb{E}\left\lbrace| h^*_{n,k} w_{n,l}|^2\right\rbrace  \nonumber \\
&=\displaystyle \sum_{n=1}^{M} \sum_{l=1 }^{K}\rho_{n,l} \mathbb{E}\left\lbrace | h^*_{n,k} w_{n,l}|^2 \right\rbrace  - \sum_{n=1}^{m} \rho_{n,k} \left| \mathbb{E}\left\lbrace h^*_{n,k} w_{n,k} \right\rbrace \right| ^2 + \sigma^2_{\mathrm{dl}}. \nonumber
\end{align}
The DL SINR $\gamma^\mathrm{dl}_{m,k}= \frac{ |h|^2 }{\mathbb{E}\left\lbrace |\upsilon_{m,k}|^2\right\rbrace}$ is equal to
\begin{equation}
\gamma^\mathrm{dl}_{m,k} = \frac{\rho_{m,k} \left| \mathbb{E}\left\lbrace h^*_{m,k} w_{m,k} \right\rbrace \right| ^2}{ \displaystyle \sum_{n=1}^{M} \sum_{l=1 }^{K}\rho_{n,l} \mathbb{E}\left\lbrace | h^*_{n,k} w_{n,l}|^2 \right\rbrace  - \sum_{n=1}^{m} \rho_{n,k} \left| \mathbb{E}\left\lbrace h^*_{n,k} w_{n,k} \right\rbrace \right| ^2 + \sigma^2_{\mathrm{dl}}}.
\end{equation}
The total SE of UE~$k$ is $\mathrm{SE}^\mathrm{dl}_k$ is equal to \eqref{wide4}, at the top of previous page. After cancellations of terms that appear in both the numerator and denominator, we obtain \eqref{wide5}, at the top of previous page and it is equal to \eqref{DLSINR_k}.

\bibliographystyle{IEEEtran}
\bibliography{IEEEabrv,referenceCellFree}

\begin{IEEEbiography}[{\includegraphics[width=1in,height=1.25in,clip,keepaspectratio]{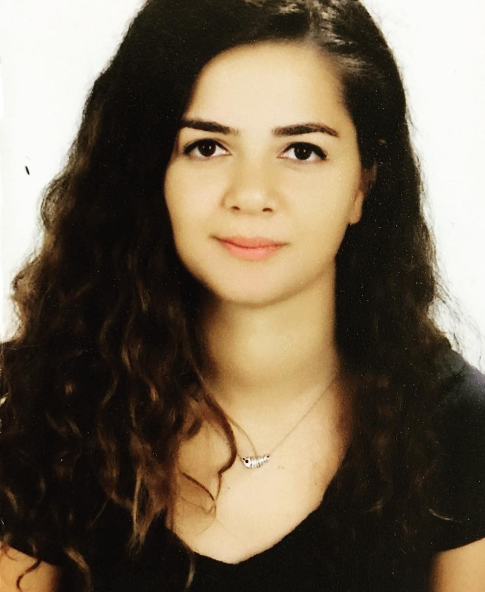}}]{\"Ozgecan \"Ozdogan}
	(S’18) received her B.Sc and M.Sc. degrees in Electronics and Communication Engineering from İzmir Institute of Technology, Turkey in
	2015 and 2017 respectively. She is currently pursuing the Ph.D. degree in communication systems at Link\"oping University, Sweden.
\end{IEEEbiography}
\vfill
\begin{IEEEbiography}[{\includegraphics[width=1in,height=1.25in,clip,keepaspectratio]{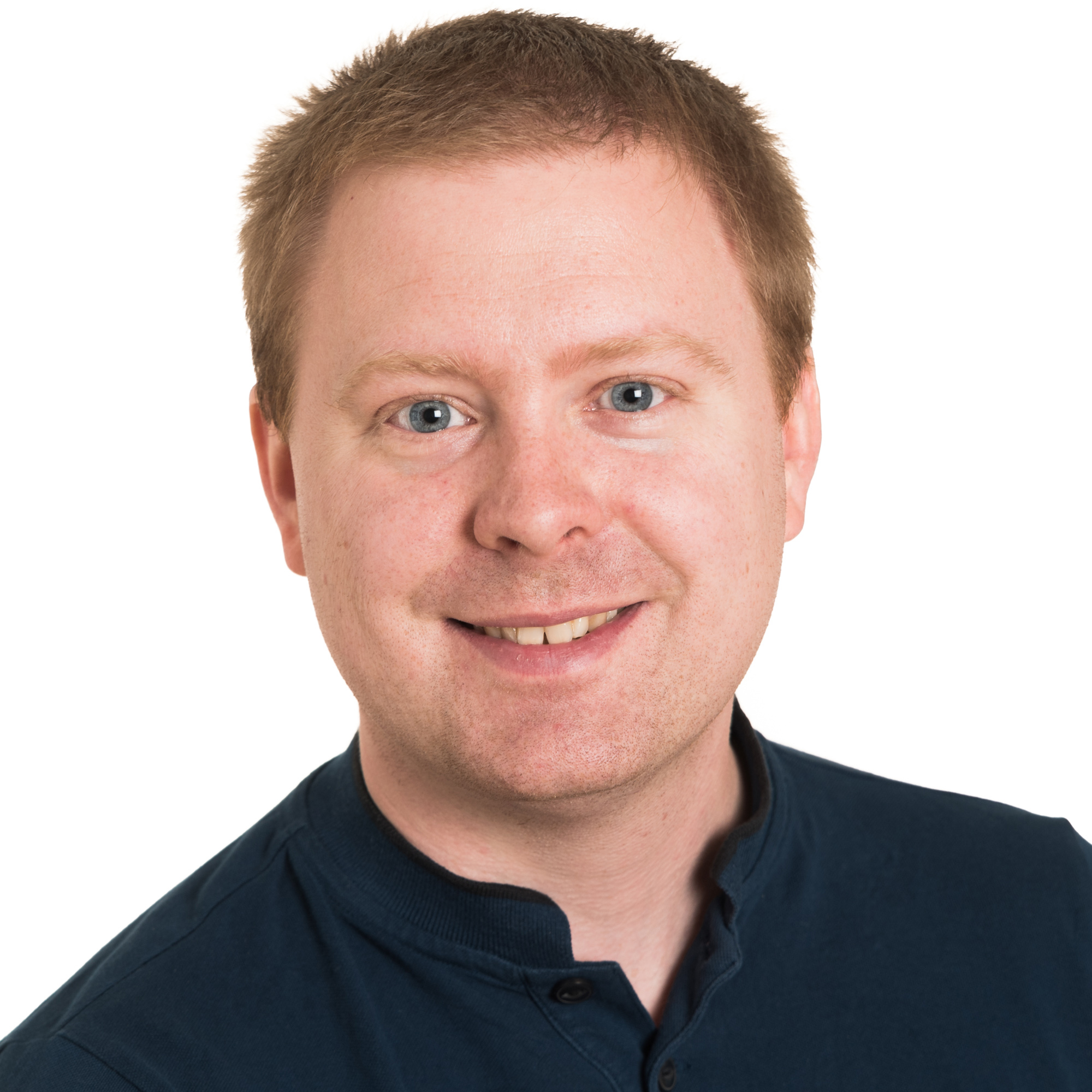}}]{Emil Bj\"ornson }Emil Bj\"ornson (S'07-M'12-SM'17) received the M.S. degree in engineering mathematics from Lund University, Sweden, in 2007, and the Ph.D. degree in telecommunications from the KTH Royal Institute of Technology, Sweden, in 2011. From 2012 to 2014, he held a joint post-doctoral position at the Alcatel-Lucent Chair on Flexible Radio, SUPELEC, France, and the KTH Royal Institute of Technology. He joined Link\"oping University, Sweden, in 2014, where he is currently an Associate Professor and a Docent with the Division of Communication Systems.
	
	He has authored the textbooks \emph{Optimal Resource Allocation in Coordinated Multi-Cell Systems} (2013) and \emph{Massive MIMO Networks: Spectral, Energy, and Hardware Efficiency} (2017). He is dedicated to reproducible research and has made a large amount of simulation code publicly available. He performs research on MIMO communications, radio resource allocation, machine learning for communications, and energy efficiency. Since 2017, he has been on the Editorial Board of the IEEE TRANSACTIONS ON COMMUNICATIONS and the IEEE TRANSACTIONS ON GREEN COMMUNICATIONS AND NETWORKING since 2016.
	
	He has performed MIMO research for over ten years and has filed more than ten MIMO related patent applications. He has received the 2014 Outstanding Young Researcher Award from IEEE ComSoc EMEA, the 2015 Ingvar Carlsson Award, the 2016 Best Ph.D. Award from EURASIP, the 2018 IEEE Marconi Prize Paper Award in Wireless Communications, the 2019 EURASIP Early Career Award, and the 2019 IEEE Communications Society Fred W. Ellersick Prize. He also co-authored papers that received Best Paper Awards at the conferences, including WCSP 2009, the IEEE CAMSAP 2011, the IEEE WCNC 2014, the IEEE ICC 2015, WCSP 2017, and the IEEE SAM 2014.
\end{IEEEbiography}
\vfill

\begin{IEEEbiography}[{\includegraphics[width=1in,height=1.25in,clip,keepaspectratio]{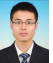}}]
	{Jiayi Zhang}(S'08--M'14) received the B.Sc. and Ph.D. degree of Communication Engineering from Beijing Jiaotong University, China in 2007 and 2014, respectively. Since 2016, he has been a Professor with School of Electronic and Information Engineering, Beijing Jiaotong University, China. From 2014 to 2016, he was a Postdoctoral Research Associate with the Department of Electronic Engineering, Tsinghua University, China. From 2014 to 2015, he was also a Humboldt Research Fellow in Institute for Digital Communications, University of Erlangen-Nuermberg, Germany. From 2012 to 2013, he was a visiting Ph.D. student at the Wireless Group, University of Southampton, United Kingdom. His current research interests include massive MIMO, cell-free massive MIMO, and performance analysis of generalized fading channels.
	
	He was recognized as an exemplary reviewer of the IEEE COMMUNICATIONS LETTERS in 2015 and 2016. He was also recognized as an exemplary reviewer of the IEEE TRANSACTIONS ON COMMUNICATIONS in 2017 and 2018. He is the leading guest editor of IEEE JOURNAL Of SELECTED AREA In COMMUNICATIONS, and serves as an Associate Editor for IEEE TRANSACTIONS ON COMMUNICATIONS, IEEE COMMUNICATIONS LETTERS and IEEE ACCESS. He received the WCSP and IEEE APCC Best Paper Award in 2017.
\end{IEEEbiography}

\end{document}